\documentclass[manuscript]{acmart}
\usepackage{booktabs} 
\usepackage{tikz}
\usepackage{booktabs} 
\usepackage{float}
\usepackage{multirow}
\usepackage{lipsum}
\usepackage{graphicx}
\usepackage{csquotes}
\usepackage{array}
\usepackage{xcolor}
\usepackage{soul}
\usepackage{multicol}
\usepackage[final]{pdfpages}
\newcolumntype{P}[1]{>{\centering\arraybackslash}p{#1}}
\newcolumntype{M}[1]{>{\centering\arraybackslash}m{#1}}
\usepackage{subfiles}
\usepackage{blindtext}
\usepackage{wrapfig}
\usepackage[utf8]{inputenc}
\usepackage{subcaption}
\usepackage{xfrac}
\usepackage{multicol}
\newcommand{\ignore}[1]{}
\definecolor{pp}{HTML}{5731ef} 

\newcolumntype{P}[1]{>{\centering\arraybackslash}p{#1}}
\newcolumntype{M}[1]{>{\centering\arraybackslash}m{#1}}
\definecolor{cerulean}{HTML}{0040FF} 

\AtBeginDocument{%
  \providecommand\BibTeX{{%
    \normalfont B\kern-0.5em{\scshape i\kern-0.25em b}\kern-0.8em\TeX}}}

\begin{document}
\setlength{\tabcolsep}{2pt}
\title[Generative AI for Monetization]{Monetizing Generative AI: YouTubers' Collective Knowledge on Earning from Generative AI Content}

\author{Shuo Niu}
\email{shniu@clarku.edu}
\orcid{https://orcid.org/0000-0002-8316-4785}
\affiliation{%
  \institution{Clark University}
  \streetaddress{950 Main St.}
  \city{Worcester}
  \state{MA}
  \country{USA}
  \postcode{01610}
}

\author{Yao Lyu}
\orcid{0000-0003-3962-4868}
\affiliation{%
  \institution{University of Michigan}
  \city{Ann Arbor}
  \state{Michigan}
  \country{USA}
}
\email{yaolyu@psu.edu}

\author{He Zhang}
\affiliation{%
  \institution{Pennsylvania State University}
  \city{University Park}
  \country{USA}}
\email{hpz5211@psu.edu}
\orcid{0000-0002-8169-1653}

\author{Na Li}
\affiliation{%
  \institution{Pennsylvania State University}
  \city{University Park}
  \country{USA}}
\email{nzl5264@psu.edu}

\author{Bumjin Kim}
\affiliation{%
  \institution{Pennsylvania State University}
  \city{University Park}
  \country{USA}}
\email{bqk5313@psu.edu}

\author{Jie Cai}
\orcid{0000-0002-0582-555X}
\affiliation{%
  \institution{Tsinghua University}
  \city{Beijing}
  \state{Beijing}
  \country{China}
}
\email{jie.cai@psu.edu}

\renewcommand{\shortauthors}{Niu, et al.}

\begin{abstract}
Generative Artificial Intelligence (GenAI) is reshaping creative labor by enabling the rapid production of text, images, and videos. On YouTube, creators are developing new ways to leverage these tools and share knowledge about how to pursue income through such strategies. However, little is known about what GenAI knowledge has been collectively constructed around monetizing GenAI as a community practice of acting both with and against algorithmically mediated platforms. We analyze 377 YouTube videos in which creators publicly promote workflows, revenue claims, and monetization strategies for GenAI-enabled content. Our analysis identifies ten shared use cases that frame AI-supported income opportunities, and examines how this GenAI knowledge repository embodies a collective effort to leverage platform infrastructures for monetization -- including advertising, direct sales, affiliate marketing, and revenue-sharing models. We further surface structural tensions in AI-mediated creative labor, including unverifiable income claims, content misappropriation, synthetic engagement practices, and shifting authorship norms. We conceptualize creators' collective understanding and adoption of GenAI in the context of monetizing creative labor, with implications for the design of creator-centered GenAI technologies and responsible platform policy.

\end{abstract}

\ccsdesc[500]{Human-centered computing~Empirical studies in collaborative and social computing}

\keywords{Generative AI; AI ethics; YouTube; content creator; monetization; affiliate marketing; advertisement; transaction; subscription}

\maketitle

\section{Introduction}
Generative Artificial Intelligence (GenAI), including large language models (LLMs) and multimedia generation tools \cite{Epstein2023GenAI, Gamage2022Deepfake}, is increasingly adopted by content creators. AI-generated content significantly influences creative work, social connections within communities, and platform and market design for sustaining creators' income. Content creators are increasingly sharing their GenAI tactics and strategies publicly, forming a body of collective knowledge about how to adopt these tools in \textit{creative labor}~\cite{Simpson2023CreativeWork} and increase revenue. While prior Human-Computer Interaction (HCI) and social computing research has primarily examined the practices related to visibility work~\cite{Bishop2019Managing, Bishop2020Algorithmic, Emerson2024Shared, Eschebach2025Playing, Verviebe2026theAlgorithm} and the use of GenAI in the creative process \cite{Hua2024GenAIUGC, Lyu2024GenAI, Anderson2025AIVideo, Sun2024CreativeWorker, Epstein2023GenAI}, the emerging community-framed practices around adopting GenAI as a tool for managing creative labor and navigating governance policies remain underexplored.

\par
On the one hand, GenAI content appears across platforms such as YouTube, TikTok, Instagram, Medium, and WordPress~\cite{Lyu2024GenAI, Sun2024CreativeWorker, Hua2024GenAIUGC, Woodruff2024GenAI, Min2024GenAI}. GenAI technologies can seamlessly convert text into a wide range of outputs, including new texts, images, audio/speech, and videos \cite{Bandi2023PowerGenAI}, offering the potential to enhance both productivity and creativity. On the other hand, GenAI content often lacks human connection \cite{Halperin2025AISoulless}, and its synthetic realism can be used to deceive audiences \cite{Satra2023GenAI} or misappropriate copyrighted content \cite{Goetze2024ATheft}.
\par
HCI and social computing research has examined how content creators collectively translate scattered experiences into shared knowledge, including the management of creative labor and their informal theories of visibility work when navigating recommendation and monetization systems~\cite{Weber2021KindArt, Chen2023MyCulture, Emerson2024Shared, Eschebach2025Playing, Bishop2020Algorithmic, MacDonald2021AlgorithmicLore, Reynolds2024User, Cotter2024Practical}. Analyzing such public discourses helps inform platform and market design under algorithmic mediation, which shapes creators' visibility and financial security~\cite{Kojah2025CreativeLabor}. Creators typically earn income through three main approaches: affiliate marketing, ad revenue sharing, and direct sales of merchandise~\cite{Simpson2023CreativeWork, Weber2021KindArt}. These monetization models are designed to reward and incentivize creative labor that reflects human creativity and originality \cite{KopfMonetization, Ye2024Monetization}. Meanwhile, YouTubers must navigate tensions between maintaining their integrity and generating income~\cite{Eschebach2025Playing}. Researchers have identified exploitative practices aimed at maximizing income, such as account trading, fake engagement, and content theft \cite{Chu2022ExploitativeMoney}. However, with the increasing adoption of GenAI within the content creator community, \textbf{little is known about creator-AI interaction from a social science perspective, particularly regarding how GenAI is publicly framed as a viable monetization pathway and how collective knowledge about GenAI affordances operates within or challenges existing monetization infrastructures.}

\par

This research is motivated by conceptualizing the collective knowledge of \textit{Generative AI for Monetization (GenAI4Money)} -- the publicly taught and circulated strategic uses of generative AI to produce income-generating content. We contribute a content analysis of a specific yet influential segment of the creator ecosystem. Rather than measuring off-screen GenAI interactions, these public demonstration videos represent a shared understanding of GenAI that reveals the emerging workflows and the community's underlying assumptions about using GenAI as a viable path to income. Based on a social science theory~\cite{Hayes2011MonetizatoinModel}, we triangulate the GenAI use cases, monetization models, and risky behaviors discussed in these videos. Through this analysis, we aim to initiate broader conversations in HCI and social computing about community-developed practices surrounding GenAI's role in creative labor and platform infrastructures~\cite{Simpson2023CreativeWork}. Specifically, we address three research questions:

\begin{itemize}
    \item RQ1. What use cases do YouTube videos demonstrate for incorporating GenAI into content creation for monetization?
    
    \item RQ2. How do YouTubers' publicly claimed GenAI4Money use cases align with existing monetization models on social media platforms?
    
    \item RQ3. What challenges emerge from using GenAI in monetizing creative labor, and how are these challenges related to the identified use cases?  
\end{itemize}

We conduct a content analysis of 377 YouTube videos produced by 47 creators, in which they publicly demonstrate the use of GenAI to monetize GenAI content. YouTube is the largest video-sharing platform with 65.3 million creators in 2025\footnote{\url{https://www.demandsage.com/youtube-creator-statistics/}}. These videos exemplify how GenAI4Money strategies function both as forms of self-presentation by YouTubers and as collective knowledge about media practices~\cite{Couldry2024Theorising, Lindlof2017Qualitative}. Analyzing publicly shared videos is a well-adopted methodology for understanding public knowledge formation~\cite{Dezuanni2024BookTok, Emerson2024Shared, Emerson2024Anther, Chen2023MyCulture, NiuTeamTrees} and for examining YouTubers' participation in creative labor and their management of relationships with algorithmic platforms~\cite{Reynolds2024User, MacDonald2021AlgorithmicLore, Cotter2024Practical}. As a result, we develop a conceptual framework that illustrates key GenAI4Money knowledge and its impacts on both platforms and creators. Building on this framework, we discuss three core trade-offs as design and policy implications to guide future HCI and social computing research in this area.

\section{Background}

\subsection{Monetizing Creative Labor}
Content creators on YouTube pursue personal development and self-expression while blending entertainment with aspirations for influence and monetization
~\cite{Burgess2018YouTube}. In social computing research, monetization is viewed as a central motive shaping creators' platform engagement~\cite{Eschebach2025Playing}. Socio-technical research emphasizes how monetization mechanisms, as part of the ``infrastructure for inspiration,'' interact with creative practices~\cite{Yang2024TheFuture, Simpson2025Infrastructures}. On the one hand, creative labor theory argues that algorithmic recommendation and reward programs shape creators' ostensive and performative practices~\cite{Simpson2023CreativeWork}. Creators continually recalibrate monetization expectations and work strategies in response to algorithmic infrastructures~\cite{Choi2023Creator}. On the other hand, creators also cope with and influence platform infrastructures, driving changes in monetization models, moderation policies, and labor workflows~\cite{Lyu2025Systematic, Partin2020BitTwitch}.

\par
To generate revenue, creators refine their production skills and transition into professional uploaders~\cite{DingBilibili}. Influencers earn income through affiliate links, sponsorships, employment, fan support, revenue sharing, deal exchanges, and knowledge monetization~\cite{Weber2021KindArt}. Creators care about their visibility within algorithmic recommendation systems, as audience growth can directly influence revenue opportunities~\cite{Bishop2019Visibility, KarizatAlgorithmicFolk}. Monetization programs reward desirable content, incentivize creators to deliver value, foster entrepreneurial identities, and strengthen ties with platforms~\cite{KopfMonetization, Annabell2025IdealInfluencer}. To maximize these benefits, creators adapt their content and distribute it across multiple platforms to expand monetization opportunities~\cite{Ma2023Creator}. Some creators directly collaborate with affiliate marketers by producing product review videos~\cite{Fitriani2020ReviewVideo}.
\par
Another line of social computing research has examined how monetary incentives can produce negative consequences. Some studies highlight that creators often prioritize financial gain over creative expression~\cite{Choi2023Creator}. The struggle to balance freedom of expression with economic pressure leads to stress, anxiety, and self-blame~\cite{Kojah2025CreativeLabor}. To increase visibility and earnings, creators adapt their content to align with advertisers' preferences~\cite{Choi2023Creator, Caplan2020Tiered, MaAlgorithmicContent}. Some engage in mass production of trending content to maximize revenue~\cite{Soha2016MonetizingMeme}. Others fail to disclose affiliate marketing activities in their videos~\cite{Rieder2023Creator, Mathur2018Endorsements}. Financial incentives have drawn sexual content creators to share explicit material~\cite{Caplan2020Tiered}.
\par
Beyond incentivizing content production, research on social computing infrastructure has identified demonetization (the restriction of content from generating revenue) as a central mechanism of content moderation. Demonetization has been applied to penalize copyright violations~\cite{Fiesler2016Copyright}, and content misaligned with advertisers' preferences may also be excluded from revenue opportunities~\cite{MaAlgorithmicContent}. 
Content moderation research in social computing research notes that such practices are often perceived as punitive and opaque~\cite{Ma2023Moderation}, producing visibility and income inequities~\cite{Dunna2022Demonetization, Caplan2020Tiered}. This frustration can lead to resistance within creator communities~\cite{Alkhatib2019StreetLevel}.
Some creators even exploit monetization systems through harmful content, fake engagement, or content theft~\cite{Chu2022ExploitativeMoney}.
\par 
Monetization on platforms such as YouTube is a double-edged sword: it motivates some creators to increase production while pressuring others to compromise self-expression to satisfy advertisers' preferences. The rise of GenAI introduces new tools that enable creators to experiment with new practices and reshape community heuristics for managing reach and revenue~\cite{Bishop2019Visibility}.

\subsection{Use of Generative AI in Content Creation}
Generative AI (GenAI) refers to \textit{``artificial intelligence systems that can create new content, such as text, images, audio, or video, rather than just analyzing or acting on existing data''} \cite{Brizuela2023GenAIReview}. Tools such as LLMs and diffusion models are capable of producing high-quality digital content, as well as informal and emotional material \cite{Epstein2023GenAI, Gamage2022Deepfake}. GenAI is increasingly adopted to support creativity~\cite{Sun2024CreativeWorker, Yildirim2022AIDesign, Hwang2022GenAI}. Online creators employ GenAI strategies such as workflow customization, cross-tool experimentation, community learning, and ethical self-regulation~\cite{Sun2024CreativeWorker, Epstein2023GenAI}. GenAI has been used to provide inspiration, access video sources, and automate highlight generation~\cite{Kim2024ASVG}. It also facilitates content modality transformation, such as converting text into platform-optimized videos~\cite{Wang2024ReelFramer}.

\par
Research has also examined the ethical challenges of GenAI in content creation and monetization. Nah et al. propose a taxonomy of 16 challenges through which GenAI may negatively impact businesses and creators \cite{Nah2023GenAI}, including the production of harmful or inappropriate content, over-reliance on AI, and copyright violations. Critics argue that AI-generated content often appears mechanical and inauthentic, lacking emotional depth and human connection \cite{Halperin2025AISoulless}. Creators may misuse GenAI to generate harmful or misleading material, and the GenAI itself may contain biases or hallucinations \cite{Satra2023GenAI}. The risk of creators becoming overly dependent on AI threatens to reduce originality and innovation \cite{Kshetri2024GenAIMarketing}. Furthermore, concerns have emerged regarding copyright infringement \cite{Soni2023GenAIMarketing}. For example, GenAI models such as DALL·E, Midjourney, and Stable Diffusion have been criticized for exploiting the labor of human artists~\cite{Goetze2024ATheft}. While these harms have been discussed in relation to their impact on human creators and creative work, there is limited understanding of how monetization demand may actively drive such uses. 

\par
Another thread of research has examined the challenges posed by GenAI from the audience's perspective. Anthropomorphic features may foster parasocial relationships \cite{Sew2025AIInfluencer}, and virtual AI avatars can enhance users' sense of connection \cite{Gamage2022Deepfake, Lu2021VTuber}. However, some users express strong aversion to the realism of AI avatars \cite{Sew2025AIInfluencer}, citing concerns about manipulation and a perceived lack of authenticity \cite{Maeda2024AIParasocial}. Skepticism also persists due to anxieties surrounding misinformation, impersonation, and deceptive representations \cite{Gamage2022Deepfake, McCosker2024Deepfake, Sew2025AIInfluencer, Maeda2024AIParasocial}. Despite these insights, there remains limited understanding of what forms of monetizable GenAI content are emerging as new norms of participation on UGC platforms.

\subsection{Creative Labor and the Infrastructural Influence of Generative AI}
Simpson and Semaan define \textit{creative labor} as \textit{``the professionalization and monetization of social media content, involving not just creation but also strategic navigation of platform demands, audience expectations, and technological systems''}~\cite{Simpson2023CreativeWork}. Contributing knowledge about GenAI4Money through public videos influences platform infrastructures and creative labor in three key aspects~\cite{Simpson2023CreativeWork}: \textit{creative expression}, \textit{professionalization}, and \textit{social connection}.

\par
\textit{Creative expression} refers to the routines involved in producing influential content, which both shape and are shaped by the algorithmic infrastructure that determines creators' visibility~\cite{Simpson2023CreativeWork}. On YouTube, creators must appear authentic~\cite{DingBilibili} and adapt their content to evolving audience tastes~\cite{BartaAuthenticityTikTok, Duffy2019Cultural}, while strategically leveraging the platform as a medium for transferring informal knowledge and hybridizing community practices~\cite{BarthelCollaborativeKnowledgeBuilding, Walker2019AScaled, Chen2023MyCulture}. Creators' communal expressions of creativity, shared through how-to videos and creative products~\cite{Dezuanni2024BookTok, NiuTeamTrees, Emerson2024Anther, Walker2019AScaled}, function as peer-created knowledge artifacts and forms of collective action that shape emerging community norms. These knowledge-sharing videos have demonstrated educational value and constitute environments for informal learning~\cite{Dubovi2019Examining, Dezuanni2024BookTok, Walker2019AScaled}. The creative process -- including planning, production, uploading, and management~\cite{Choi2023Creator, Kim2024ASVG} -- has also been turned into public tutorials that teach creators how to maximize rewards through monetization~\cite{KopfMonetization, Annabell2025IdealInfluencer}. Recent studies further suggest that GenAI may help alleviate aspects of this creative labor~\cite{Choi2023Creator, Kim2024ASVG}.

\par
\textit{Professionalization} refers to creators achieving micro-celebrity status by developing technical and professional expertise~\cite{DingBilibili}. The tiered system that rewards different levels of professionalism is a central component of YouTube's monetization infrastructure~\cite{Postigo2014DigitalLabor}. Creators' professionalization is strongly shaped by algorithmic recommendation systems~\cite{Verviebe2026theAlgorithm}. Creators share collective knowledge about algorithmic experiences, which translates scattered experiences into shared heuristics for managing reach and revenue~\cite{Bishop2019Visibility, Cotter2024Practical}. Experienced creators also share strategies for increasing visibility, helping others reduce algorithmic invisibility~\cite{Bishop2020Algorithmic, MacDonald2021AlgorithmicLore}. These collective interpretations of YouTube's algorithms shape common strategies and concerns regarding platform governance~\cite{Reynolds2024User}. The emergence of GenAI4Money may therefore influence the three key monetization models -- \textit{advertising}, \textit{subscription}, and \textit{transaction}~\cite{Hayes2011MonetizatoinModel}. These infrastructures~\cite{Duffy2021Precarities, Annabell2025IdealInfluencer}, traditionally shaped by algorithmic visibility, are now challenged by rapidly produced AI-generated content and new GenAI-supported strategies for engaging with recommendation algorithms.

\par

\textit{Social connection} in creative labor refers to the effort to connect with others and build community~\cite{Simpson2023CreativeWork}. Creators' learning, synthesizing, and public translation of ideas shape what knowledge is communicated within the community and how it reaches viewers~\cite{Xia2022Millions, Emerson2024Shared}. However, other research has found that platform incentives influence what knowledge is produced and circulated within the community~\cite{Eschebach2025Playing}. Creators must maintain their identities and pursue visibility by cultivating parasocial relationships and securing audience support~\cite{WohnParasocialInteraction, Fitriani2020ReviewVideo}. For creators who provide services, sustaining such connections and managing impressions can impose significant emotional burdens~\cite{Munoz2022Freelance, Foong2020Freelance}. With the integration of GenAI, it is therefore critical to examine how GenAI may translate these practices into problematic forms of content creation (such as copyright infringement and harassment~\cite{ThomasVSPHarassment}), and how such content affects social interactions between creators and fans.

\par
In summary, the rise of GenAI introduces both efficiencies and potential transformations in creative labor. However, comprehensive studies examining how content creators collectively form knowledge about GenAI in creative expression, navigating monetization infrastructures, and building community connections remain largely absent from current HCI and social computing literature. Drawing on the social media monetization models proposed by Hayes and Graybeal~\cite{Hayes2011MonetizatoinModel}, this study examines the collective knowledge shared through GenAI4Money videos, offering a conceptual understanding of how GenAI is promoted by YouTube creators for monetization purposes.

\section{Data Collection and Screening Process}


We examine collective knowledge on GenAI4Money by analyzing YouTube videos created by content creators who specialize in monetizing GenAI content. The analysis serves three main purposes. First, understanding this collective knowledge reveals the symbolic actions surrounding GenAI use and explains how GenAI becomes collectively accepted within socially and culturally embedded media practices~\cite{Couldry2024Theorising, Lindlof2017Qualitative}. Such use cases may influence other creators' content production. Second, publicly shared demonstration videos provide a distinctive basis for understanding the new norms that may emerge around GenAI use and the monetization of such content~\cite{Couldry2024Theorising}. Third, this emerging body of knowledge about GenAI can inform platform governance and encourage improvements in monetization policies that support healthy and ethical uses of GenAI.

\par

\begin{figure}[!h]
  \centering
  \includegraphics[width=0.95\linewidth]{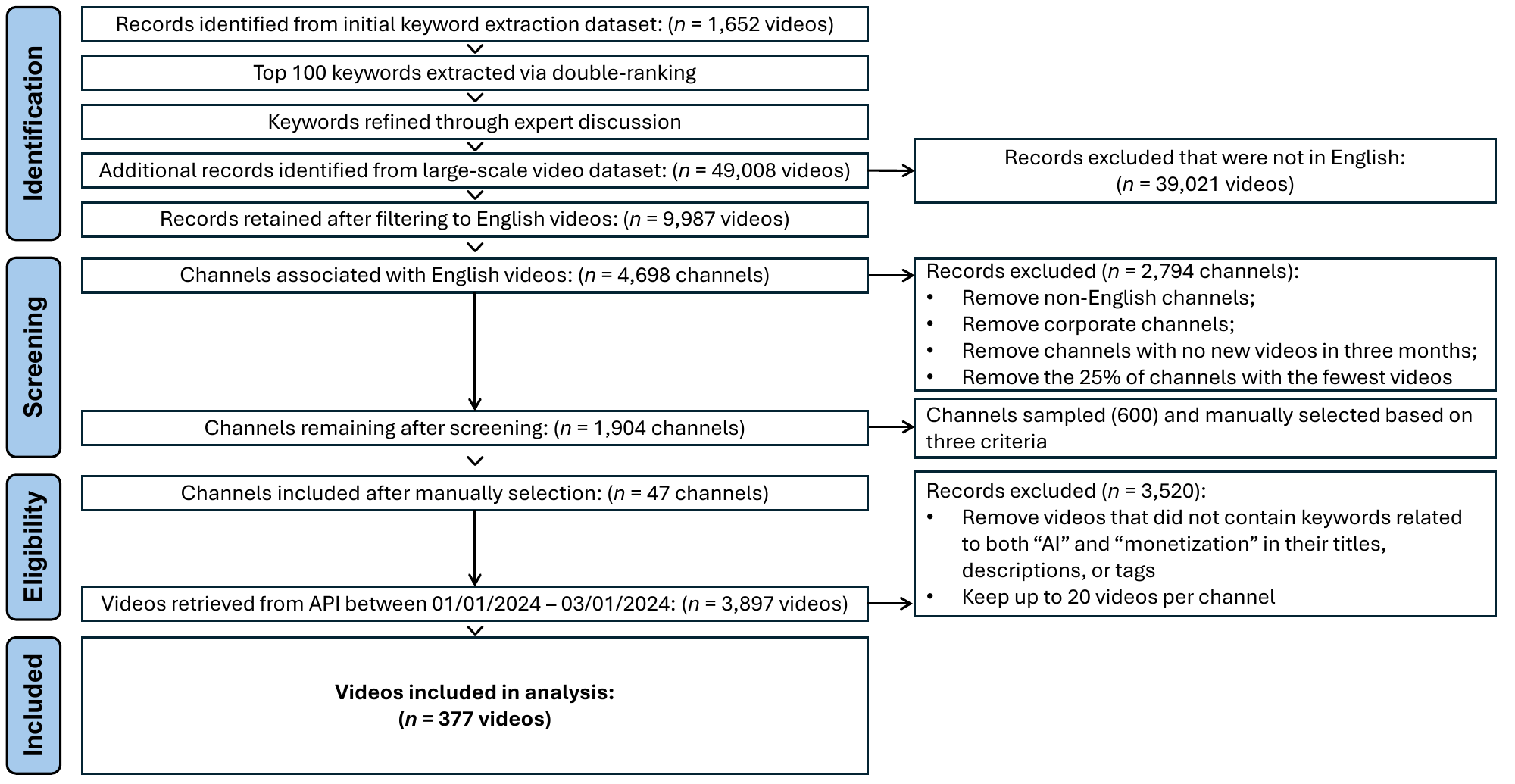}
  \caption{Data Collection and Screening Process Workflow}
  \Description{Workflow Diagram: The figure shows the data collection and screening workflow from initial keyword extraction to final video inclusion. Starting with 1,652 videos and adding 49,008 more, non-English videos are removed, leaving 9,987 English videos across 4,698 channels. After excluding corporate, inactive, and low-volume channels, 1,904 channels remain. From 600 sampled channels, 47 are manually selected. API retrieval yields 3,897 videos, from which those lacking both “AI” and “monetization” keywords and those exceeding 20 per channel are removed. The final dataset includes 377 videos for analysis.}
  \label{fig.data-collection}
\end{figure}

The data collection and filtering process is illustrated in \autoref{fig.data-collection}. This process involved three steps. First, we collected preliminary data to explore relevant video topics and identify effective search keywords. Second, we filtered for channels that primarily focus on GenAI monetization and gathered their videos. Third, we sampled and selected videos that explicitly address strategies for monetizing GenAI content.

\par

In the first step, we used the YouTube Data API\footnote{\url{https://developers.google.com/youtube/v3}} to collect a sample of 1,652 videos to identify common terms related to monetizing AI content. We searched for ``AI Marketing,'' extracted tags of the videos, and applied a double-ranking approach~\cite{Wang2016SearchKeywords} to generate a keyword corpus. Four authors manually screened the top 100 keywords related to Gen AI tools (e.g., ChatGPT, Midjourney) and monetization (e.g., affiliate marketing, passive income). These keywords were then used to search a larger dataset of 49,008 videos from the YouTube API, published between December 1, 2023, and March 1, 2024. Our initial review revealed that, besides tutorials on GenAI monetization, many videos focused on AI news, business promotions of GenAI, general AI tool introductions without monetization, or personal opinions on AI's influence. We also observed that YouTubers tend to focus on similar topics across their channels~\cite{Burgess2018YouTube}. Therefore, the next step aimed to filter videos showcasing monetizing GenAI content from channels specialized in this topic.

\par

In step 2, we extracted the list of channels from the videos in step 1 and programmatically excluded channels with non-English characters in their names or descriptions, as well as company channels such as ``Adobe,'' ``Google Cloud,'' and ``Yahoo Finance.'' We then removed the 25\% of channels with the fewest videos, as well as those without any videos published in the past three months. From the remaining channels, we sampled 600 and manually evaluated them based on the three criteria below. We focus on YouTubers who specialize in GenAI monetization because they are at the leading edge of GenAI adoption. As video-creation skill-sharing is common within the YouTube community~\cite{YouTubeParticipatoryCulture}, leading YouTubers' practices often reveal emerging monetization logics that may later diffuse across broader creator communities. Two authors assessed each channel, resolving any disagreements through discussion. This process yielded 47 channels for step 3. All these channels met the criteria:

\begin{itemize}
    \item[(1)] Independent Content Creators: We included channels operated by independent creators and excluded those affiliated with profit-making organizations, including AI companies, by verifying the channels' profile pages.  
    \item[(2)] Focus on GenAI Content Creation and Monetization: We analyzed the titles, descriptions, and content of both popular and recent videos. Channels were included only if more than half of their videos discussed both GenAI and monetization.  
    \item[(3)] Language Criterion: We manually reviewed video content to confirm that only channels using English were included.  
\end{itemize}

In the third step, we collected all videos published by the selected channels between January 1 and March 1, 2024, totaling 3,897 videos. We applied keyword-based filtering using the compiled list of relevant terms, including only videos whose titles, descriptions, and tags contained keywords related to both ``AI'' and ``monetization.'' We sampled up to 20 videos per channel. Compared to complete random sampling, this approach ensured that we had sufficient videos for analysis while also including content from smaller channels, thereby preventing the over-representation of larger channels. After programmatic filtering, four researchers manually reviewed the videos, including only those that provided instructions for creating AI-generated content (e.g., videos, images) and explicitly mentioned how the content could be monetized. We excluded advertisements for AI tools that lacked walkthroughs or technical details. Each video was annotated by at least two researchers, and any disagreements were resolved through discussion with the four authors. 
\par
Our data collection process yielded 377 videos that form the corpus for our qualitative analysis. The videos were produced by 47 channels, with each channel contributing an average of 2.89 videos ($SD = 1.67$). The average duration is 12.05 minutes. In total, they have accumulated 22,408,488 views, 875,838 likes, and 37,181 comments. All of the data are publicly available, and we did not process any personally identifiable information; therefore, this research was reviewed and exempted by the IRB.

\section{Research Questions and Analysis Method}
Our video analysis focuses on understanding the collective knowledge of GenAI in relation to the three monetization models: \textit{advertisement}, \textit{subscription}, and \textit{transaction}~\cite{Hayes2011MonetizatoinModel}. We begin by examining the use cases publicly shared about creating GenAI content for monetization (RQ1). Building on this foundation, we then investigate how these use cases align with different monetization models (RQ2). To identify the unique challenges posed to content-sharing platforms, we also annotate potential issues that arise when monetizing GenAI content (RQ3).

\subsection{RQ1: Categorize GenAI4Money Use Cases}
RQ1 categorizes the GenAI4Money use cases, identifying what GenAI knowledge has been developed for creative labor and how it is strategically designed to maximize income generation. To address this question, we employed a grounded theory approach~\cite{charmaz2015grounded}, following three stages: open coding, axial coding, and selective coding. In the open coding stage, we analyzed 200 videos (50 per author), taking notes on key concepts related to GenAI usage and the types of content generated. During the axial coding phase, we identified relationships among these notes. Using affinity diagramming, we collaboratively reviewed and grouped the use cases based on similar content types, resulting in 10 distinct categories. In the selective coding phase, we refined these categories through three rounds of annotation (30 videos per round, 90 videos in total) to improve category definitions and calculate inter-rater agreement. The finalized GenAI4Money use cases are summarized in \autoref{tab:Categories}.

\subsection{RQ2: Examine GenAI4Money Monetization Models}

RQ2 examines how GenAI content is suggested for monetization based on Hayes and Graybeal's theory of social media monetization~\cite{Hayes2011MonetizatoinModel}. In this framework, the \textit{advertisement model} generates revenue through ads, banners, or affiliate marketing by directing traffic to advertisers. The \textit{subscription model} charges users for access, often through a freemium structure in which basic content is free while premium features are paywalled. The \textit{transaction model} generates income through direct product or service sales from creators to consumers. In this study, we adopt these definitions and adapt them to the context of YouTube (see \autoref{tab:Categories}) -- specifically, we examine whether GenAI content redirects users to advertisers, increases traffic to creators' channels, or promotes products and services for viewers to purchase.

\subsection{RQ3: Identify GenAI4Money Challenges}
RQ3 investigates the challenges associated with GenAI4Money on content-sharing platforms. During our analysis for RQ1, we also documented challenges related to GenAI knowledge by referencing the 16 issues identified by Nah et al. \cite{Nah2023GenAI}. Although Nah et al.'s framework addresses the broader societal and business implications of GenAI, it must be contextualized within content creation to provide a clearer understanding of the specific challenges faced by creator communities. We adopted a similar coding approach as described in RQ1. We began by taking notes on potential GenAI-related harms observed in the 200 videos, guided by \cite{Nah2023GenAI}, and identifying instances of statements and usage demonstrations that exemplify the harm. We then used affinity diagramming to discuss emerging patterns and inductively develop themes based on recurring patterns that appeared across multiple videos and were pertinent to GenAI4Money. From this analysis, we identified four key GenAI4Money challenges that were prevalent in our dataset: \textit{synthetic human activity}, \textit{misappropriation}, \textit{explicit content}, and \textit{non-verification} (see \autoref{tab:Categories} for definitions).

\par
During independent data annotation, researchers examine the segments in which GenAI use is demonstrated and review the GenAI actions and YouTubers' verbal explanations that constitute evidence for each challenge category. For \textit{synthetic human activity}, we annotate whether the YouTuber showcases AI-generated realistic human figures or language (e.g., comments or reviews) that are generally perceived as human activities. For \textit{misappropriation}, we mark whether the YouTuber directly uses others' textual, image, or video content as AI input to generate shareable content. For \textit{explicit content}, we evaluate whether the YouTuber presents that GenAI can produce violent or sexual material. For \textit{non-verification}, we annotate whether the YouTuber generates an entire knowledge- or information-based text and directly uses it for content creation, without indicating any further validation or fact-checking.

\begin{table}[!h]
\centering
\small
\begin{tabular}{|p{0.02\linewidth}|P{0.19\linewidth}|p{0.75\linewidth}|}
\hline
 & \textbf{Category} & \textbf{Definition} \\ \hline

\multirow{12}{*}{\begin{tabular}[c]{@{}c@{}}\rotatebox{90}{Use Case}\end{tabular}} 
& Blog &
Write or rewrite blogs, microblogs, or online articles with tools like ChatGPT. \\ \cline{2-3}

& E-book & 
Create e-books with GenAI and sell on e-book marketplace like Amazon. \\ \cline{2-3}

& Graphic Design & 
Use tools like Midjourney or Leonardo.AI to create merchandise, stock images, or products (e.g., T-shirts, stickers, cover designs, coloring books, print-on-demand). \\ \cline{2-3}

& Influencer & 
Create a social media account with AI-generated human photos. \\ \cline{2-3}

& Newsletter & 
Write emails, Facebook ads, or newsletters with tools like ChatGPT or Gemini. \\ \cline{2-3}

& Product Promotion Video & 
Create AI videos directly promoting products or services. \\ \cline{2-3}

& SEO & 
Optimize titles, tags, or descriptions with GenAI tools to improve search engine rankings. \\ \cline{2-3}

& Trending Video & 
Create AI videos related to public interests or educational topics. \\ \cline{2-3}

& Video Reformatting & 
Reformat videos for multiple platforms or diversify audiences (e.g., create short-form videos or translate them into different languages). \\ \cline{2-3}

& Web Design & 
Use AI-generated code or templates to design or create websites. \\ \hline

\multirow{3}{*}{\begin{tabular}[c]{@{}c@{}}\rotatebox{90}{Model}\\ \end{tabular}} 
& Advertising & 
The video mentions adding links to GenAI content to earn money from affiliate advertisers. \\ \cline{2-3}

& Subscription & 
The video mentions publishing GenAI content to earn through revenue-sharing programs. \\ \cline{2-3}

& Transaction & 
The video mentions selling AI-generated products or offering freelancing services using GenAI. \\  
\hline

\multirow{6}{*}{\begin{tabular}[c]{@{}c@{}}\rotatebox{90}{Challenge}\end{tabular}} 
& Synthetic Human Activity & 
The video demonstrates using GenAI to create content simulating real people's activities, such as realistic human figures or users' comments and reviews. \\ \cline{2-3}

& Misappropriation & 
The video shows the modification of others' original content (e.g., videos or blogs) in GenAI creations. \\ \cline{2-3}

& \begin{tabular}[c]{@{}c@{}}Explicit Content\end{tabular} & 
The video shows using GenAI to create violent, offensive, or erotic content. \\ \cline{2-3}

& Non-Verification & 
The YouTuber demonstrates using GenAI to generate entire pieces of knowledge or information without further editing or validation. \\ \hline

\end{tabular}%
\caption{Descriptions and definitions of GenAI4Money use cases, models, and challenges.}
\label{tab:Categories}
\end{table}

\subsection{Final Data Annotation}

The finalized codebook is presented in \autoref{tab:Categories}. For all three research questions, we selected all applicable categories for each video. In the final selective coding round involving 30 videos, we calculated inter-rater agreement using Krippendorff's alpha \cite{krippendorff2004reliability} with Jaccard distance. The resulting scores were 0.58 for GenAI4Money use cases (a lower score due to the presence of 10 possible categories), 0.89 for monetization models, and 0.61 for GenAI challenges. After finalizing the codebook, the three authors of this paper independently annotated a total of 287 videos (including recoding the set of 200 videos used for initial exploration and the remaining 87 videos) divided among them.

\section{Results}
\begin{figure*}[htp]
    \centering
    \includegraphics[width=0.75\textwidth]{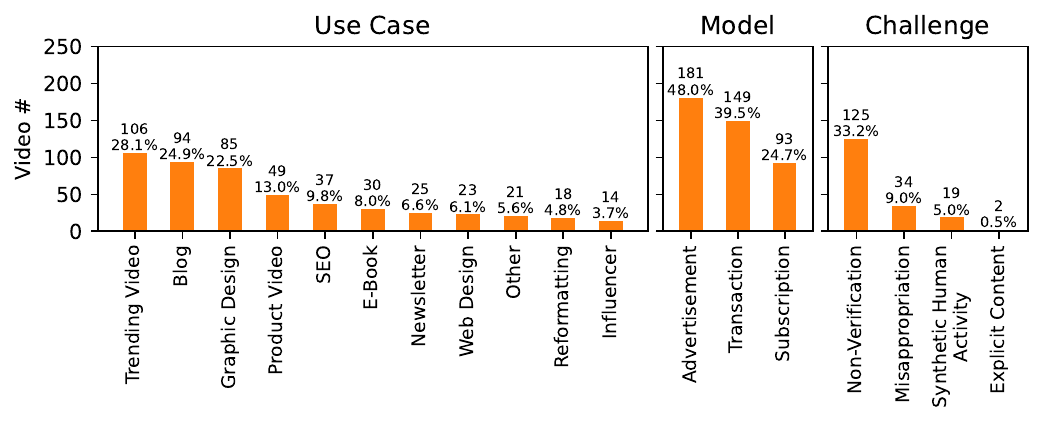}
    \caption{The distribution of YouTube videos in each use case, model, and challenge subcategory.}

    \label{fig:distribution}
    \Description{Distribution Bar Charts: This figure shows a series of bar charts comparing the number and percentage of videos across categories. The first chart presents use cases such as trending video, blog writing, graphic design, promotion videos, SEO, e-books, newsletters, web design, reformatting, and influencer creation. The second chart displays the distribution of monetization models (advertisement, transaction, subscription). The third chart shows challenges like non-verification, misappropriation, synthetic human activity, and explicit content. Together, the bars emphasize which categories appear most frequently.}
\end{figure*}

\begin{figure}[!h]
\centering
    \begin{tabular}{ll}
        \begin{subfigure}[t]{.36\textwidth}
        \includegraphics[width=\textwidth]{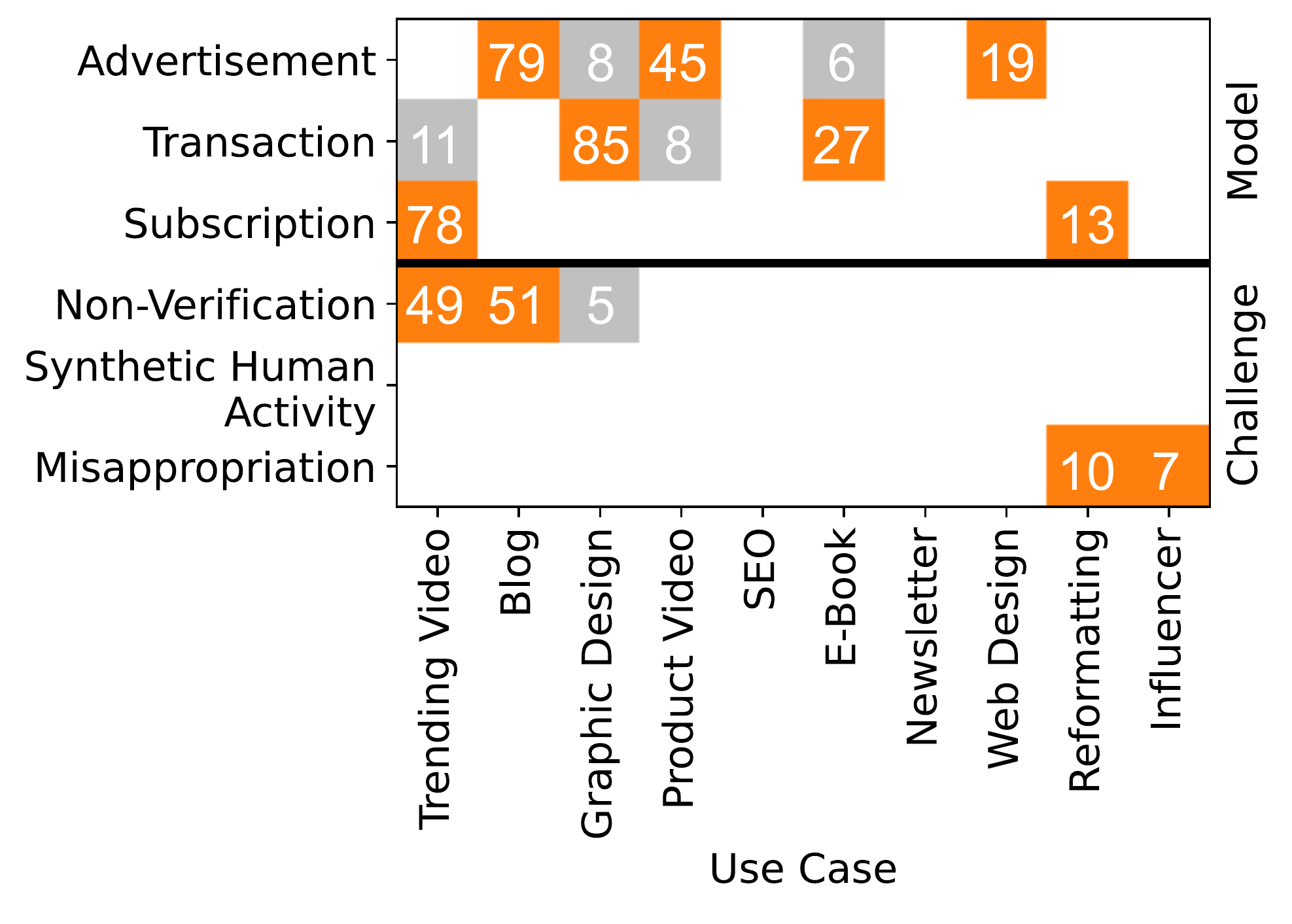}
        \end{subfigure}
        &
        \begin{subfigure}[t]{.58\textwidth}
        \includegraphics[width=\textwidth]{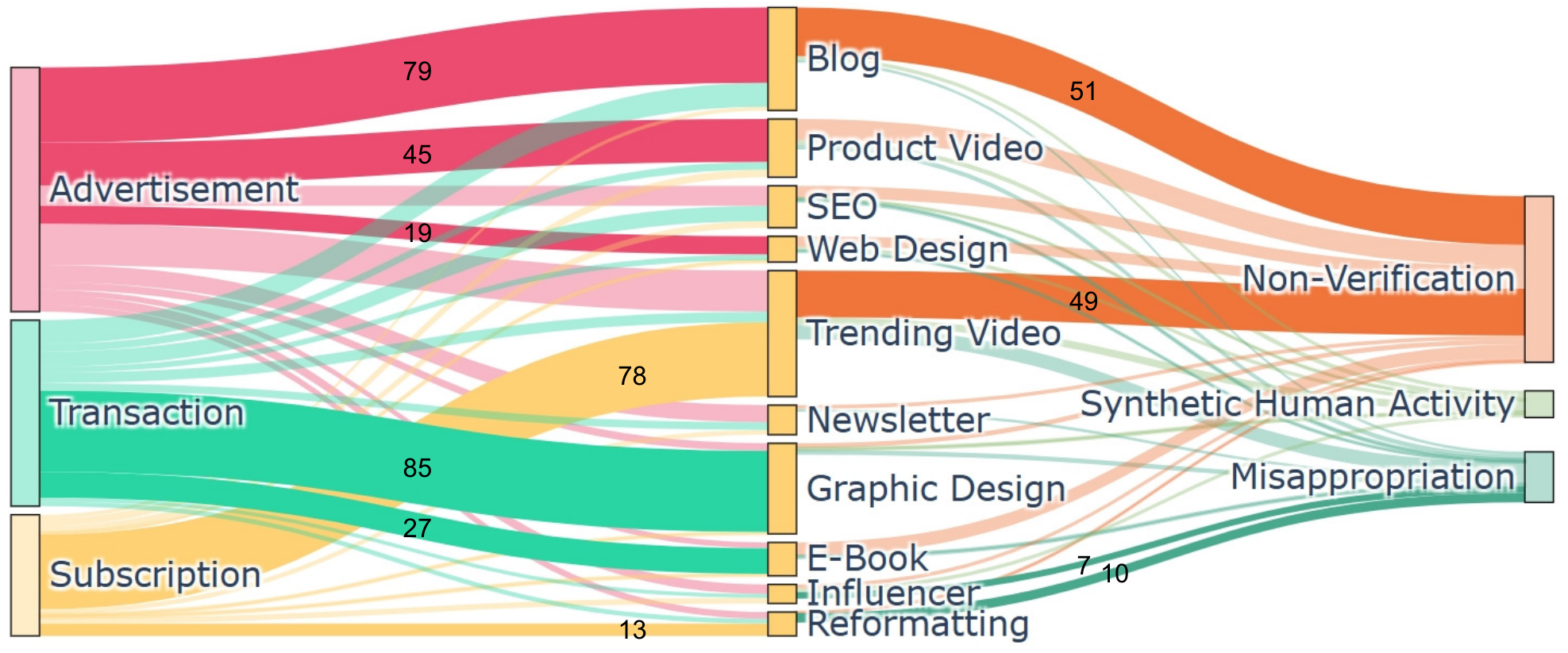}
        \end{subfigure}
        \\
    \end{tabular}
\caption{Associations among monetization use cases, models, and ethical issues. Numbers indicate the video counts that contain both categories.}

\Description{Association Diagram: This image depicts statistically significant associations between use cases, monetization models, and ethical challenges. It uses a grid or network-style layout where lines or highlighted cells indicate strong links. For example, between blogs and advertisement models, or between reformatting and misappropriation. The graphic communicates that certain GenAI monetization practices cluster together and share risks.
}
\label{fig:association}
\end{figure}

The names, counts, and percentages of videos in each category are presented in \autoref{fig:distribution}. We conducted a Chi-square test to examine the relationships among models, challenge categories, and the 10 use cases ($\alpha=0.05$, with Bonferroni correction). Statistically significant associations are visualized in \autoref{fig:association}.

\subsection{RQ1: GenAI4Money Use Cases}

\subsubsection{Trending Video} 
\begin{figure}[!h]
\centering
    \begin{tabular}{lll}
        \begin{subfigure}[t]{.32\textwidth}
        \includegraphics[width=\textwidth]{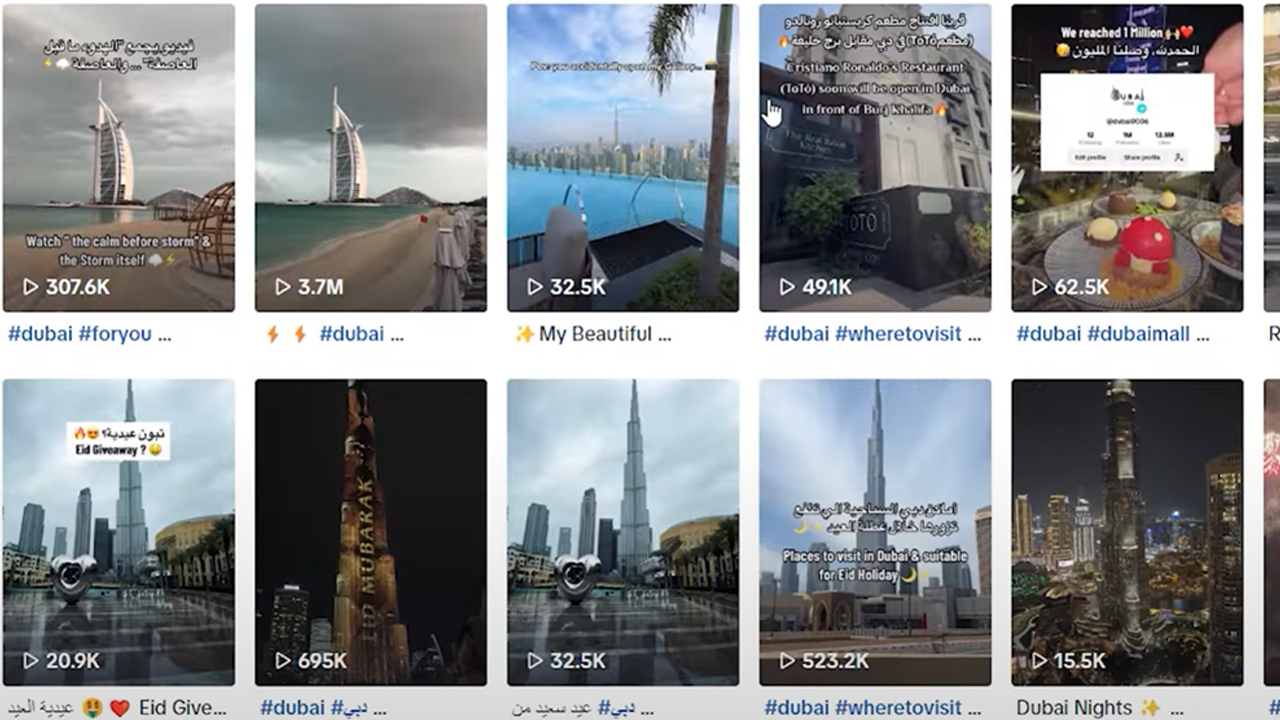}
        \caption{(a) Travel is one of the popular video ``niches.''} 
        \label{fig.strategy1.1}
        \end{subfigure}
        &
        \begin{subfigure}[t]{.32\textwidth}
        \includegraphics[width=\textwidth]{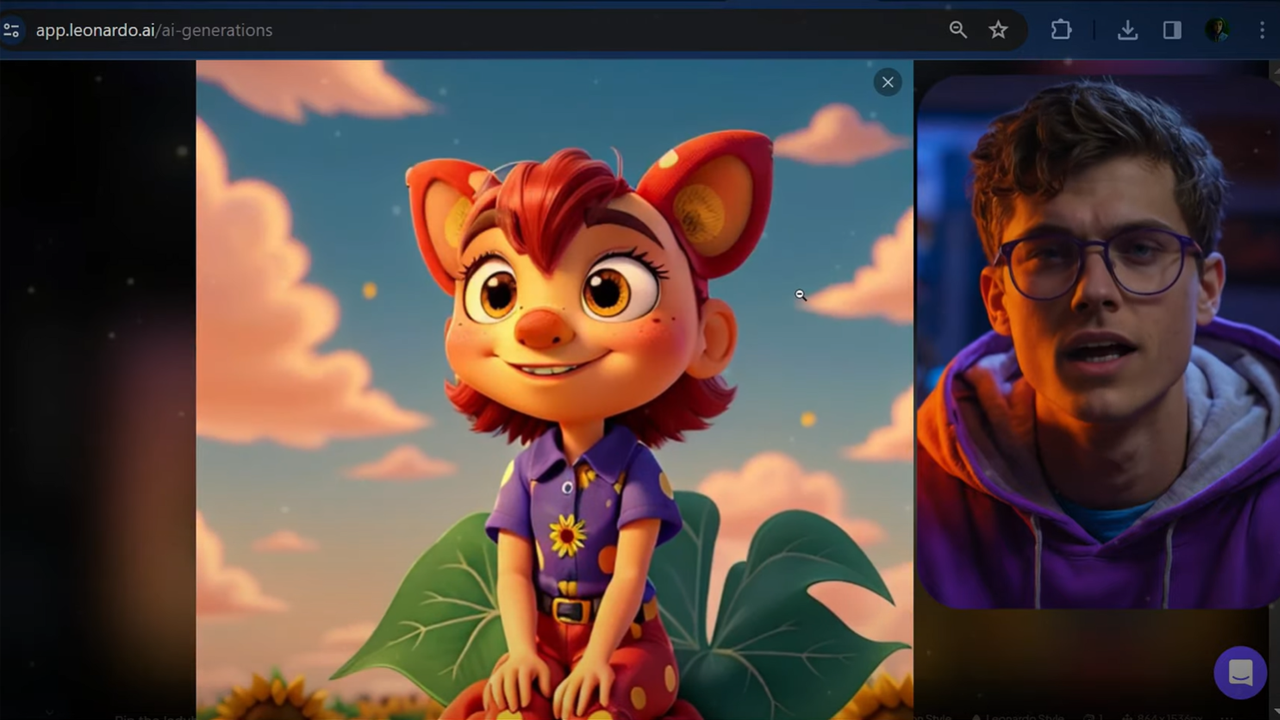}
        \caption{(b) A kid story video created with Leonardo.AI.} 
        \label{fig.strategy1.2}
        \end{subfigure}
        &
        \begin{subfigure}[t]{.32\textwidth}
        \includegraphics[width=\textwidth]{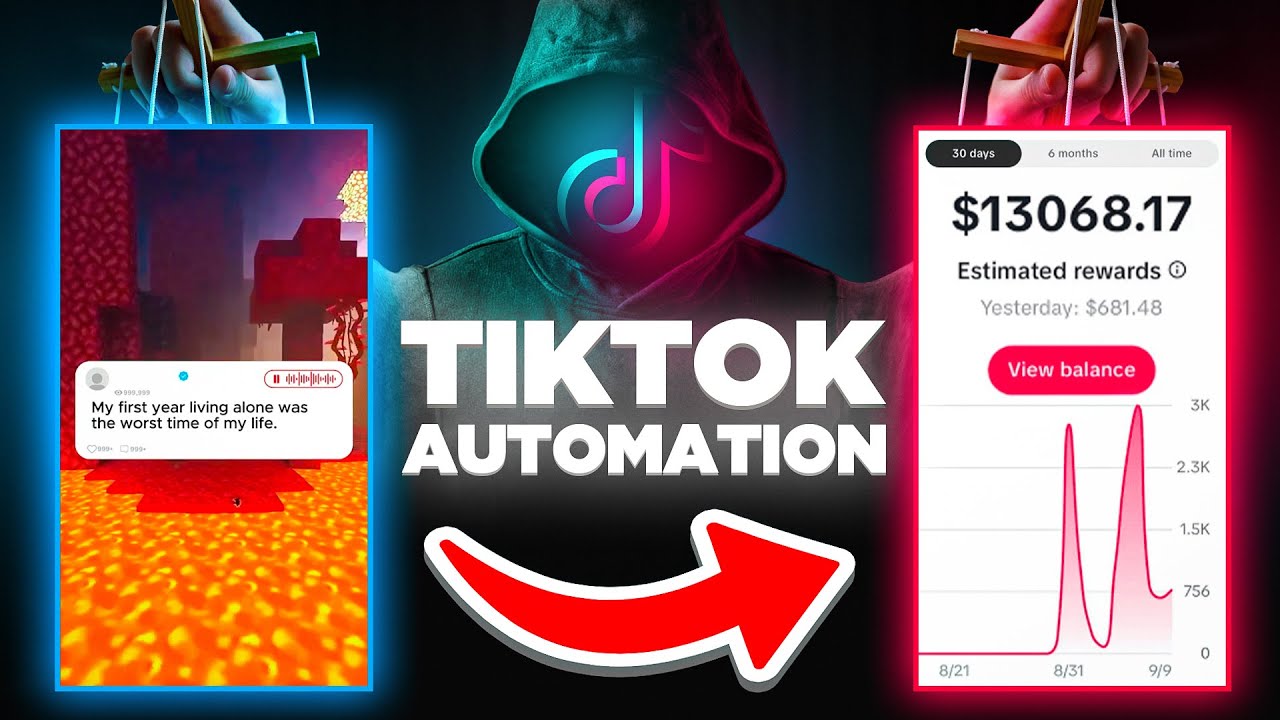}
        \caption{(c) TikTok creativity program for monetization.}  
        \label{fig.strategy1.3}
        \end{subfigure}
    \end{tabular}
\caption{Examples of monetization using AI-generated trending videos.}
\Description{This image shows a series of subimages with different GenAI for monetization scenarios. (1) a: Travel Niche Demonstration: This image shows a YouTube tutorial slide listing recommended AI-generated video niches. The example displayed focuses on the travel genre, showing a set of category tiles or text entries that suggest creating AI-written or AI-assembled travel guides and country or city overviews as monetizable video content. (2) b: AI-Generated Children's Story Video: This screenshot features a children's story video created entirely with AI tools. It contains bright, cartoon-like illustrations, produced with Leonardo.AI, paired with a script generated by Google Bard. The characters and background have a Disney-inspired style, demonstrating how simple text prompts can produce visually appealing videos for young audiences. (3) c: TikTok Creativity Program Panel: The image shows an interface for the TikTok Creativity Program, including metrics or eligibility information. It highlights how creators may earn monetization bonuses by uploading videos that meet certain performance thresholds, connecting AI-generated content to potential earnings. 
}
\label{fig.strategy1}
\end{figure}

The most common use case demonstrated in the knowledge videos involves creating trending videos with GenAI technologies ($N=106,28.1\%$). YouTubers build knowledge around how to use AI tools such as ChatGPT to generate video scripts, synthesize voice through text-to-speech technologies, and create visuals using AI-generated or stock footage, all to produce content that can generate revenue through monetization programs. YouTubers often suggest video niches that can be efficiently produced using various AI tools. These niches cover general knowledge (e.g., health, travel, pets), entertainment (e.g., children's stories), or inspirational quotes (e.g., celebrity or motivational quotes). For example, the video shown in \autoref{fig.strategy1.1} introduces 14 niche ideas suited for AI-generated videos, such as country/city guides, productivity tips, and spirituality. YouTubers often suggest the potential of using GenAI to produce various forms of ``faceless,'' anonymous content. \autoref{fig.strategy1.2} illustrates a kid short story in which the script is generated by Google Bard and the images are produced by Leonardo.AI; these materials are then assembled into a Disney-style video for children. Many of these videos conclude by introducing monetization opportunities -- such as the TikTok Creativity Program or the YouTube Partner Program (\autoref{fig.strategy1.3}).

\subsubsection{Blog} 
\begin{figure}[!h]
\centering
    \begin{tabular}{lll}
        \begin{subfigure}[t]{.32\textwidth}
        \includegraphics[width=\linewidth]{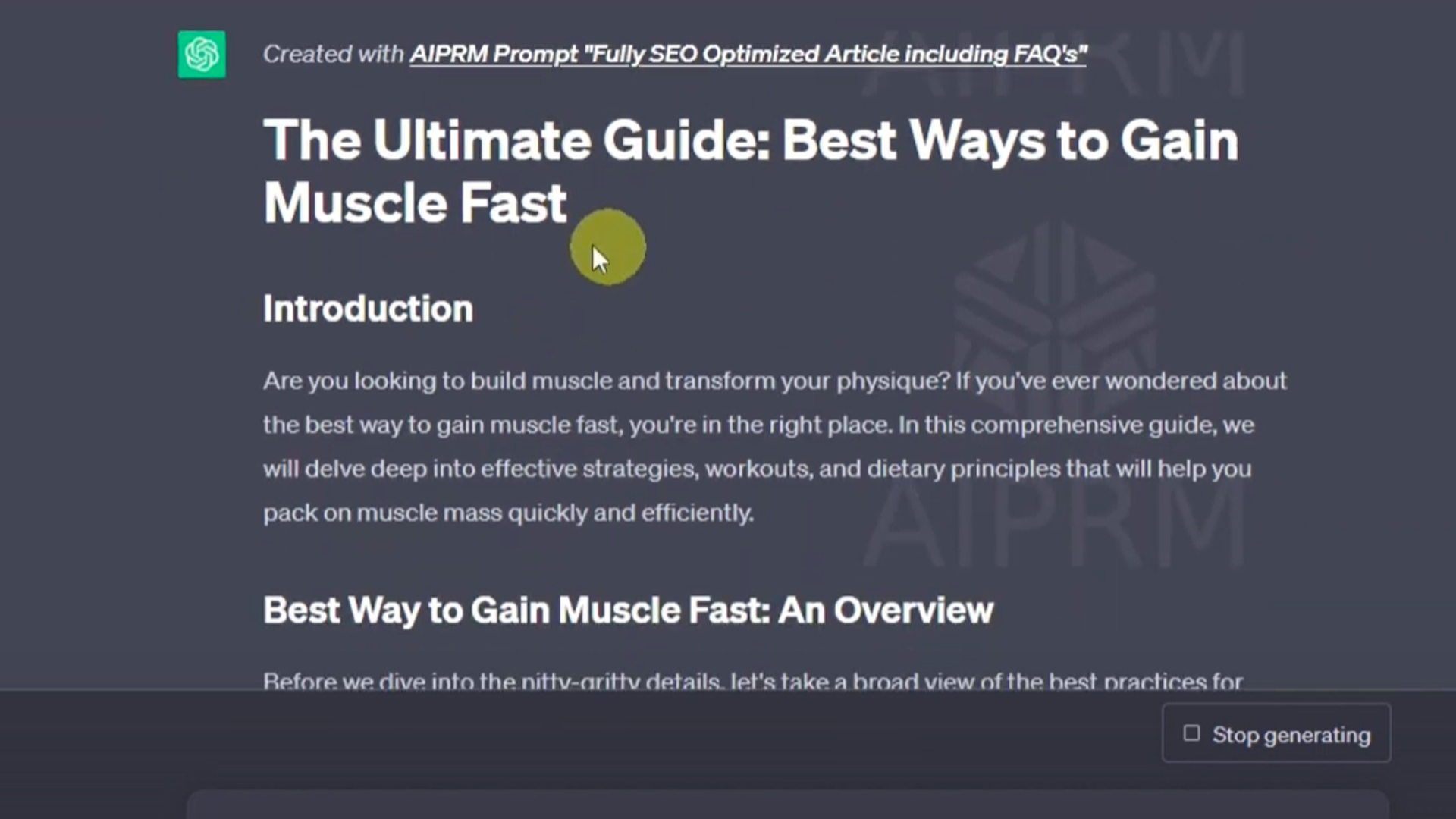}
        \caption{(a) Use ChatGPT to write a blog related to muscle building.} 
        \label{fig.F1}
        \end{subfigure}
        &
        \begin{subfigure}[t]{.32\textwidth}
        \includegraphics[width=\linewidth]{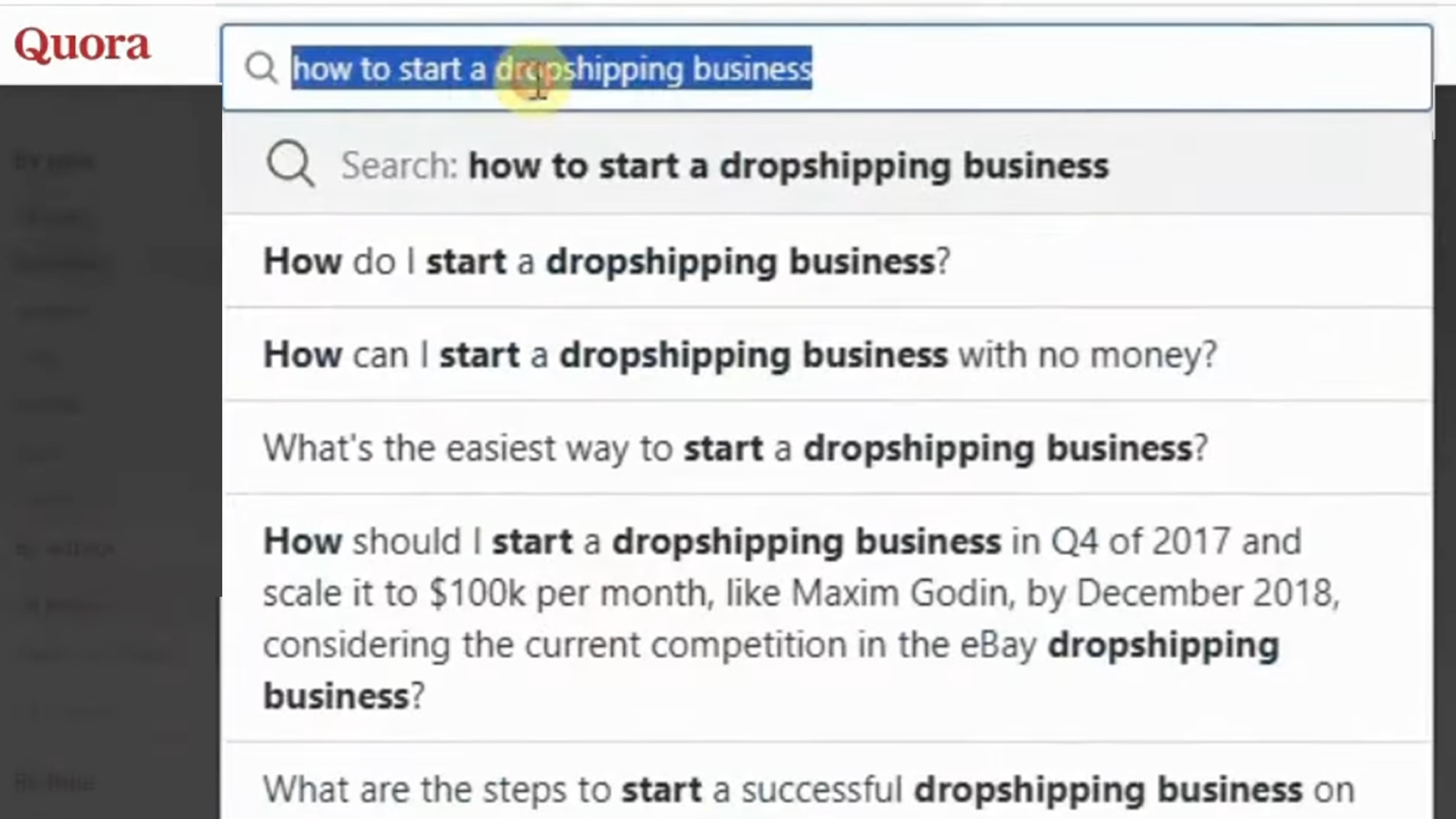}
        \caption{(b) Write Quora answer with Google Bard.} 
        \label{fig.F2}
        \end{subfigure}
        &      
        \begin{subfigure}[t]{.32\textwidth}
        \includegraphics[width=\linewidth]{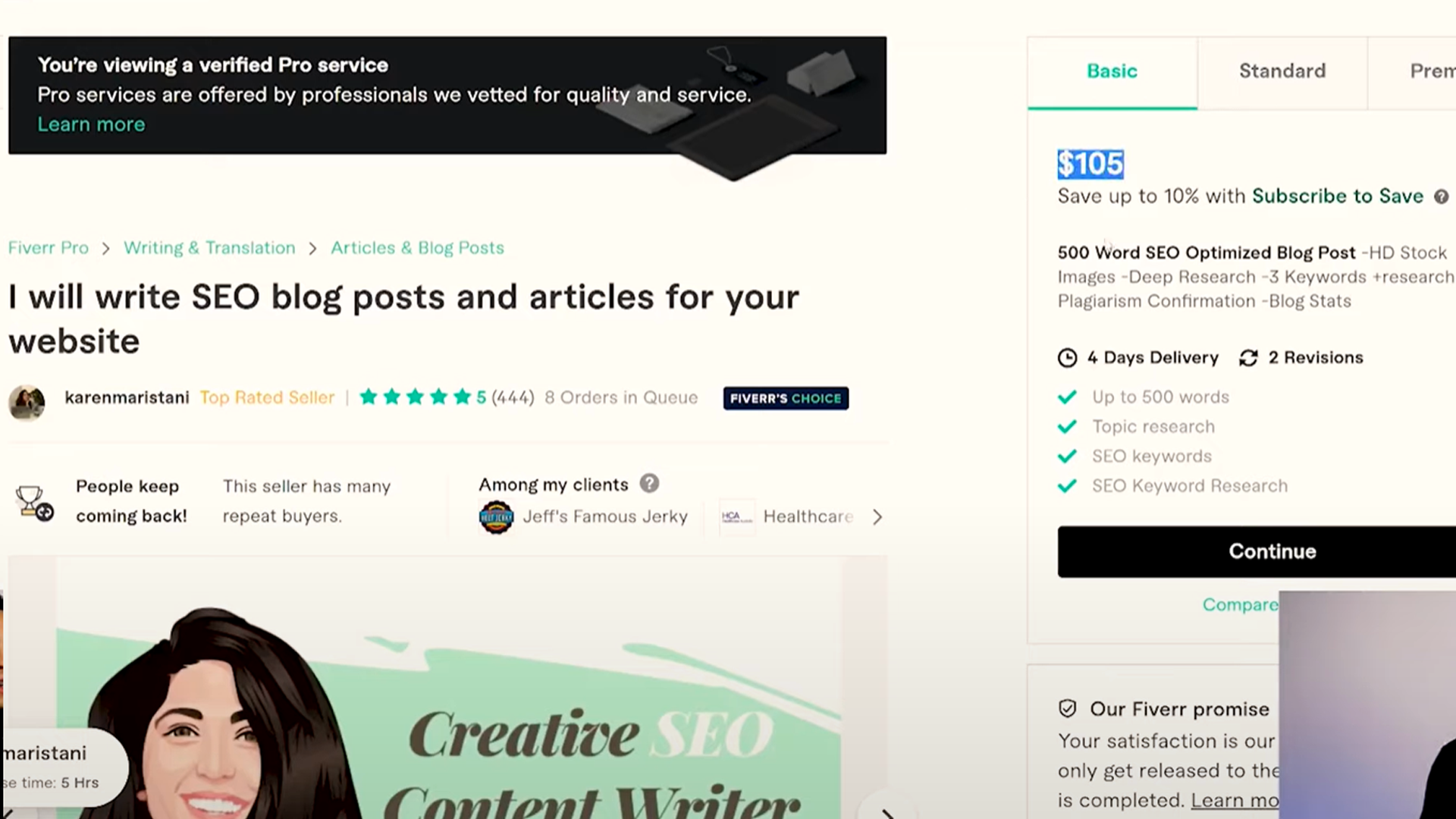}
        \caption{(c) Offer AI SEO blog writing on Fiverr.} 
        \label{fig.F3}
        \end{subfigure}
    \end{tabular}
\caption{Examples of monetization using AI-generated blogs.}
\Description{This image shows Examples of Monetization with AI-generated Blogs. (1) a: ChatGPT-Written Muscle-Building Blog: This image shows a blog article draft about muscle building displayed in a text editor. The article content is fully generated by ChatGPT, demonstrating how creators can rapidly produce health-related written content and insert affiliate links. (2) b: AI-Generated Quora Answer. This screenshot shows a Quora answer field filled with text generated by Google Bard. The response is polished and promotional in tone, including product references or links intended to drive traffic toward affiliate offerings. (3) c: Fiverr Service Offering for AI Blog Writing. This image shows a Fiverr freelancer listing that advertises an AI-powered SEO blog-writing service. The page includes a thumbnail, title, and pricing tiers, illustrating how creators can sell GenAI writing labor as a gig.}
\label{fig.strategy1}
\end{figure}

Writing blogs with GenAI is another popular suggested use case ($N=94, 24.9\%$). YouTubers develop knowledge about monetizing AI-generated blogs, product reviews, and posts on social Q\&A sites. Many highlight that GenAI blogs can embed affiliate links and optimize search engine rankings (SEO). AI tools such as ChatGPT and Jasper AI facilitate the quick creation of product demonstrations, engaging content, and personalized suggestions to attract customers. For example, \autoref{fig.F1} demonstrates how to create a blog related to muscle building and insert a link to a seller's website for a legal steroid product. Some YouTubers recommend using GenAI to answer related questions on Quora by writing an article with Google Bard while embedding product links (\autoref{fig.F2}). Additionally, some creators mention that they can sell their GenAI blog writing as a service on platforms like Fiverr (\autoref{fig.F3}).

\subsubsection{Graphic Design}
\begin{figure}[!h]
\centering
    \begin{tabular}{lll}
        \begin{subfigure}[t]{.32\textwidth}
        \includegraphics[width=\linewidth]{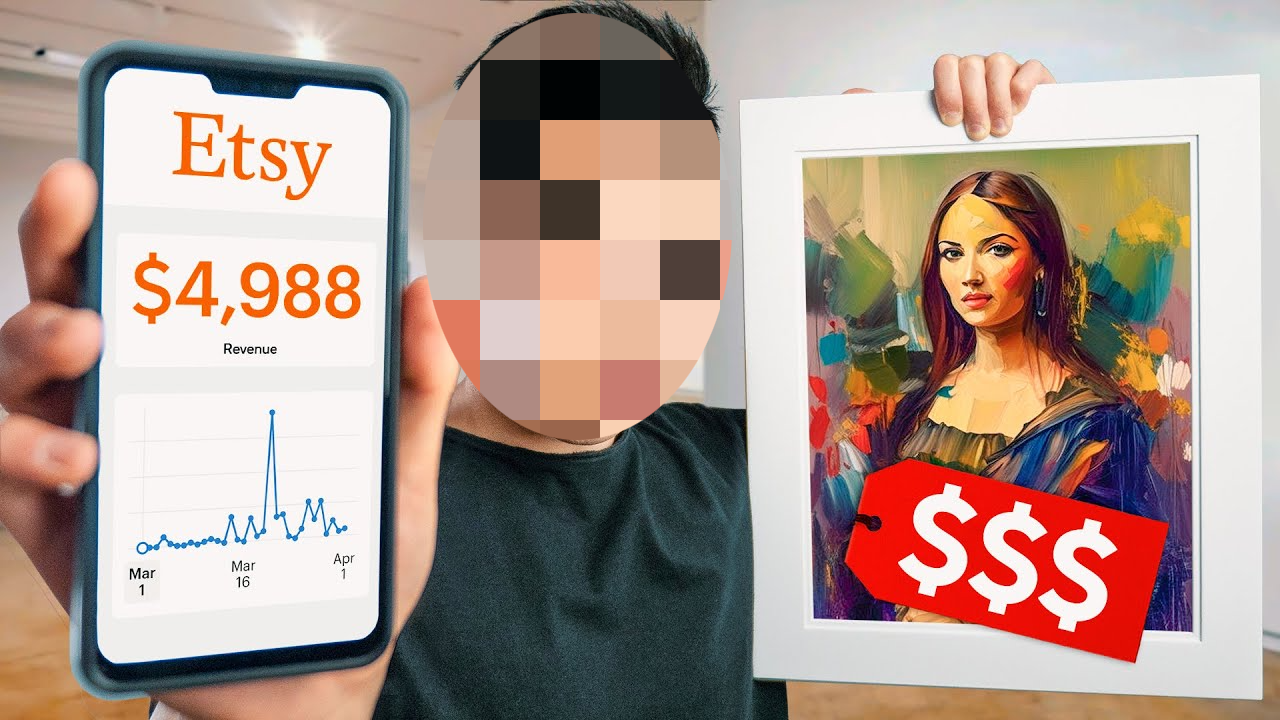}
        \caption{(a) Monetizing by selling AI art on Esty.} 
        \label{fig.GF-design1}
        \end{subfigure}
        &
        
        \begin{subfigure}[t]{.32\textwidth}
        \includegraphics[width=\textwidth]{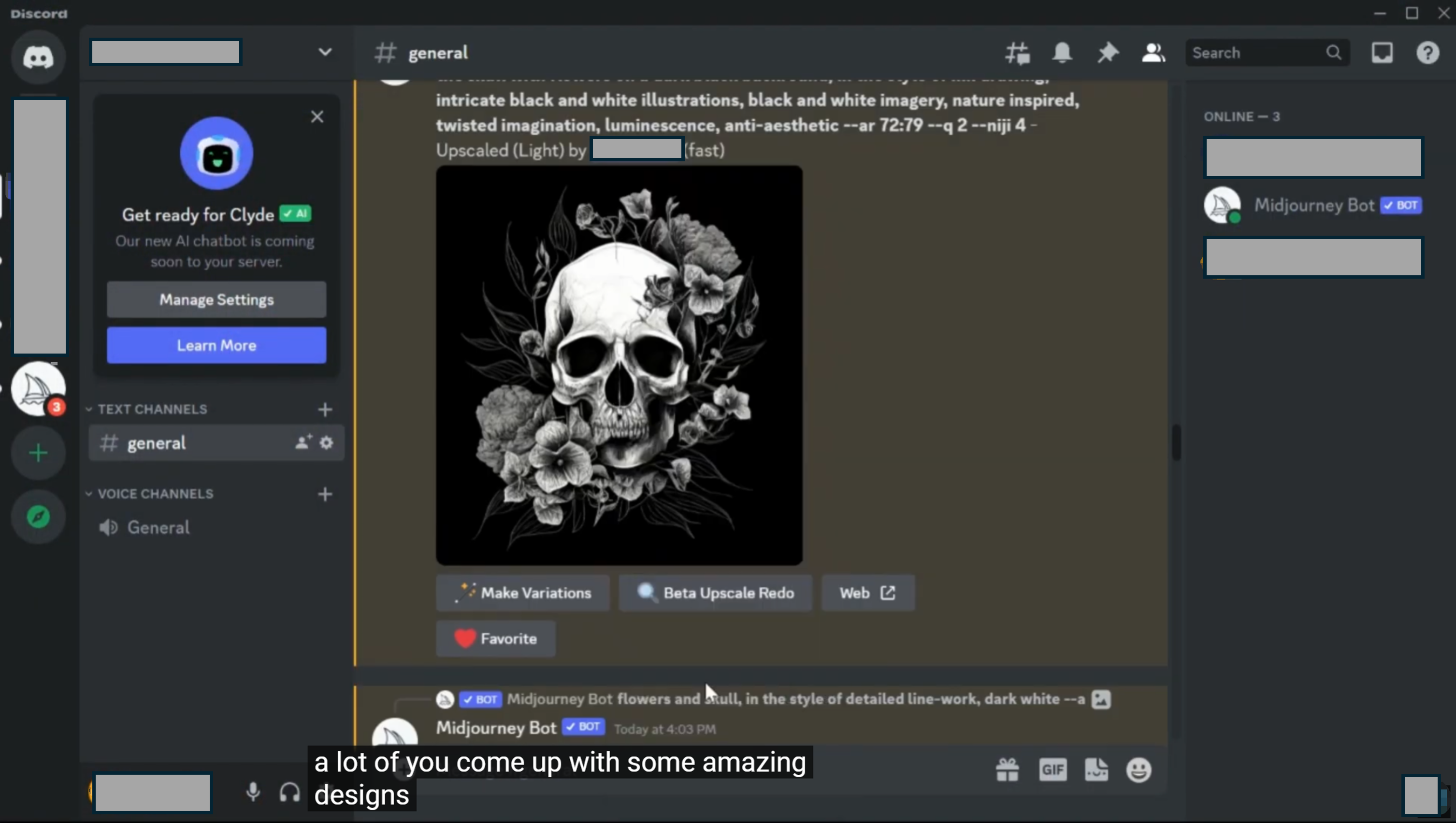}
        \caption{(b) Generate designs with MidJourney.} 
        \label{fig.GF-design3}
        \end{subfigure}

        &
        
        \begin{subfigure}[t]{.32\textwidth}
        \includegraphics[width=\textwidth]{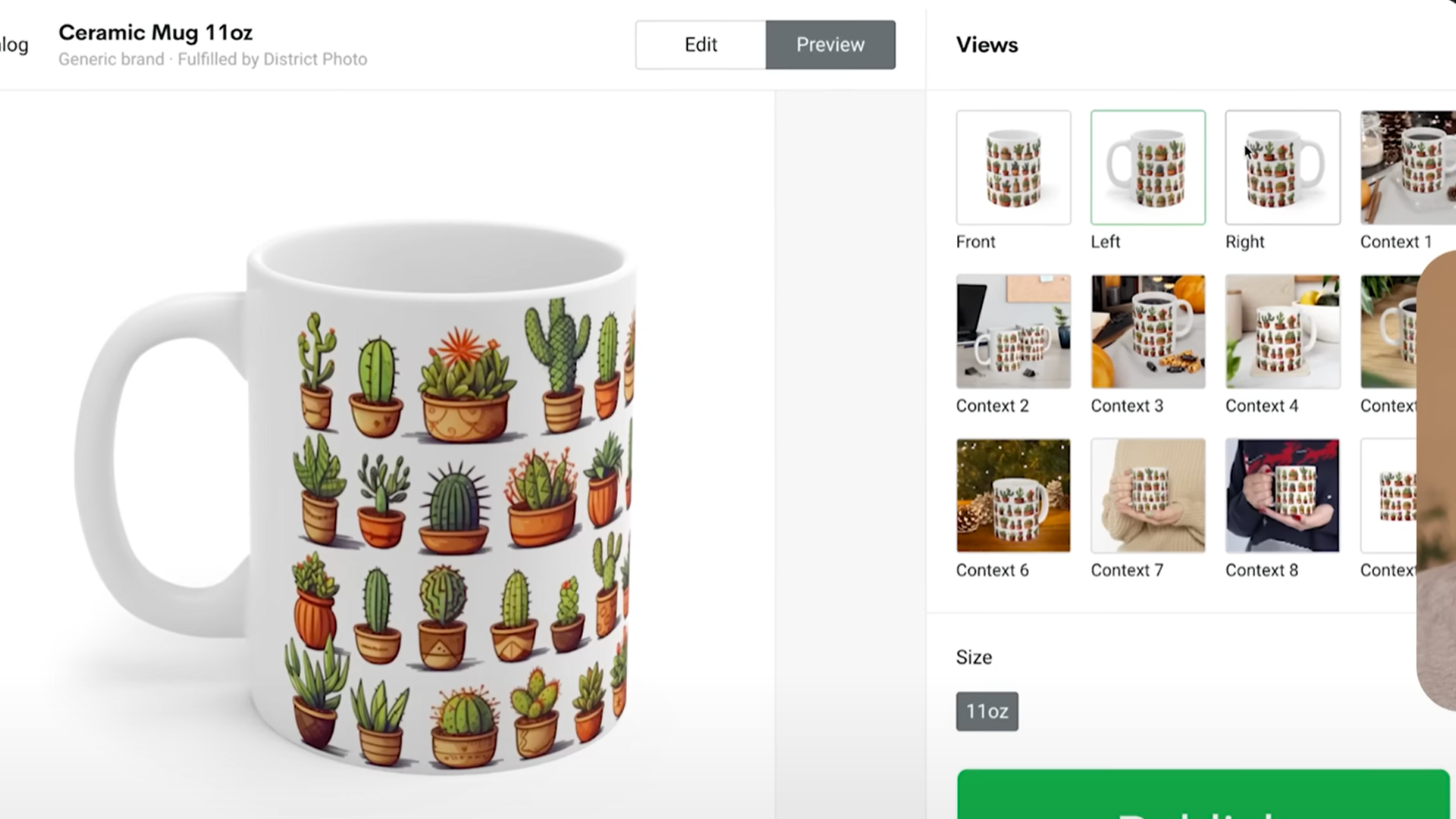}
        \caption{(c) Mugs with AI-generated design for sale.} 
        \label{fig.GF-design4}
        \end{subfigure}
    
    \end{tabular}
\caption{Examples of monetization using AI-generated graphic designs.}

\Description{This image shows Examples of Monetization with AI-generated Graphic Designs. (1) a: Selling AI Art on Etsy. This screenshot shows an Etsy storefront filled with digital art products or clipart bundles generated by AI. The preview images are neatly arranged, showing stylized illustrations meant for customers who purchase downloadable art assets. (2) b: MidJourney T-Shirt Designs. The image presents a collage of T-shirt graphics produced using MidJourney. The designs follow popular aesthetic trends and are intended for sellers to upload onto print-on-demand marketplaces. (3) c: Mugs with AI-Generated Prints. This picture shows white mugs printed with visually striking AI-generated artwork. The mugs are displayed in a mock-up style, demonstrating how AI illustrations can be applied to physical merchandise.
}
\label{fig.ExampleGF-design}
\end{figure}

A subset of YouTubers ($N=85,22.5\%$) share GenAI methods about creating graphic designs for sale as printable products or digital stock. These merchandise items includes T-shirts, stickers, book covers, coloring books, stock images, and clipart. Creators frequently demonstrate GenAI knowledge about tools such as DALL·E, MidJourney, and Stable Diffusion, which transform text prompts into visual AI-generated content. These graphics are subsequently sold either as digital downloads or through handcraft marketplace platforms such as Etsy (\autoref{fig.GF-design1}). Many YouTubers emphasize the importance of tracking and aligning with current design trends to increase sales. Theses strategies often involve analyzing top-selling products for inspiration, inputting relevant keywords into GenAI tools to generate new visuals, and producing them through print-on-demand services. For example, one YouTuber utilizes Alura to analyze popular Etsy designs and uses MidJourney to create similar T-shirt graphics (\autoref{fig.GF-design3}). \autoref{fig.GF-design4} presents an example of print-on-demand services used to apply GenAI designs onto mugs.


\begin{figure}[!h]
\centering
    \begin{tabular}{lll}
        \begin{subfigure}[t]{.32\textwidth}
        \includegraphics[width=\textwidth]{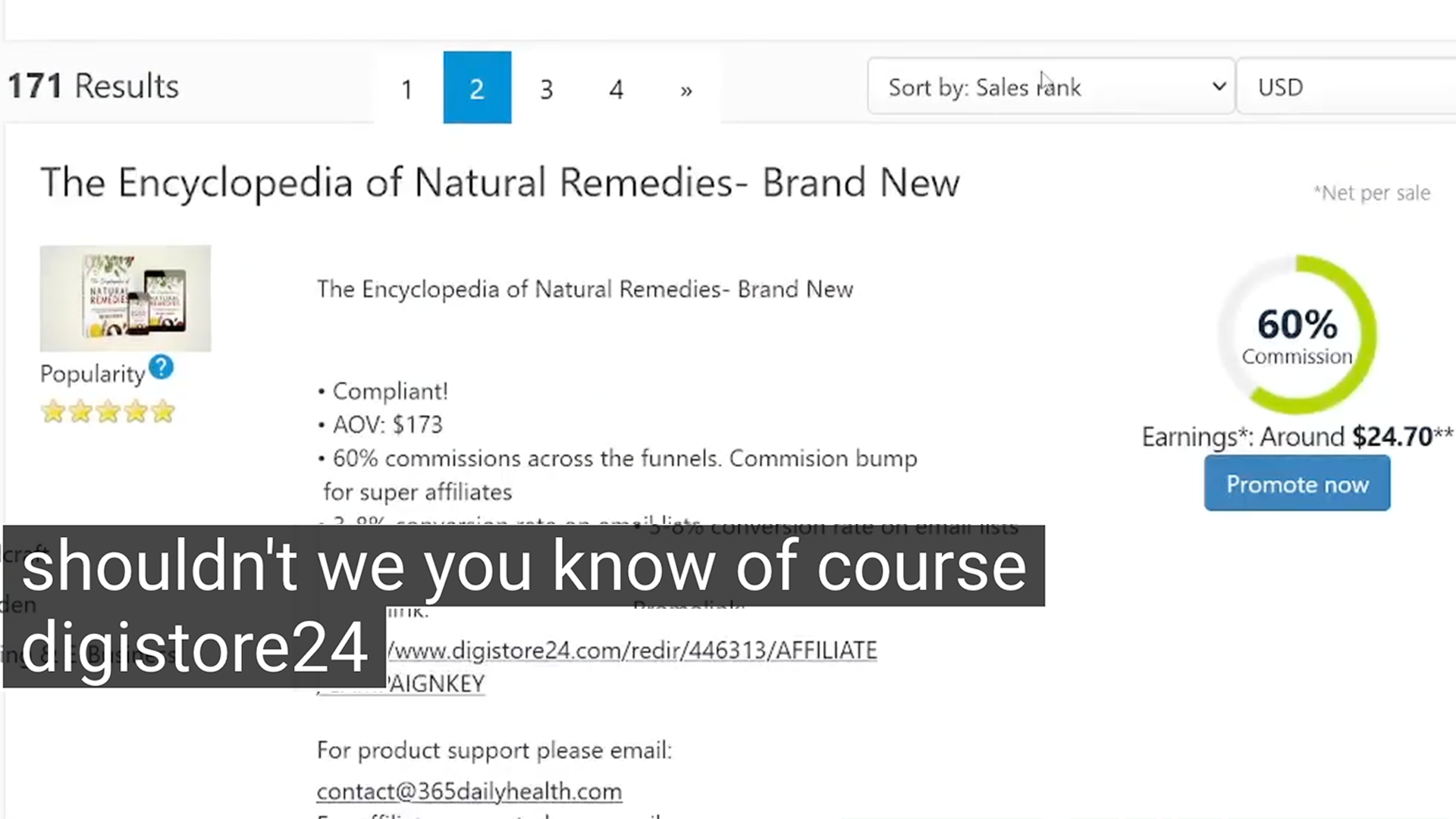}
        \caption{(a) Get product info on Digistore24.} 
        \label{fig.PPV2}
        \end{subfigure}
        &
        \begin{subfigure}[t]{.32\textwidth}
        \includegraphics[width=\textwidth]{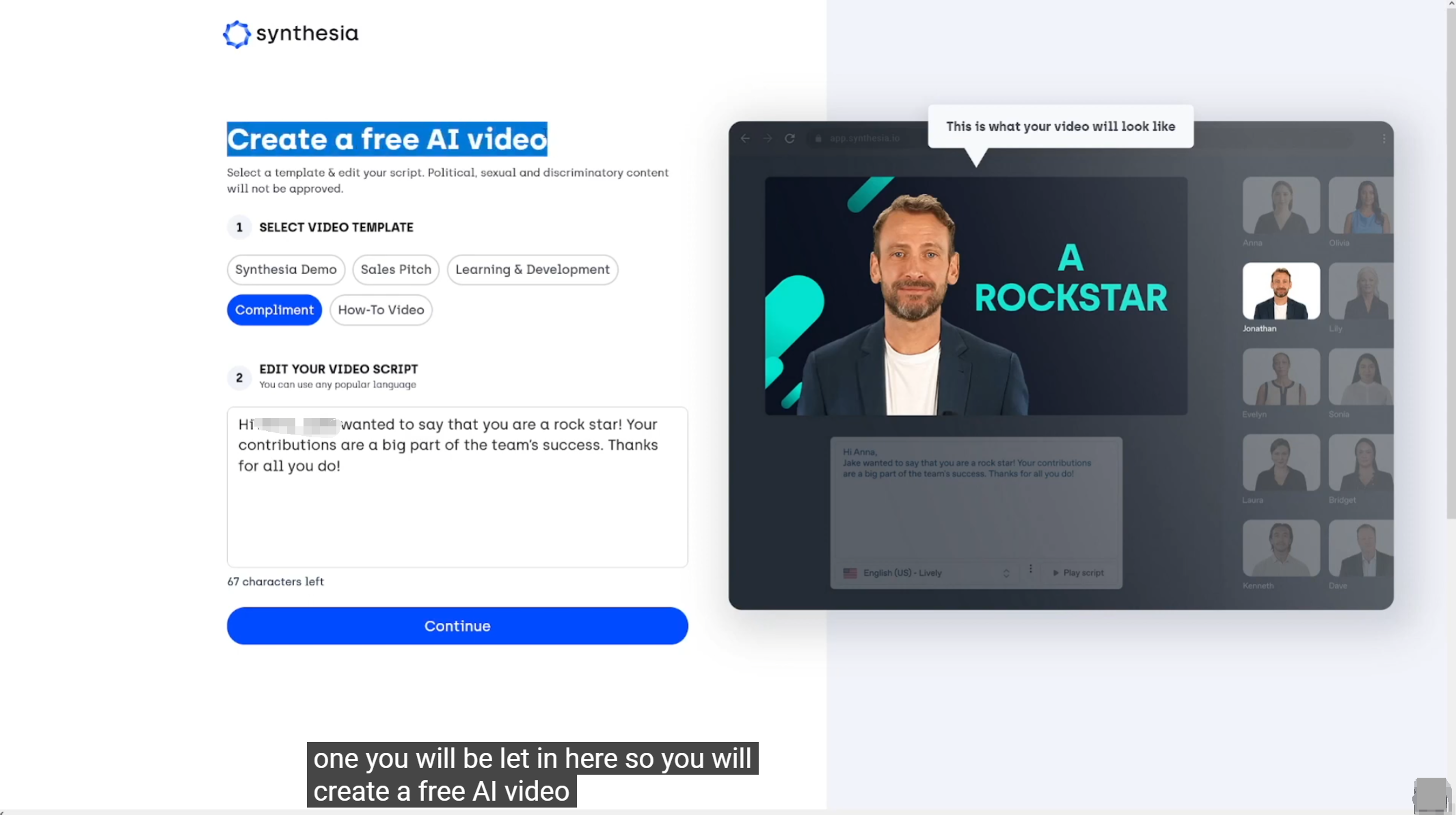}
        \caption{(b) Create human avatar on Synthesia.} 
        \label{fig.PPV3}
        \end{subfigure}
        &
        \begin{subfigure}[t]{.32\textwidth}
        \includegraphics[width=\textwidth]{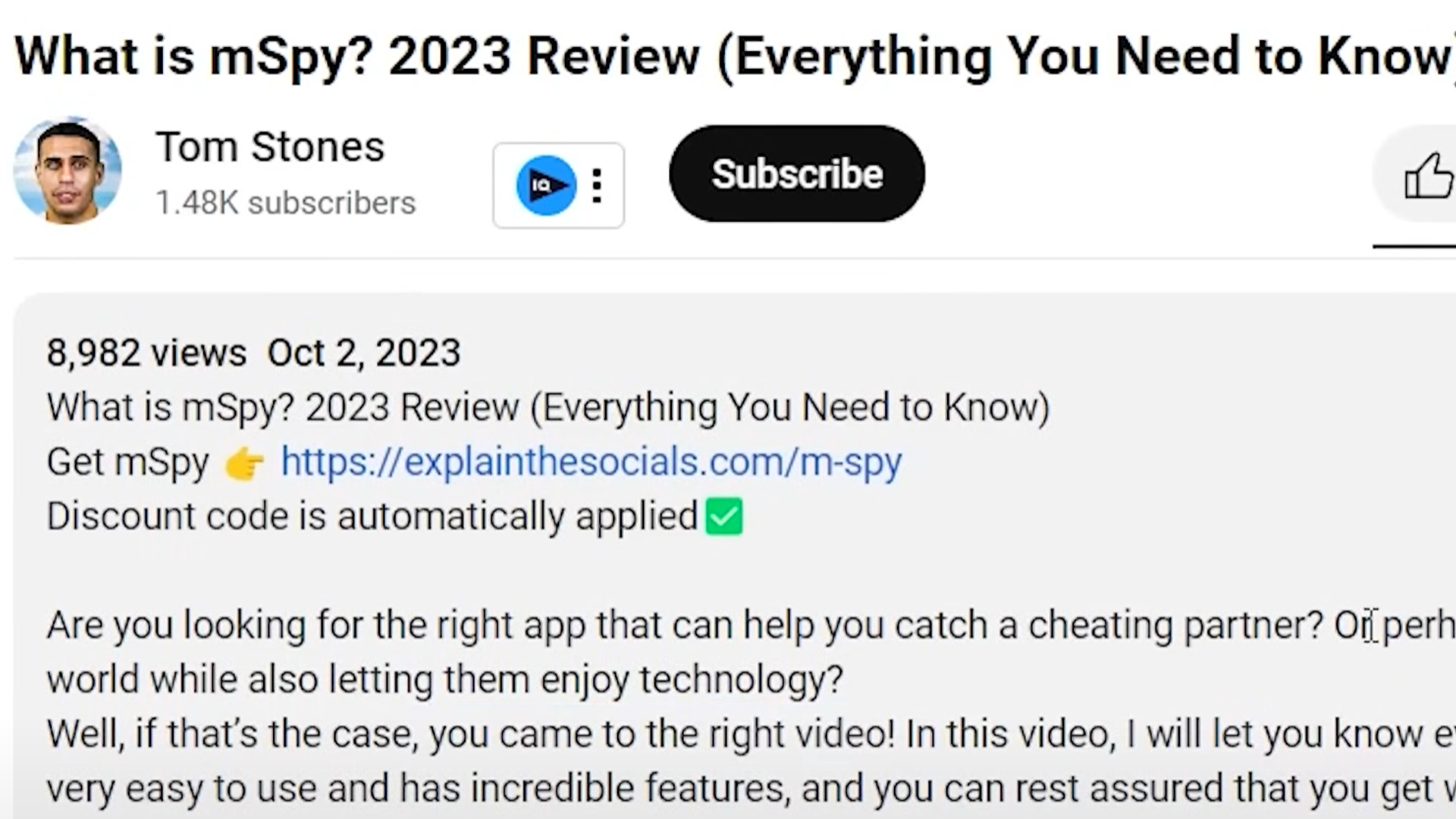}
        \caption{(c) An AI-generated software review video.} 
        \label{fig.PPV4}
        \end{subfigure}
    \end{tabular}
\caption{Examples of monetization with AI-generated promotion videos.}
\Description{This image shows Examples of Monetization with AI-generated promotion videos and AI-powered SEO and E-Book. (1) a: Digistore24 Product List. The screenshot displays the Digistore24 affiliate marketplace interface, where creators can browse digital products along with commission percentages. It highlights high-paying items that can be promoted through AI-generated content. (2) b: Synthesia AI Human Avatar. This image shows a virtual human avatar generated by Synthesia. The avatar appears lifelike, with natural facial movements and professional attire, intended to speak AI-generated product marketing scripts on the creator's behalf. (3) c: AI-Generated Software Review Video. The screenshot shows a video player window where an AI-generated narrator presents a software review. The visuals include interface captures or stock imagery, illustrating how creators automate promotional video production.}
\end{figure}
\subsubsection{Product Promotion Video}
Out of 377 videos, 49 (13.0\%) focus on introducing how to create product promotion videos to viewers. This use case illustrates how YouTubers develop new ways to use GenAI to directly promote products. Similar to the creation of trending videos, YouTubers share practices using ChatGPT to generate product description scripts, image-generating tools to produce visuals, and AI voices for voiceovers. However, unlike videos centered on general trending topics, these promotional videos emphasize identifying profitable affiliate links and converting product descriptions into engaging video content. For instance, one YouTuber demonstrates making a video promoting fitness-related products. They use ChatGPT to brainstorm content ideas and identifies affiliate products on Digistore24 (\autoref{fig.PPV2}). They then draft scripts with ChatGPT that include product details and avatar-based narratives, and employs Synthesia to create an AI talking avatar to present the product (\autoref{fig.PPV3}). YouTubers often recommend publishing these videos on platforms such as YouTube, TikTok, and Instagram to maximize reach. Some videos are styled as product reviews or personal experiences. For example, certain creators suggest using GenAI to generate product reviews, embedding affiliate links within the video descriptions (\autoref{fig.PPV4}).


\subsubsection{SEO}
\begin{figure}[!h]
\centering
    \begin{tabular}{l}
        \begin{subfigure}[t]{0.4\textwidth}
            \centering            \includegraphics[width=\textwidth]{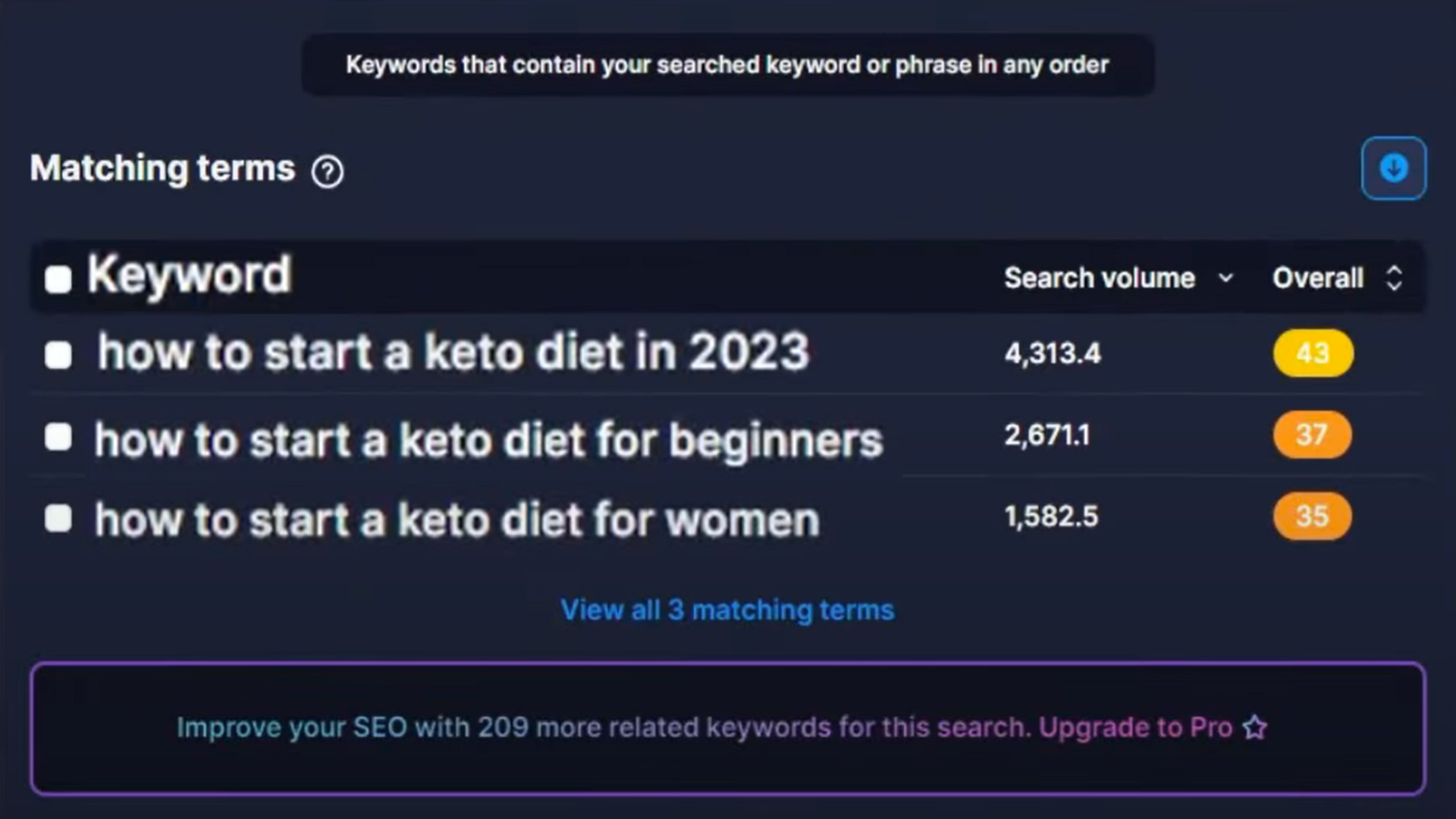}
        \end{subfigure}
    \end{tabular}
\caption{Get SEO keywords list using ChatGPT.}
\Description{This image shows getting SEO keyword lists using ChatGPT. There is a ChatGPT dialogue box displaying a list of SEO keywords generated to help creators optimize the discoverability of their videos or blog posts. }
\label{fig.SEO1}
\end{figure}

YouTubers suggest using GenAI for search engine optimization (SEO) with social media content ($N=37,9.8\%$). This use case highlights how GenAI-based SEO tools help creators optimize content with search-friendly keywords to boost visibility in search engines and recommendation algorithms. For many creators, gaining greater exposure to online audiences is essential for monetization. Tools like ChatGPT, SurferSEO, and Jasper AI are commonly introduced for this purpose. SEO applies to various content formats, including blogs and video descriptions. For example, one YouTuber demonstrates how to use VidIQ to identify relevant keywords for a promotional video (\autoref{fig.SEO1}).

\subsubsection{E-Book} 
\begin{figure}[!h]
\centering
    \begin{tabular}{ll}
        \begin{subfigure}[t]{0.4\textwidth}
            \centering
            \includegraphics[width=\textwidth]{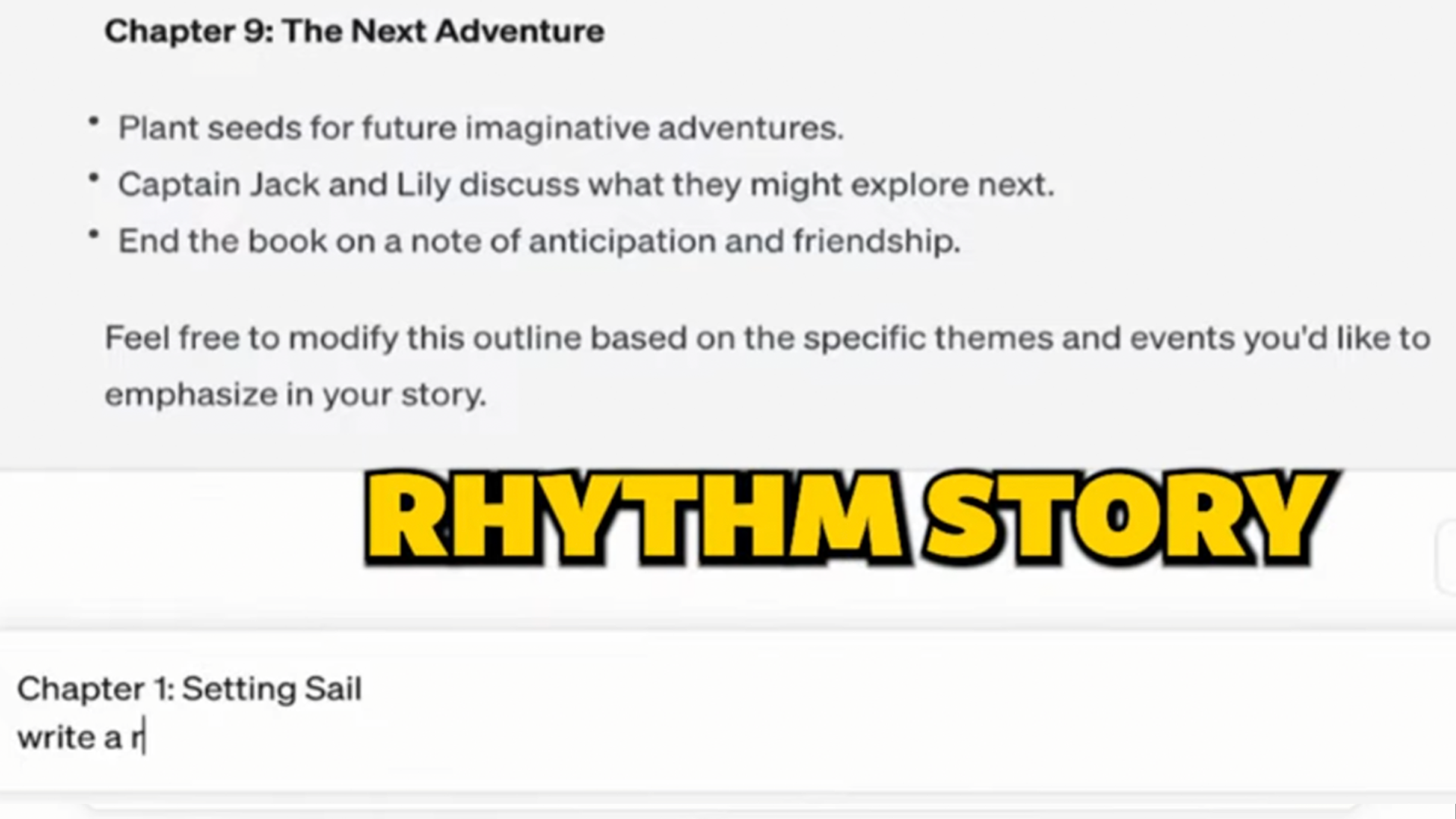}
            \caption{(a) Create text content using ChatGPT.} 
            \label{fig.Ebook1}
        \end{subfigure}
            &
        \begin{subfigure}[t]{0.4\textwidth}
            \centering
            \includegraphics[width=\textwidth]{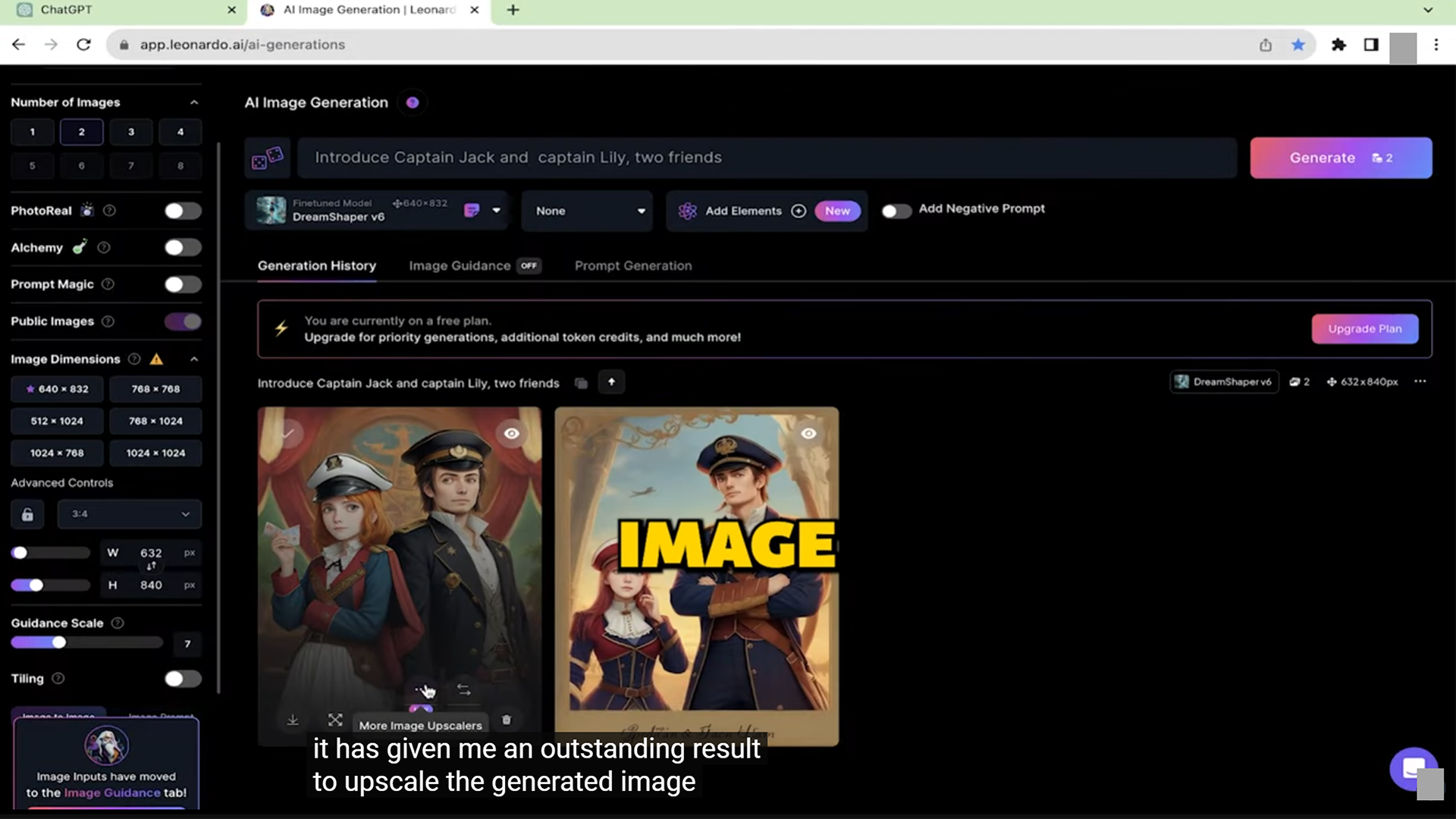}
            \caption{(b) Generate images for e-book using Leonardo.AI.} 
            \label{fig.Ebook2}
        \end{subfigure}
    \end{tabular}
\caption{Examples of monetization using AI-powered e-books.}

\Description{This image shows Examples of Monetization with AI-powered E-Books. (1) a: ChatGPT-Generated E-Book Text. The screenshot shows ChatGPT outputting story outlines, chapter summaries, or full paragraphs of text for a children's e-book, demonstrating how creators can produce entire manuscripts automatically. (2) b: Leonardo.AI E-Book Illustrations. This picture shows illustrated scenes or characters generated in Leonardo.AI for the corresponding children's e-book. The images are stylized and colorful, matching the narrative prompts.
}
\end{figure}

Some YouTubers ($N=30,8.0\%$) share knowledge about how to write e-books using GenAI. They explain how tools like ChatGPT can generate book ideas, develop chapters, and even create complete content. For example, one video shows how to use AI to create children's e-books. The creator uses ChatGPT to generate story ideas, outlines, chapter content, and the book cover design (\autoref{fig.Ebook1}). They then use Leonardo.AI to create illustrations based on the story prompts (\autoref{fig.Ebook2}). YouTubers often claim that these ebooks can be sold via Kindle Direct Publishing (KDP) programs.

\subsubsection{Newsletter} 
\begin{figure}[!h]
    \centering
    \begin{tabular}{l}
    \begin{subfigure}[t]{0.4\textwidth}
        \centering
        \includegraphics[width=\textwidth]{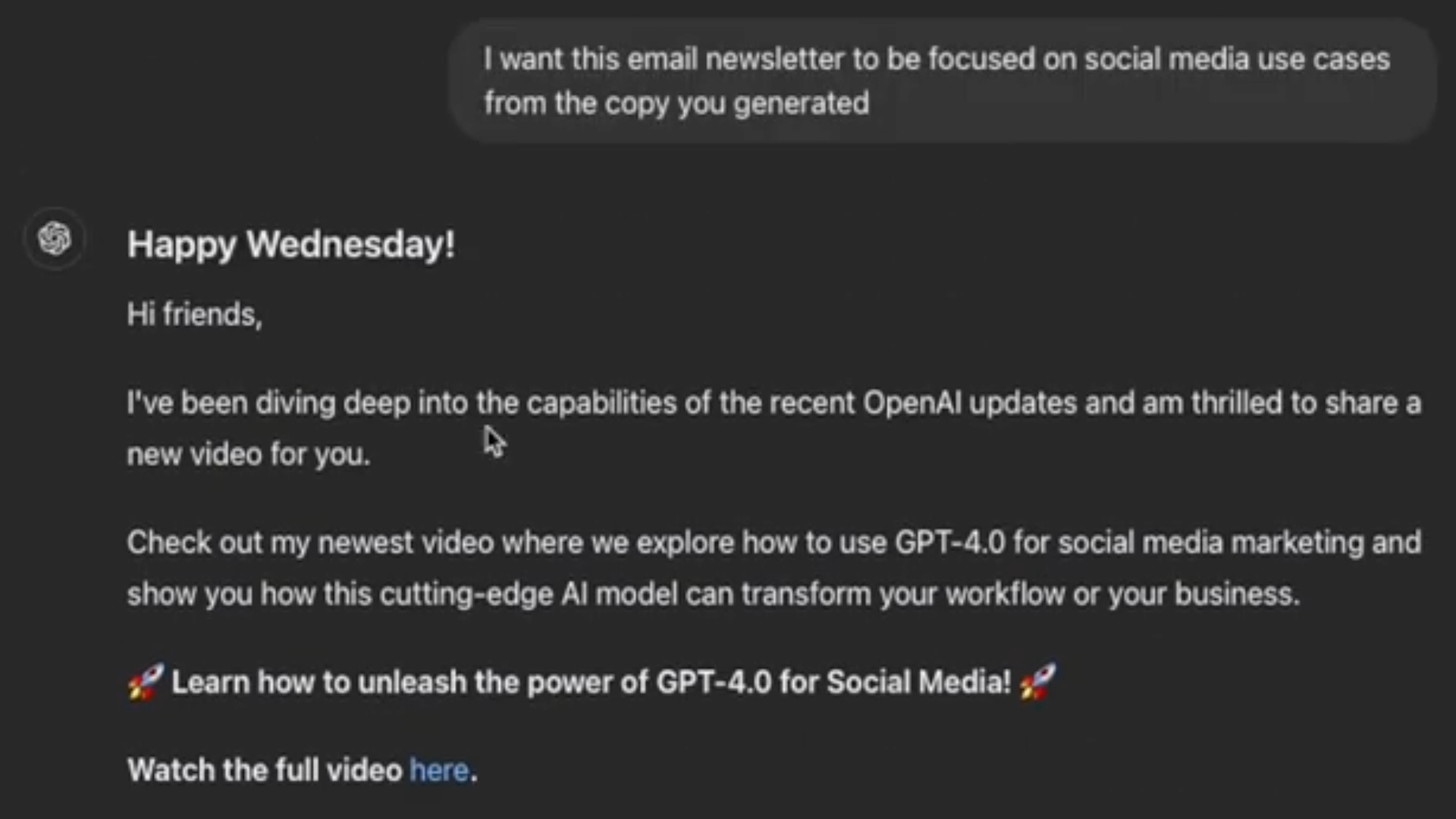}
    \end{subfigure}
    \end{tabular}
    \caption{Write customized newsletter with ChatGPT.}
    \Description{This image shows Examples of AI-written newsletters. This screenshot displays a newsletter editor filled with professionally formatted promotional messaging generated by ChatGPT. The text includes subject lines, preheaders, and branded messaging crafted automatically from prompts. }
    \label{fig.Newsletter1}
\end{figure}

Videos highlight the use case of creating newsletters to generate income ($N=25,6.6\%$). These newsletters -- typically distributed via email or social media ads -- allow content creators to generate revenue by promoting their own channels or those of affiliate marketers. Creators show how to use tools like ChatGPT to brainstorm, generate, automate, and edit engaging newsletters that attract customers. One example shows a YouTuber using ChatGPT to create email newsletters (\autoref{fig.Newsletter1}). They feed ChatGPT video transcripts and business details, generating personalized content with matching tone, subject lines, and preheader text. These AI-generated newsletters are often better formatted and more visually appealing, making them more effective for sharing YouTubers' updates or small business news.

\subsubsection{Web Design} 
\begin{figure}[!h]
    \centering
    \begin{tabular}{l}
    \begin{subfigure}[t]{0.4\textwidth}
        \centering
        \includegraphics[width=\textwidth]{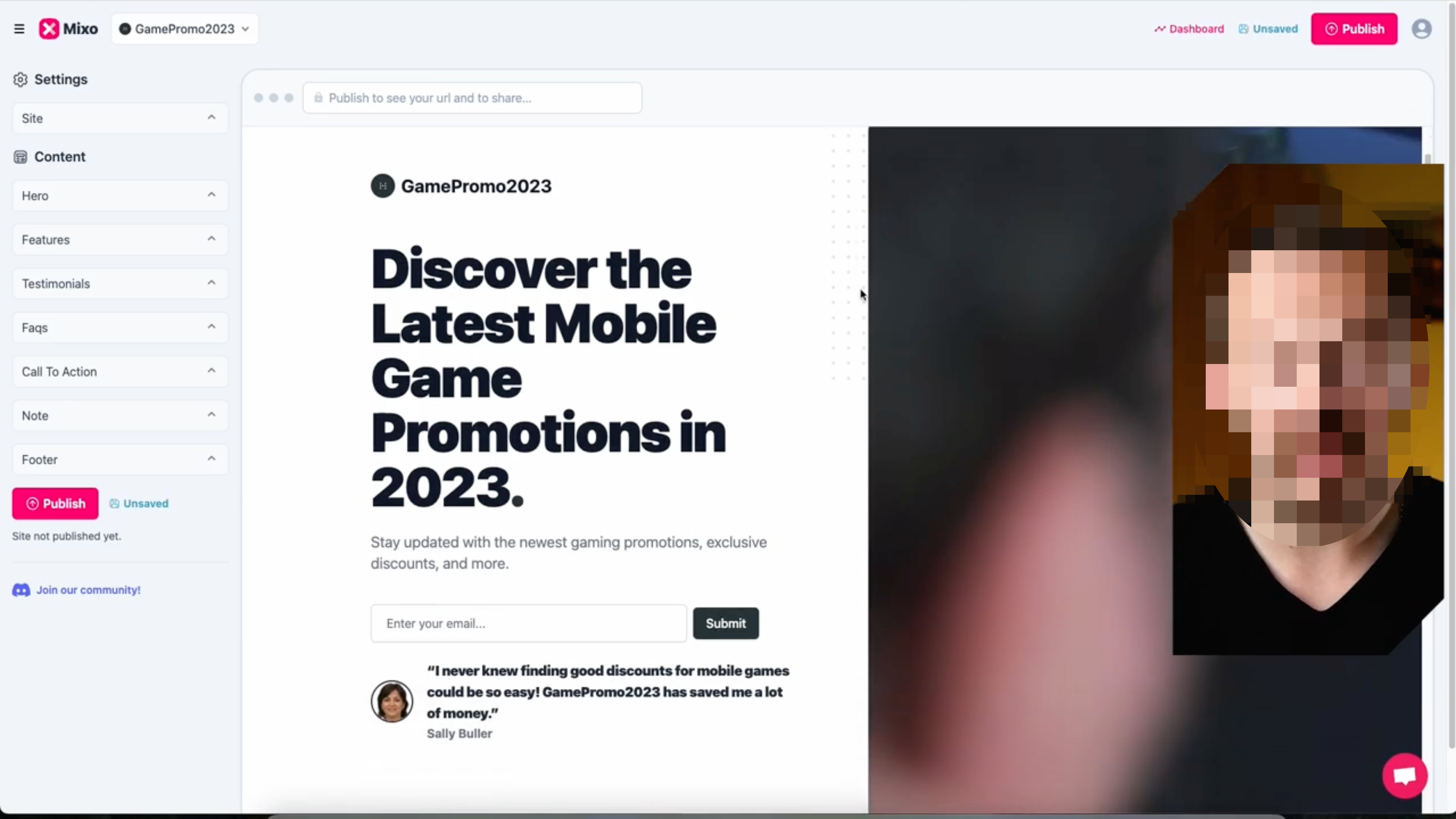}
    \end{subfigure}
    \end{tabular}
    \caption{Create websites with Mixo.}
    \Description{This image shows Examples of AI-built websites. It shows a website generated by Mixo.ai, including a hero section, sample text, and AI-selected graphics. The page resembles a simple landing page for a game or product, created entirely from a short prompt. }
    \label{fig.Web1}
\end{figure}

Creating websites is another use case demonstrated in the videos to generate income ($N=23,6.1\%$). These videos build knowledge about how to design websites for affiliate marketing, including web logic, frameworks, and content, using GenAI. Tools like MidJourney are used for website design, and some creators also use platforms like Mixo.ai to build websites from scratch (\autoref{fig.Web1}). For example, one YouTuber shows how Mixo.ai generates websites for mobile games from a simple prompt, providing both content and images. They also use ChatGPT to create game descriptions and articles. The website is later used to promote mobile games and earn a commission.

\subsubsection{Video Reformatting}  
\begin{figure}[!h]
    \centering
    \begin{tabular}{l}
    \begin{subfigure}[t]{0.4\textwidth}
        \centering
        \includegraphics[width=\textwidth]{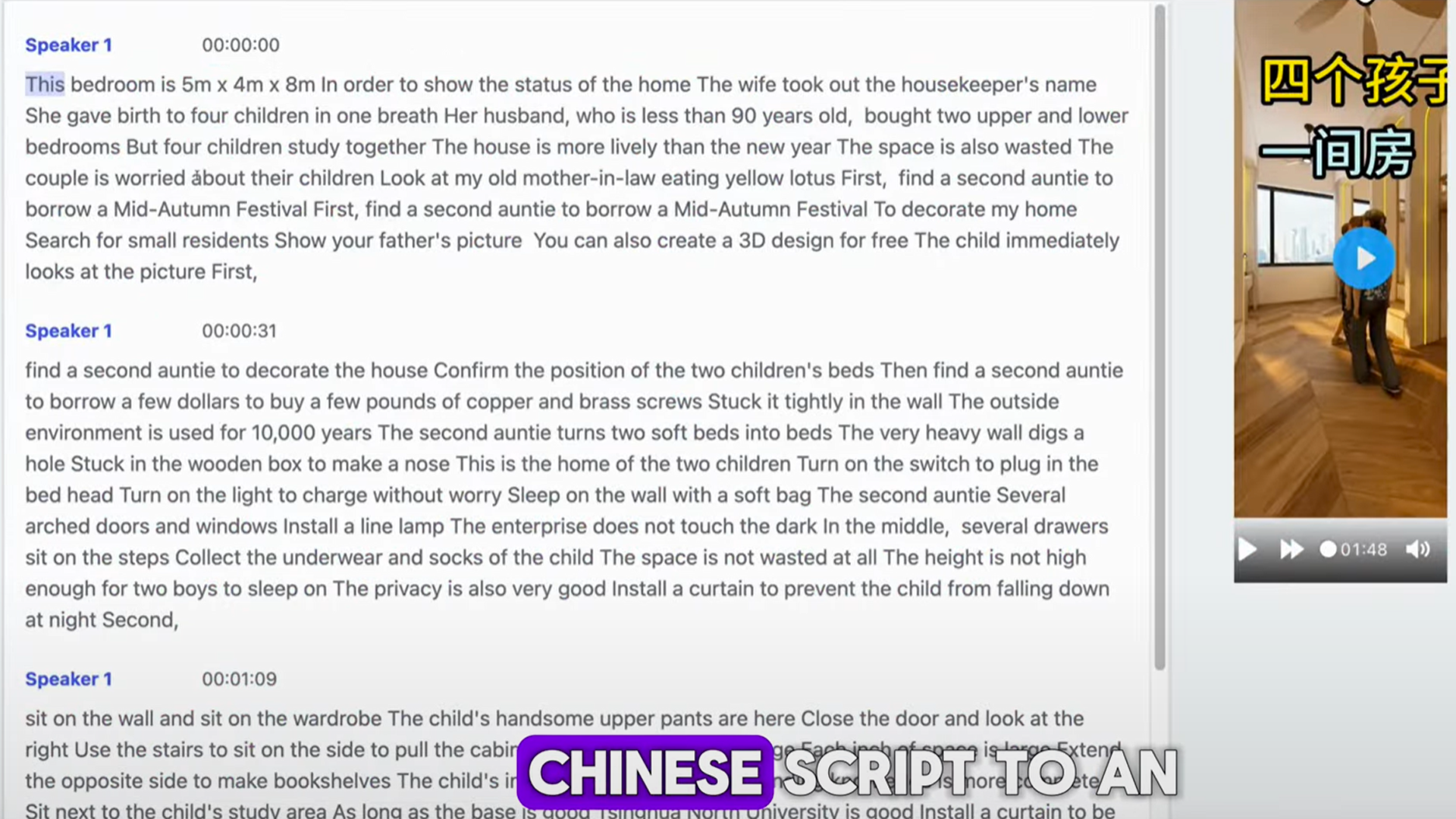}
    \end{subfigure}
    \end{tabular}
    \caption{Reformatting a video in a different language.}
    \Description{This image shows Examples of Video in the Translation/Reformatting Workflow. The screenshot shows an interface where a video originally recorded in Chinese is being translated, re-narrated, or reframed using AI tools. Text boxes or timeline markers indicate automated transcription and voice regeneration.}
    \label{fig.VRF1}
\end{figure}

A smaller group of YouTubers ($N=18, 4.8\%$) showcase using GenAI to reformat existing content. They demonstrate how these tools can clip, translate, and mimic styles, facilitating content sharing across platforms and language barriers. While some creators transform their own content for knowledge explanations, others imply using content from other creators while attempting to avoid copyright infringement. For example, one creator suggests downloading videos from TikTok, YouTube, and Instagram in Chinese, transcribing and translating them with Oris AI, and rewriting scripts using ChatGPT (\autoref{fig.VRF1}). New voiceovers are generated with Eleven Labs. Some videos present this use case as a way to inspire creators to repurpose their own content. However, others suggest that GenAI can be used to modify others' content for publication on one's own channel.

\subsubsection{Social Media Influencer} 
\begin{figure}[!h]
    \centering
    \begin{tabular}{ll}
    \begin{subfigure}[t]{0.4\textwidth}
        \centering
        \includegraphics[width=\textwidth]{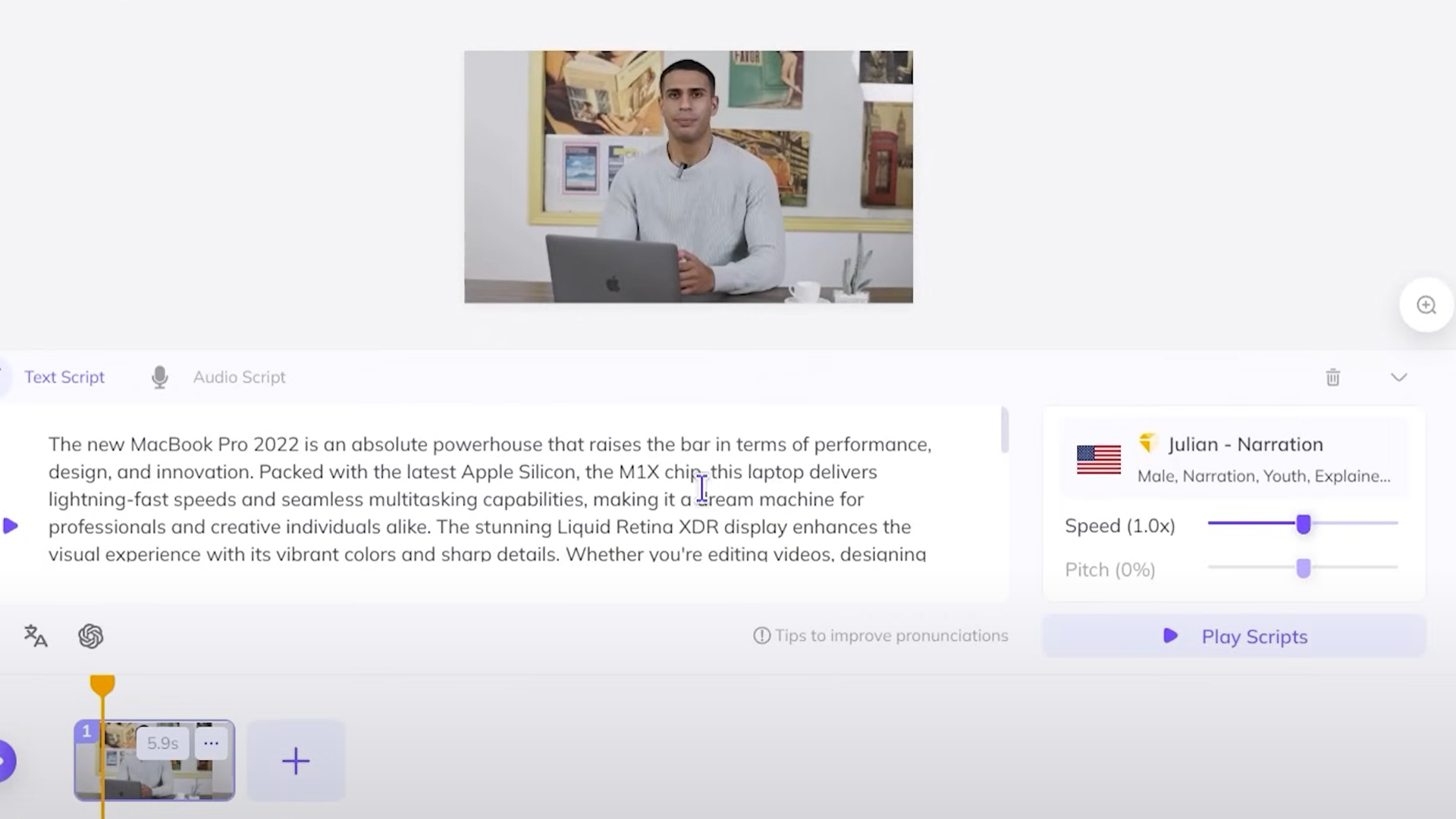}
        \caption{(a) Create influencer avatar with HeyGen.} 
        \label{fig.SMI1}
    \end{subfigure}
    &
    \begin{subfigure}[t]{0.4\textwidth}
        \centering
        \includegraphics[width=\textwidth]{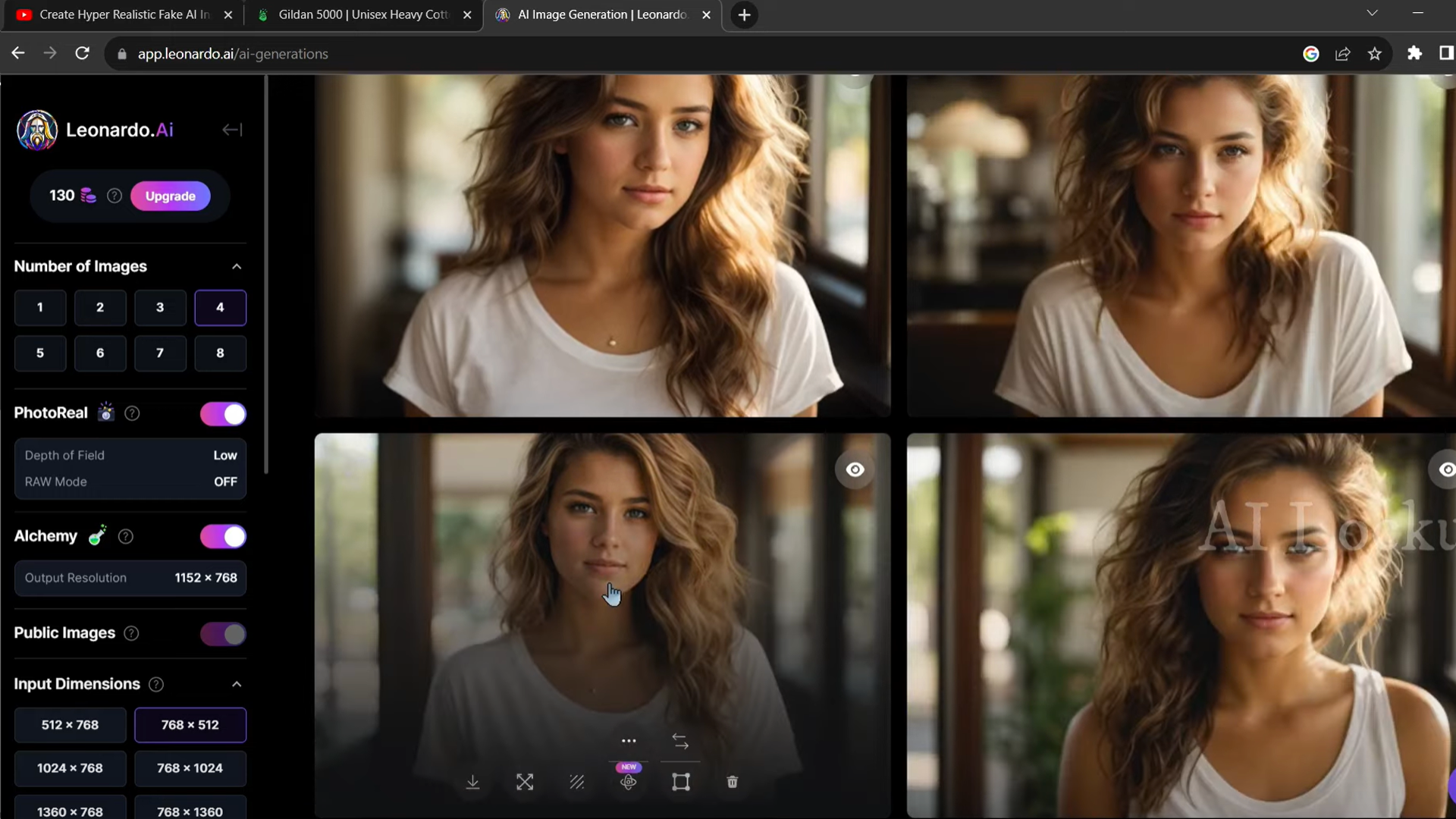}
        \caption{(b) Create an AI influencer with Leonardo.AI.} 
        \label{fig.SMI2}
    \end{subfigure}
    \end{tabular}
    \caption{Examples of monetization using AI-generated influencers.}
    \Description{This image shows Examples of Monetization with AI-generated Influencers. (1) a: HeyGen AI Influencer Avatar. This image shows a realistic AI-generated avatar standing in what resembles an office or studio background. The avatar is designed to mimic a human influencer, complete with lip-synced speech and natural gestures. (2) b: Leonardo.AI Hyper-Realistic Influencer Portrait. This portrait features a highly polished, human-like image of a virtual influencer created with Leonardo.AI. The figure has fashion model–like styling, meant for use in product endorsements.}
    \label{fig.SMI}
\end{figure}

A few YouTubers ($N=14,3.7\%$) share a use case for creating virtual influencers on social media platforms using GenAI. Using GenAI tools to create influencers involves designing virtual avatars with attractive personalities or professional knowledge to attract a fan base. For example, one YouTuber selects HeyGen AI to access pre-made avatars, including human-like profiles and realistic environment settings (e.g., an office). They then uses HeyGen's text-to-video feature to input video scripts, which are transformed into complete videos with AI-generated voice-overs and lip-syncing (\autoref{fig.SMI1}). In another video, a YouTuber suggests creating a hyper-realistic AI influencer with Leonardo.AI to promote products (\autoref{fig.SMI2}).

\subsection{RQ2: GenAI4Money Models}
We categorized how creators build knowledge and develop practices for using GenAI to influence the monetization models of \textit{Advertisement}, \textit{Subscription}, and \textit{Transaction}~\cite{Hayes2011MonetizatoinModel}. This section discusses these monetization models and their associations with the identified use cases (\autoref{fig:association}).

\subsubsection{Advertisement}
Advertisement is the most frequently mentioned monetization model, appearing in about half of the videos ($N=181, 48.0\%$). In our study, the advertisement model refers to using AI-generated content to direct users to affiliate links and earn profits from the product marketer. The Chi-square analysis indicates a significant positive association between the \textit{Advertisement} model and \textit{Blog} ($\chi^2=65.13$, $p<0.001$), \textit{Product Promotion Video} ($\chi^2=43.34$, $p<0.001$), and \textit{Web Design} ($\chi^2=11.75$, $p<0.001$) use cases, suggesting that these forms of AI-generated content are better suited for this monetization model. 
\par
\begin{figure}[!h]
\centering
    \begin{tabular}{lll}
        \begin{subfigure}[t]{.4\textwidth}
        \includegraphics[width=\textwidth]{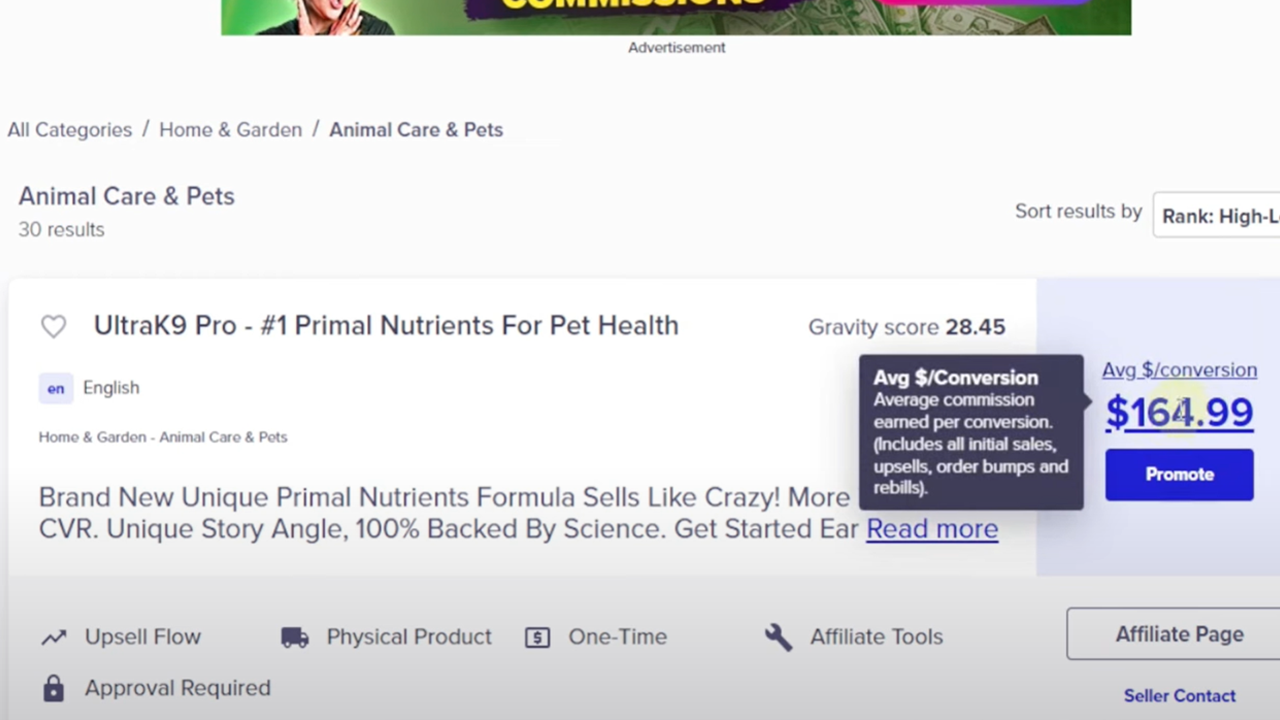}
        \caption{(a) Get promotion link on Passively.com.} 
        \label{fig.model1.1}
        \end{subfigure}
        &
        \begin{subfigure}[t]{.4\textwidth}
        \includegraphics[width=\textwidth]{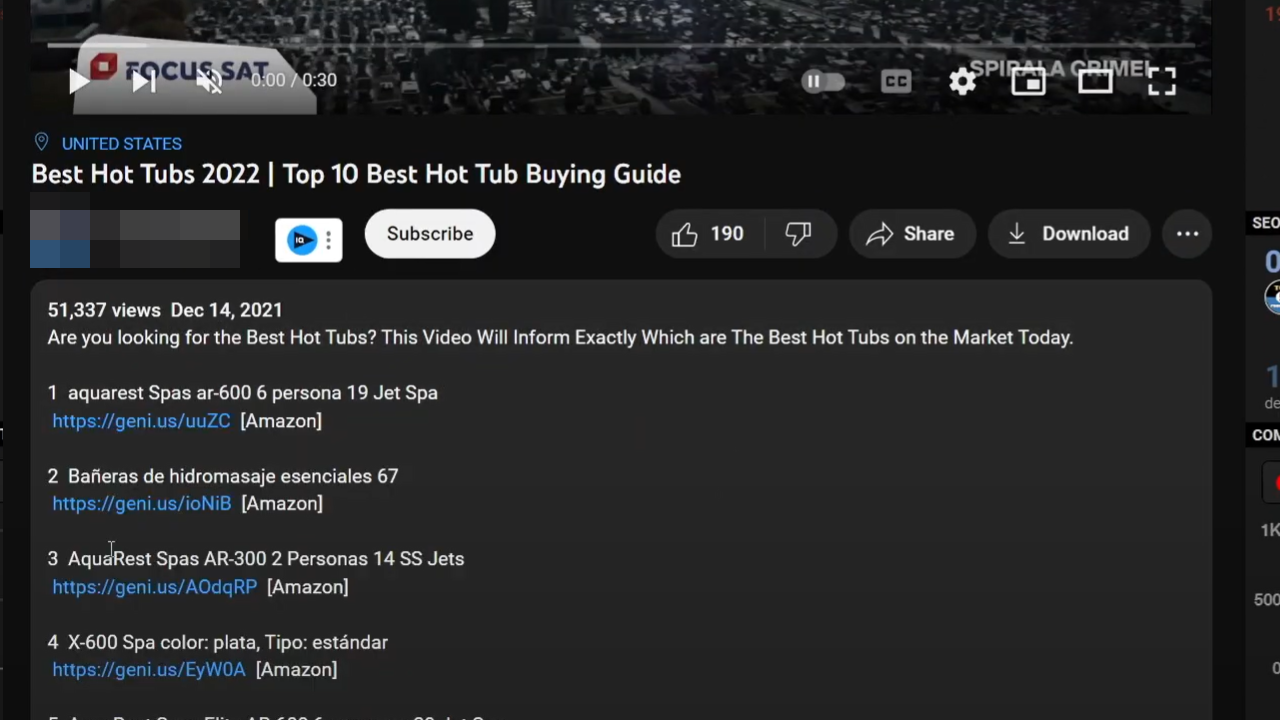}
        \caption{(b) Add links under AI videos on YouTube.} 
        \label{fig.model1.2}
        \end{subfigure}
    \end{tabular}
\caption{Examples of monetization in the advertisement model.}
\Description{This image shows Examples of Monetization in the Advertisement Model.(1) a: Affiliate Link on Passively.com. This screenshot shows a dashboard page on Passively.com with an option to “Get Promotion Link.” The interface highlights the simple process for creators to obtain affiliate URLs used in monetized AI content. (2) b: Affiliate Links in YouTube Video Description. This image shows a YouTube video description box containing multiple affiliate URLs placed under an AI-generated video. The links are presented as product recommendations or resources.
}
\label{fig.model1}
\end{figure}

In the \textit{Blog} use case, YouTubers recommend affiliate marketplaces such as Clickbank.com, Amazon Associates, and ProfitPassively.com for earning commissions (e.g., \autoref{fig.model1.1}) and embedding affiliate links in blog content. Similarly, \textit{Web Design} videos use AI-generated blogging templates to build affiliate marketing websites and link them to product promoters. The \textit{Product Promotion Video} use case also features various affiliate marketplaces, but typically involves creating videos styled as buying guides or customer reviews. In these cases, affiliate links are often embedded in the video description (\autoref{fig.model1.2}).
\par
Videos in the GenAI4Money model indicate that GenAI tools are framed as an efficient means of creating content tailored to specific product types and designed to attract viewers -- ultimately encouraging clicks on embedded affiliate links and generating commissions. This model positions GenAI as an attention-capturing tool, where seemingly plausible and detailed online content, aligned with users' ``niche interests,'' is generated by AI to serve as a hook that redirects potential traffic to affiliate link URLs.

\subsubsection{Transaction}
The transaction model, the second most publicly taught strategy in the videos, accounts for over one-third of the sample ($N=149, 39.5\%$). In this model, creators suggest practices of selling products made with GenAI or offering services facilitated by GenAI, enabling them to earn profits directly from customers \cite{Hayes2011MonetizatoinModel}. The Chi-square test indicates that this model is significantly associated with the \textit{Graphic Design} ($\chi^2=167.93$, $p<0.001$) and \textit{E-Book} ($\chi^2=34.75$, $p<0.001$) use cases. 

\par
\begin{figure}[!h]
\centering
    \begin{tabular}{lll}
        
        
        \begin{subfigure}[t]{.4\textwidth}
        \includegraphics[width=\linewidth]{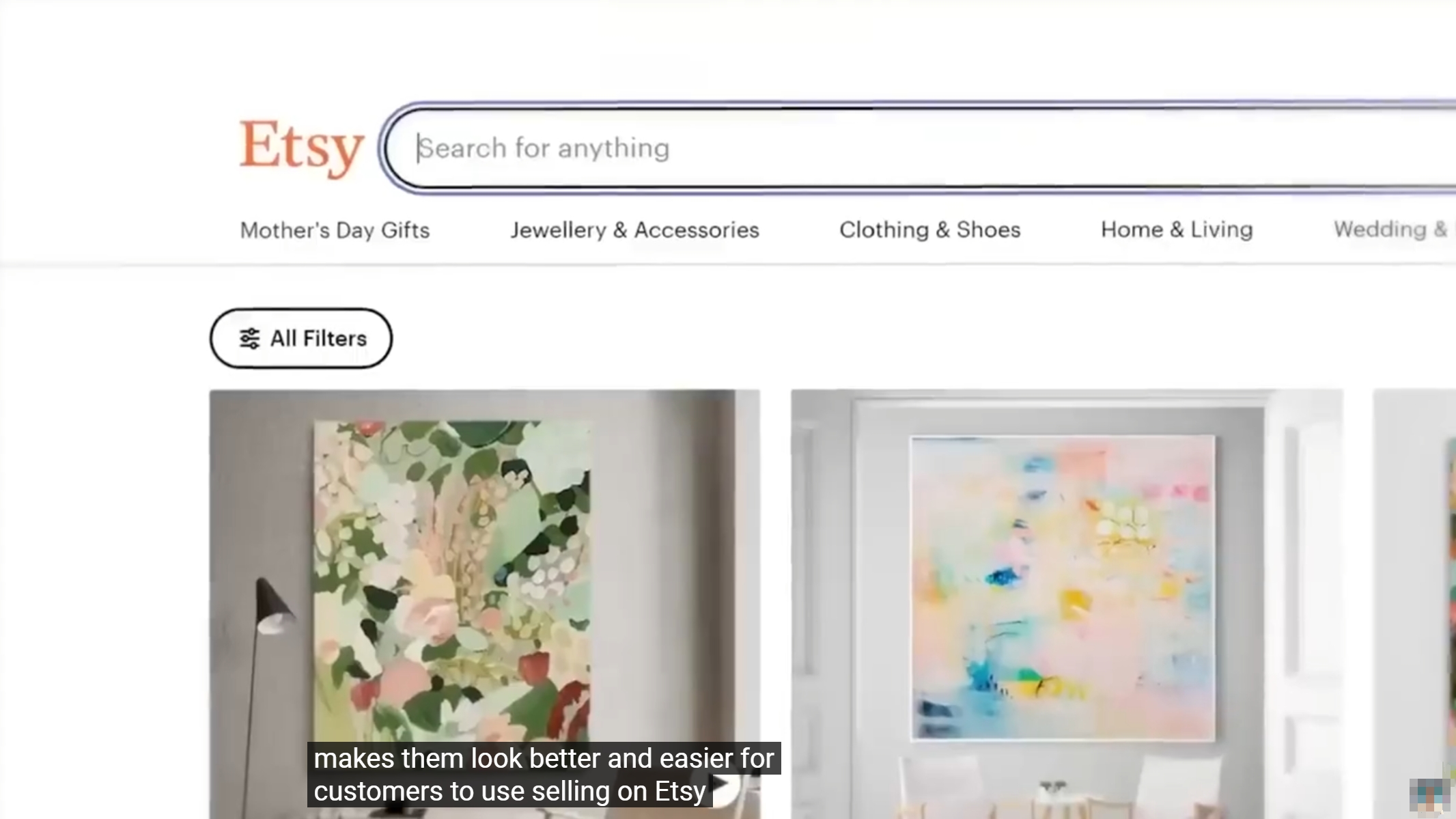}
        \caption{(a) Sell AI-generated designs on Esty.} 
        \label{fig.B4} 
        \end{subfigure}
        &
        \begin{subfigure}[t]{.4\textwidth}
        \includegraphics[width=\linewidth]{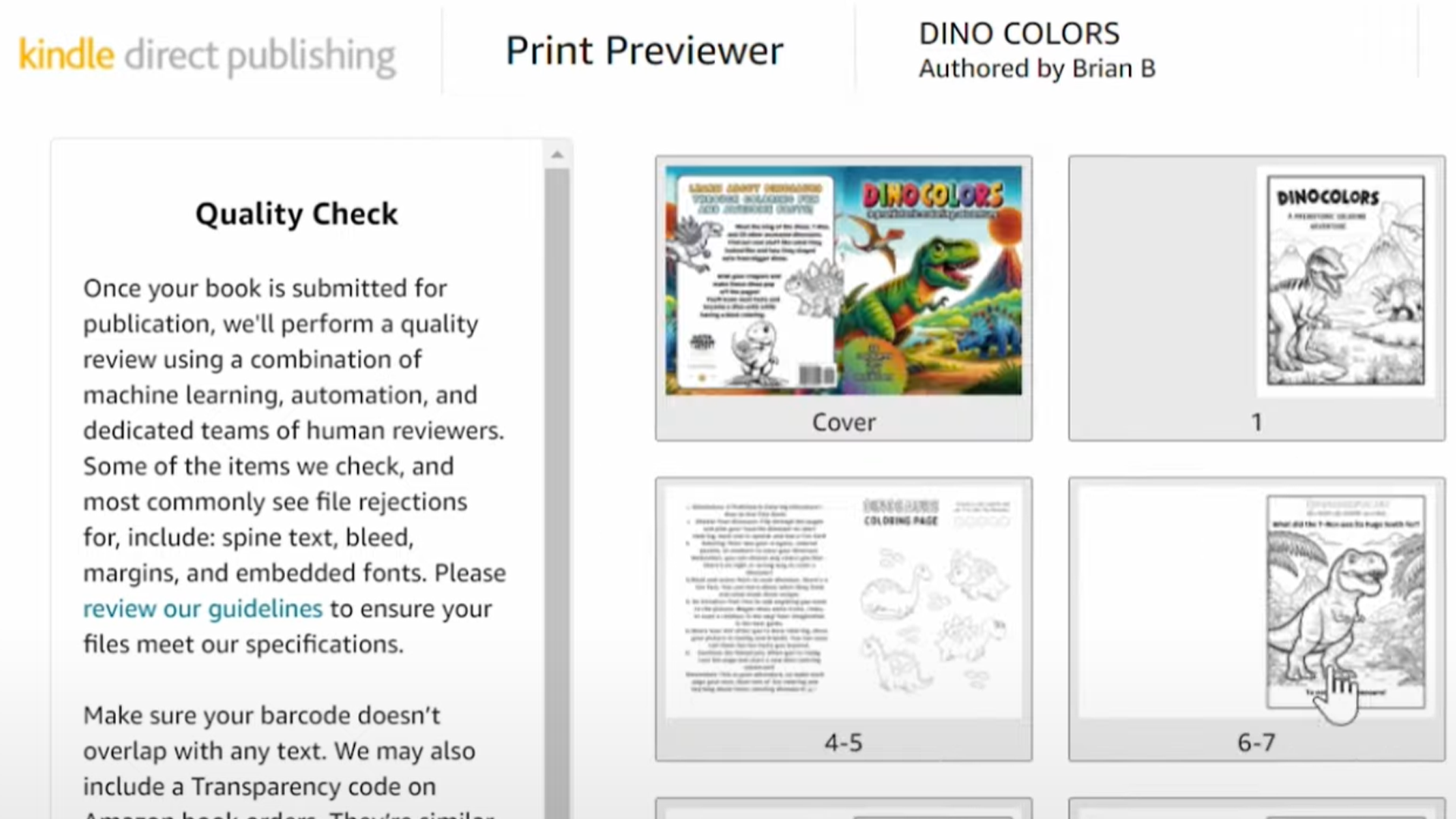}
        \caption{(b) Sell an GenAI storybook on Kindle Direct Publishing.} 
        \label{fig.B5} 
        \end{subfigure}        
    \end{tabular}
\caption{Examples of monetization in the transaction model.}
\Description{This image shows Examples of Monetization in the Transaction Model.  (1) a: Etsy Listing for AI-Generated Designs. This image shows an Etsy shop page featuring digital clipart or printable designs created with GenAI. The layout includes product thumbnails, pricing, and a purchase button. (2) b: Kindle Direct Publishing E-Book Upload. The screenshot shows the Kindle Direct Publishing interface where an AI-generated children's book—complete with AI-written text and illustrations—is being prepared for publishing.}
\label{fig.ExampleB}
\end{figure}

In the \textit{Graphic Design} use case, YouTubers teach how to use AI tools to create digital or merchandise products for direct sale on platforms such as Etsy, Redbubble, Amazon, and Shopify. Some also recommend offering freelance services or gig work on platforms like Fiverr. For example, one YouTuber illustrates how to profit from digital art by showcasing watercolor clipart sales and using MidJourney to generate similar content for Etsy (\autoref{fig.B4}). Similarly, creators can use AI tools to write \textit{E-Books}, even without subject-matter expertise, and sell them through digital publishing platforms like Kindle or as physical prints on Amazon (e.g., \autoref{fig.B5}).
\par
GenAI is introduced as a design tool, while print-on-demand services facilitate the production of AI-generated paintings and e-books. Use cases within this GenAI4Money model often publicly promote such GenAI products as ``creative work'' on platforms originally designed for creative professionals (e.g., Etsy) or independent authors (e.g., Kindle Direct Publishing).

\subsubsection{Subscription}
Around one-fourth of the sampled videos share knowledge about using GenAI for the \textit{Subscription} model ($N=93,24.7\%$). In this monetization model, YouTubers generate revenue through paid memberships or revenue-sharing programs offered by social media platforms. These GenAI4Money techniques focus on the mass creation and uploading of GenAI content, noting that creators can potentially earn cumulative income. A Chi-square test reveals significant associations between the \textit{Subscription} model and the use cases of \textit{Trending Video} ($\chi^2=189.88$, $p<0.001$) and \textit{Reformatting} ($\chi^2=23.00$, $p<0.001$).

\par

\begin{figure}[!h]
\centering
    \begin{tabular}{ll}
        \begin{subfigure}[t]{.4\textwidth}
        \includegraphics[width=\linewidth]{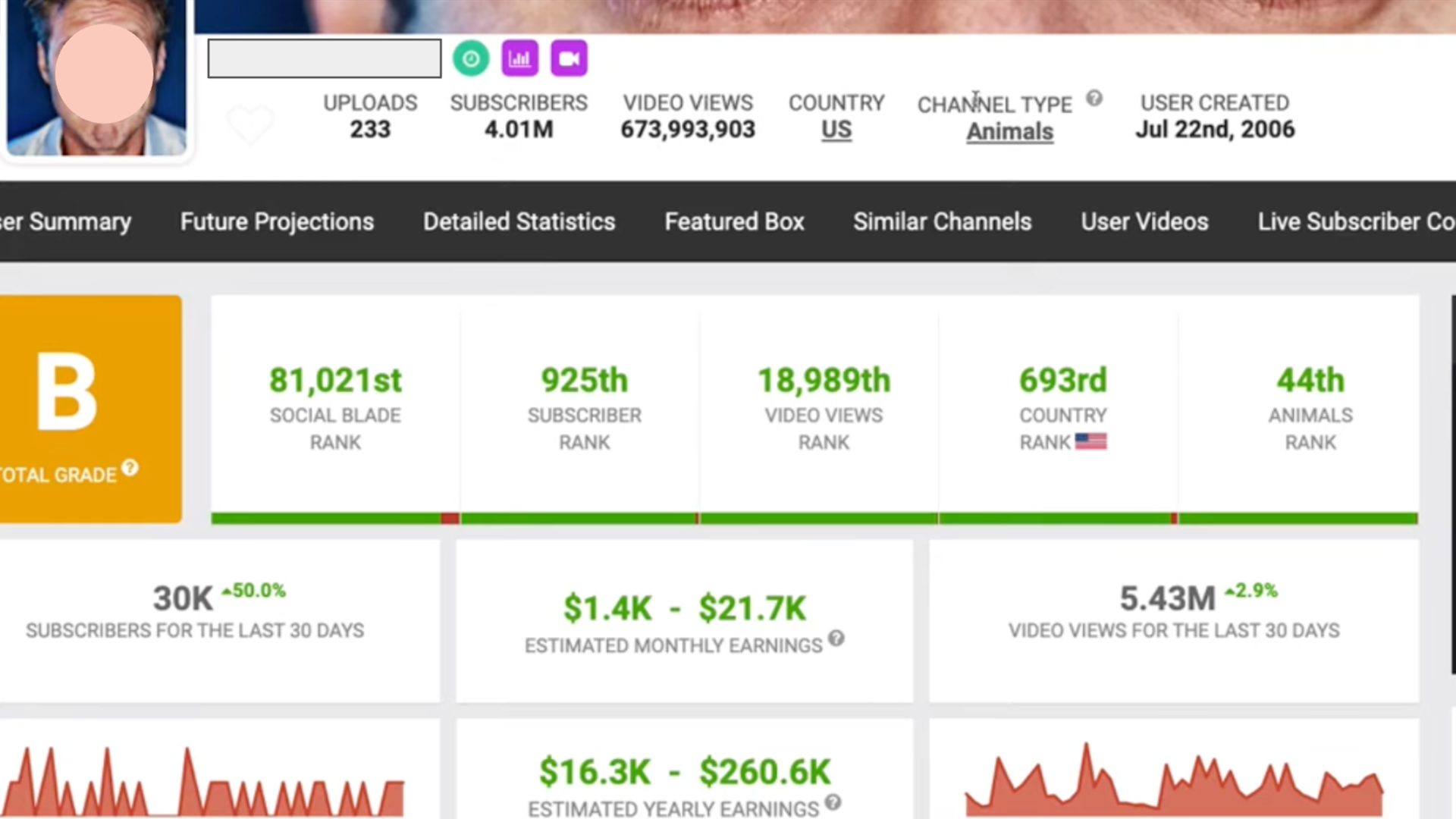}
        \caption{(a) YouTuber's subscription rate and income.} 
        \label{fig.C0}
        \end{subfigure}
        &
        
        \begin{subfigure}[t]{.4\textwidth}
        \includegraphics[width=\linewidth]{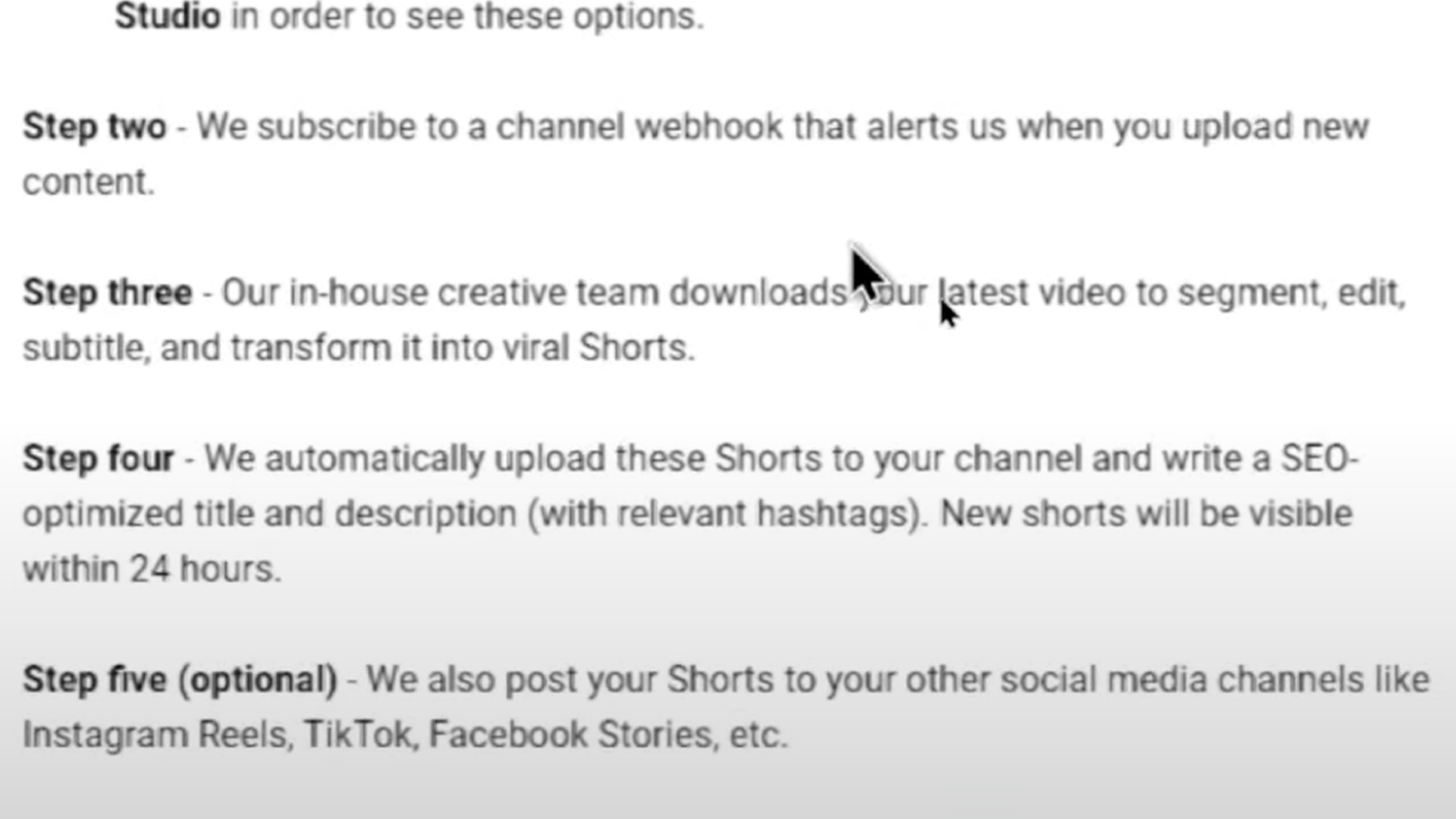}
        \caption{(b) A YouTuber suggests modifying and sharing popular videos.} 
        \label{fig.C1}
        \end{subfigure}
        
    \end{tabular}
\caption{Examples of monetization in the subscription model.}
\Description{This image shows Examples of Monetization in the Subscription Model. (1) a: YouTube Monetization Dashboard. This image shows a creator's YouTube analytics page highlighting subscriber count, video views, and estimated revenue. It demonstrates income generated from mass-produced AI content uploads. (2) b: Workflow for Reposting Popular Videos. This screenshot shows a dashboard or editing interface where a creator is modifying and republishing popular videos from other channels. It highlights the process of using AI tools to reformat and reuse trending content for monetization.}
\label{fig.ExampleC}
\end{figure}
Creators demonstrate AI tools such as ChatGPT and InVideo AI to generate \textit{Trending Videos} or \textit{Reformat} existing content. These use cases often align with trending topics to optimize chances of being recommended by algorithms, thereby increasing traffic and revenue. For example, one video shows a narrator using GenAI to create a YouTube channel with numerous simply made AI videos and displaying the earnings achieved (\autoref{fig.C0}). In a \textit{Reformatting} example, a creator monitors uploads from popular channels and uses GenAI to produce short versions, which are then uploaded to a different channel for monetization (\autoref{fig.C1}).

\par
The subscription-based GenAI4Money model illustrates how creators develop new practices that leverage GenAI to produce videos with greater potential for popularity. Techniques include aligning GenAI content with trending topics on video-sharing platforms or adapting already popular content. These practices reflect how platform monetization models influence creators' content choices and their intent to increase viewership and subscriptions through the strategic use of GenAI.

\subsection{RQ3: GenAI4Money Challenges}
For RQ3, we annotate four challenges and analyze the associations between GenAI4Money use cases and the three most common challenge categories within the collective GenAI4Money knowledge (\autoref{fig:association}).

\subsubsection{Non-Verification} 
Non-verification is the most common challenge type ($N=125,33.2\%$). In the demonstration, YouTubers often demonstrate excessive trust in the answers provided by GenAI \cite{Nah2023GenAI}, relying on these AI tools as experts to generate complete content or suggestions without further validation. Chi-square analysis reveals significant associations between this ethical subcategory and the use cases of \textit{Trending Video} ($\chi^2=11.37$, $p<0.001$) and \textit{Blog} ($\chi^2=25.15$, $p<0.001$). 
\par
\begin{figure}[!h]
\centering
    \begin{tabular}{ll}
        \begin{subfigure}[t]{.4\textwidth}
        \includegraphics[width=\textwidth]{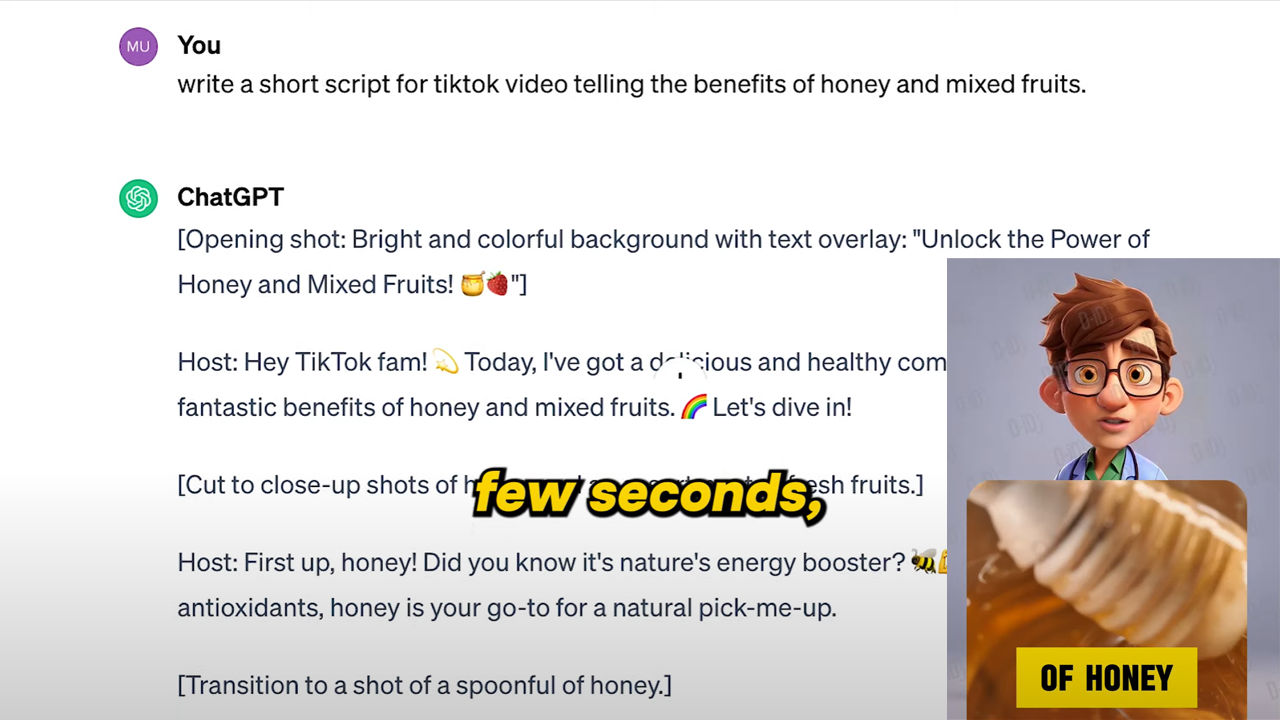}
        \caption{(a) Generate script on food with ChatGPT.} 
        \label{fig.ethic1.1}
        \end{subfigure}
        &
        \begin{subfigure}[t]{.4\textwidth}
        \includegraphics[width=\textwidth]{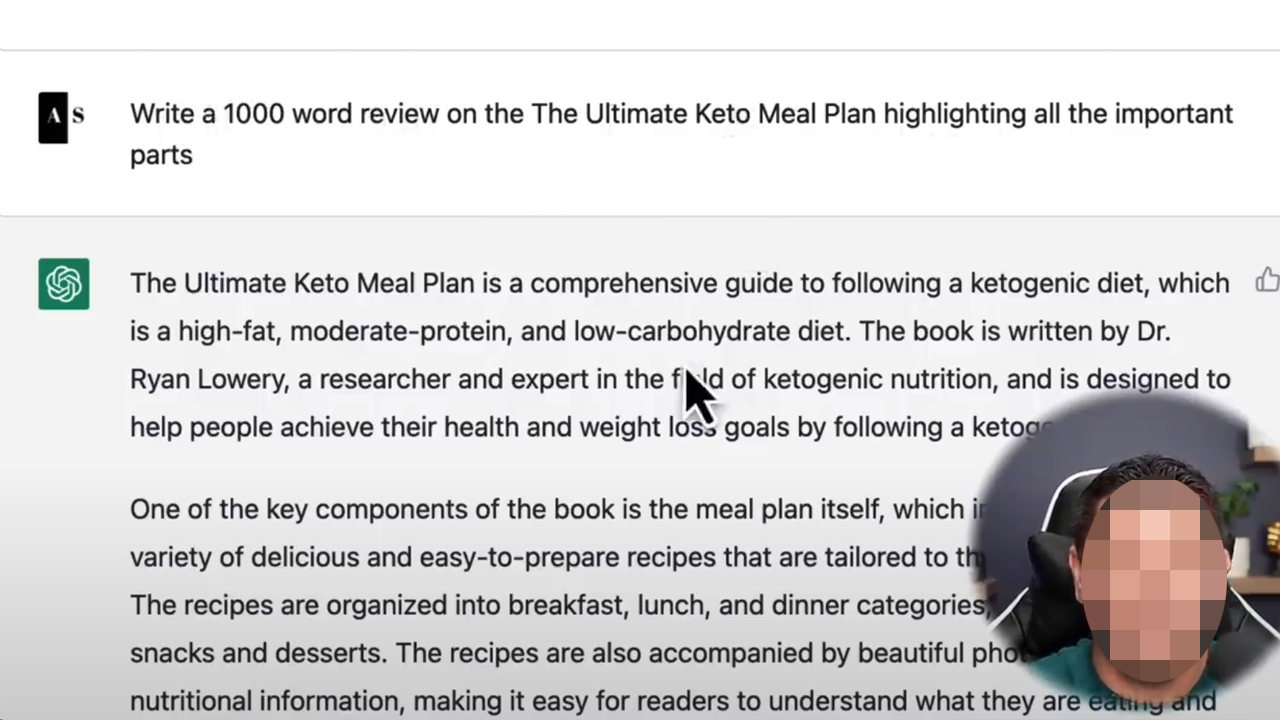}
        \caption{(b) Write review on a book with ChatGPT.} 
        \label{fig.ethic1.2}
        \end{subfigure}
    \end{tabular}
\caption{Examples of the non-verification challenge.}

\Description{This image shows Examples of the Non-Verification Challenge. (1) a: ChatGPT Script for Food and Nutrition Video. This image shows ChatGPT outputting a short script describing health benefits of honey and fruit. The script is polished and ready to be voiced over for an educational short-form video. (2) b: AI-Generated Book Review Blog. This screenshot displays a book review article fully written by ChatGPT. The content appears complete and publication-ready, without signs of human revision.}
\label{fig.ethic1}
\end{figure}

In the \textit{Trending Video} use case, YouTubers often show using LLM-recommended topics without regard for their own expertise in the niche (e.g., health, pets, travel). Educational and how-to videos may rely entirely on scripts generated by GenAI. For example, in \autoref{fig.ethic1.1}, a YouTuber demonstrates how to use ChatGPT to script a short-form video about honey and mixed fruit that introduces their nutritional benefits, which we annotate as demonstrating the non-verification challenge because the entire knowledge-based script was generated by ChatGPT without checking. Similarly, in the \textit{Blog} use case, creators demonstrate how GenAI can generate complete blog posts that include both AI-produced text and images, as illustrated in \autoref{fig.ethic1.2}.
\par
Such use of GenAI for social media content demonstrates over-reliance on AI-generated material, which can be inaccurate or misleading. Creating content with GenAI may result in content creators passing on AI mistakes or errors to viewers through AI-generated blogs and videos.

\subsubsection{Misappropriation}

In our data, 34 videos (9.0\%) exhibit traits where original content created by others, which could be protected under copyright laws and regulations, is used as input for GenAI. Chi-square tests reveal significant associations with the \textit{Reformatting} ($\chi^2=49.89$, $p<0.001$) and \textit{Influencer} ($\chi^2=29.76$, $p<0.001$) use cases.
\par
\begin{figure}[!h]
    \centering
    \begin{tabular}{lll}
    \begin{subfigure}[t]{0.4\textwidth}
        \centering
        \includegraphics[width=\linewidth]{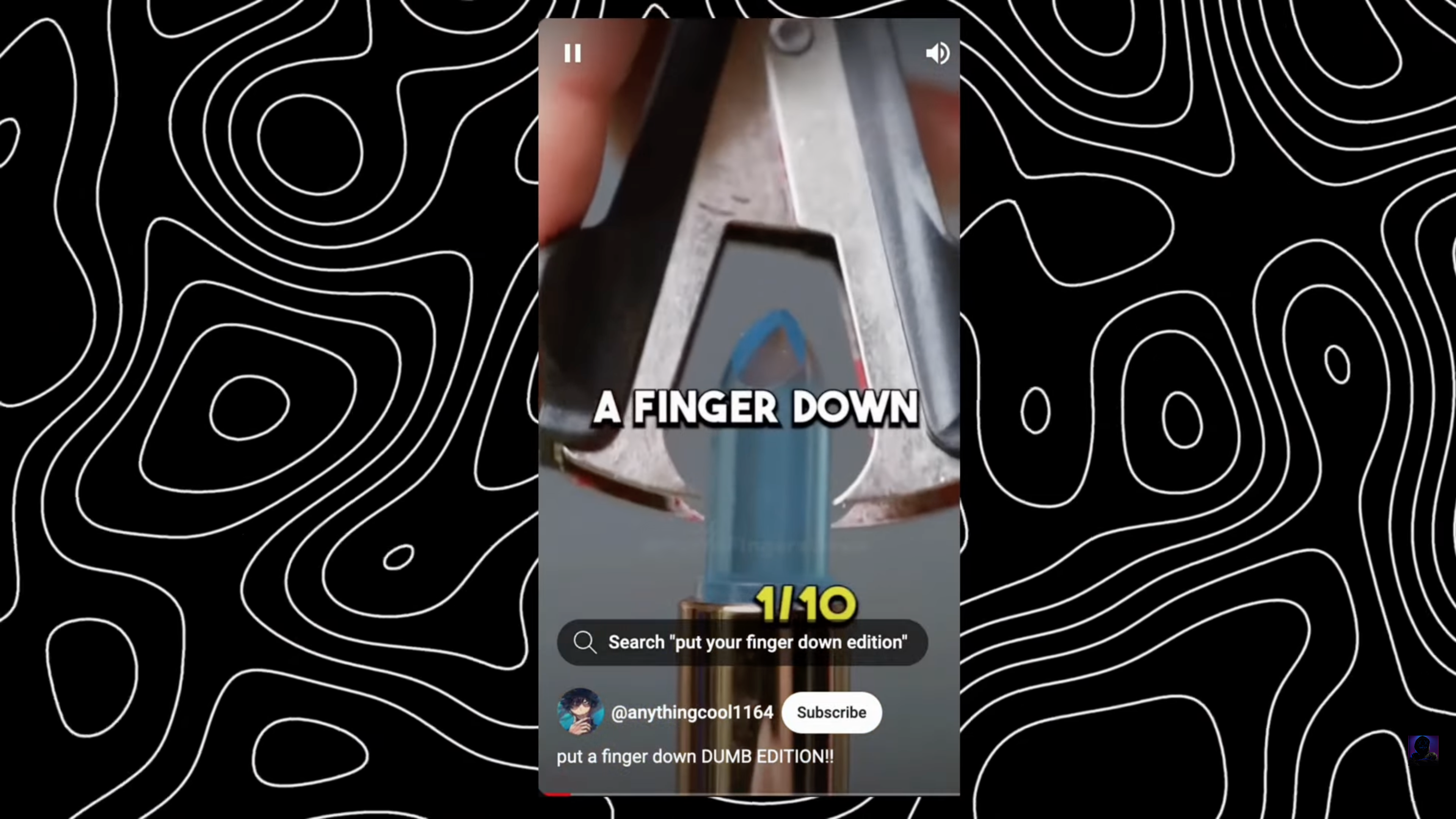}
        \caption{(a) Modifying the ``Put Your Fingers Down'' video.} 
        \label{fig.Copy1}
    \end{subfigure}
        &
    \begin{subfigure}[t]{0.4\textwidth}
        \centering
        \includegraphics[width=\linewidth]{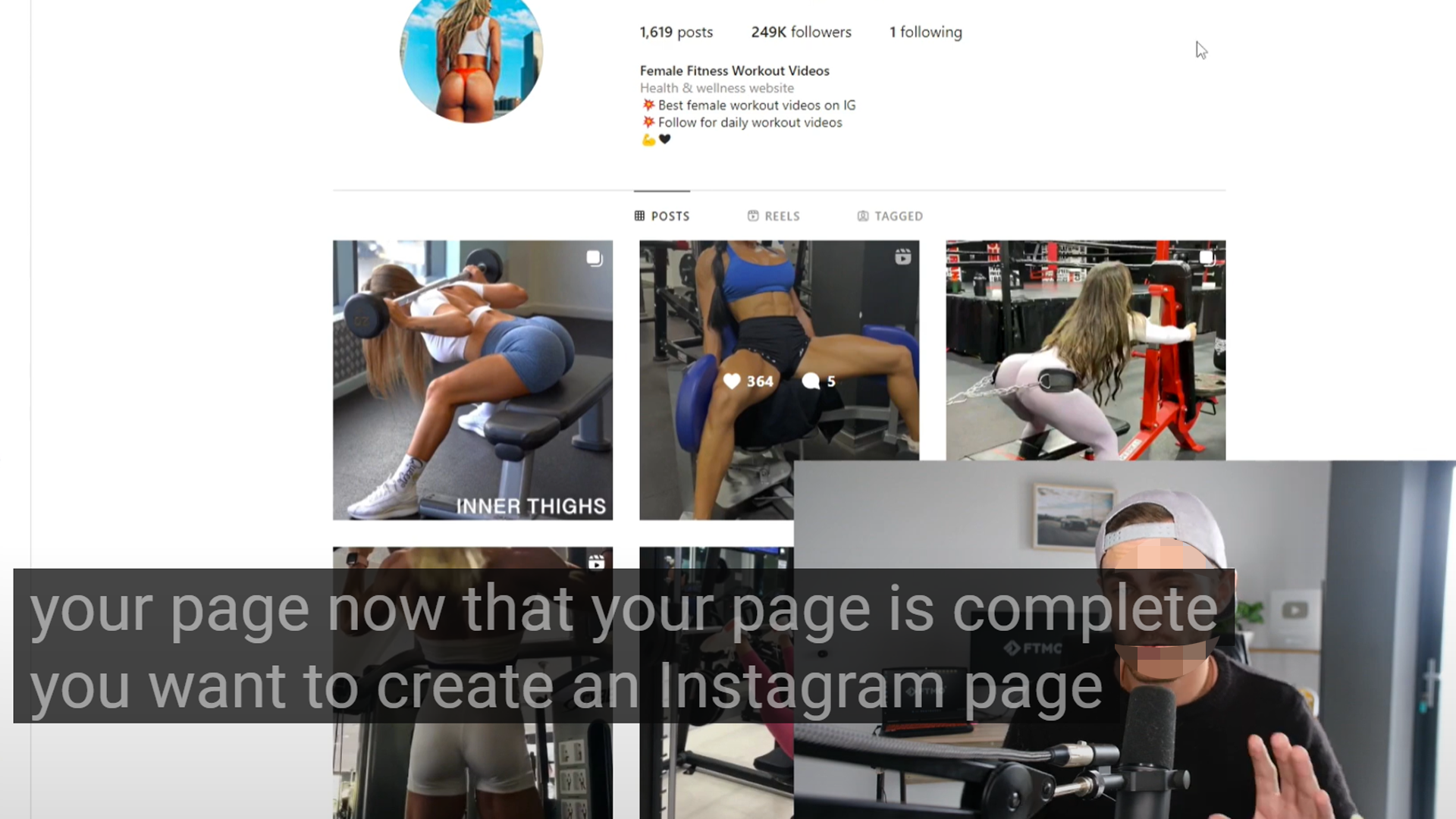}
        \caption{(b) Use ChatGPT to convert online images to a Instagram post.} 
        \label{fig.Copy2}
    \end{subfigure}
    \end{tabular}
    \caption{Examples of the misappropriation challenge.}
    \label{fig.Example-Copy-right}
    \Description{This image shows Examples of the Misappropriation Challenge. (1) a: Modified “Put Your Fingers Down” Video. This image shows a video in the “Put Your Fingers Down” trend, where a creator uses AI tools to rewrite prompts and add AI-generated voiceovers while retaining original clips sourced from other people's videos. (2) b: Instagram Post Made from Others' Images. This screenshot shows a workflow where a creator uses someone else's online photos, has ChatGPT generate captions or text overlays, and assembles the output as a new Instagram fitness post.}
\end{figure}

YouTubers present methods for modifying and repurposing content created by others, then presenting the results as their own. For example, one YouTuber demonstrates how to build a faceless channel by producing trending videos in the ``Put Your Fingers Down'' niche (a popular TikTok trend, shown in \autoref{fig.Copy1}). This video is a case of misappropriation because the creator mentions using visual content composed of clips taken from others' original videos, whereas they use ChatGPT to transform the scripts and incorporate AI-generated voiceovers. In another \textit{Influencer} example, a YouTuber advises using others' images to prompt ChatGPT and then combining the resulting AI-generated text and images to launch an Instagram account focused on workouts (\autoref{fig.Copy2}).
\par
In the GenAI4Money context, the profit-driven repurposing of others' content may undermine original creators' rights and diminish their opportunities for visibility and popularity \cite{Nah2023GenAI}. GenAI is often presented as a tool for circumventing copyright restrictions. YouTubers frequently publicly assert that GenAI sufficiently transforms the content to avoid copyright strikes under YouTube's content moderation policies.

\subsubsection{Synthetic Human Activity} 
\begin{figure}[!h]
    \centering
    \begin{tabular}{ll}
    \begin{subfigure}[t]{0.4\textwidth}
        \centering
        \includegraphics[width=\linewidth]{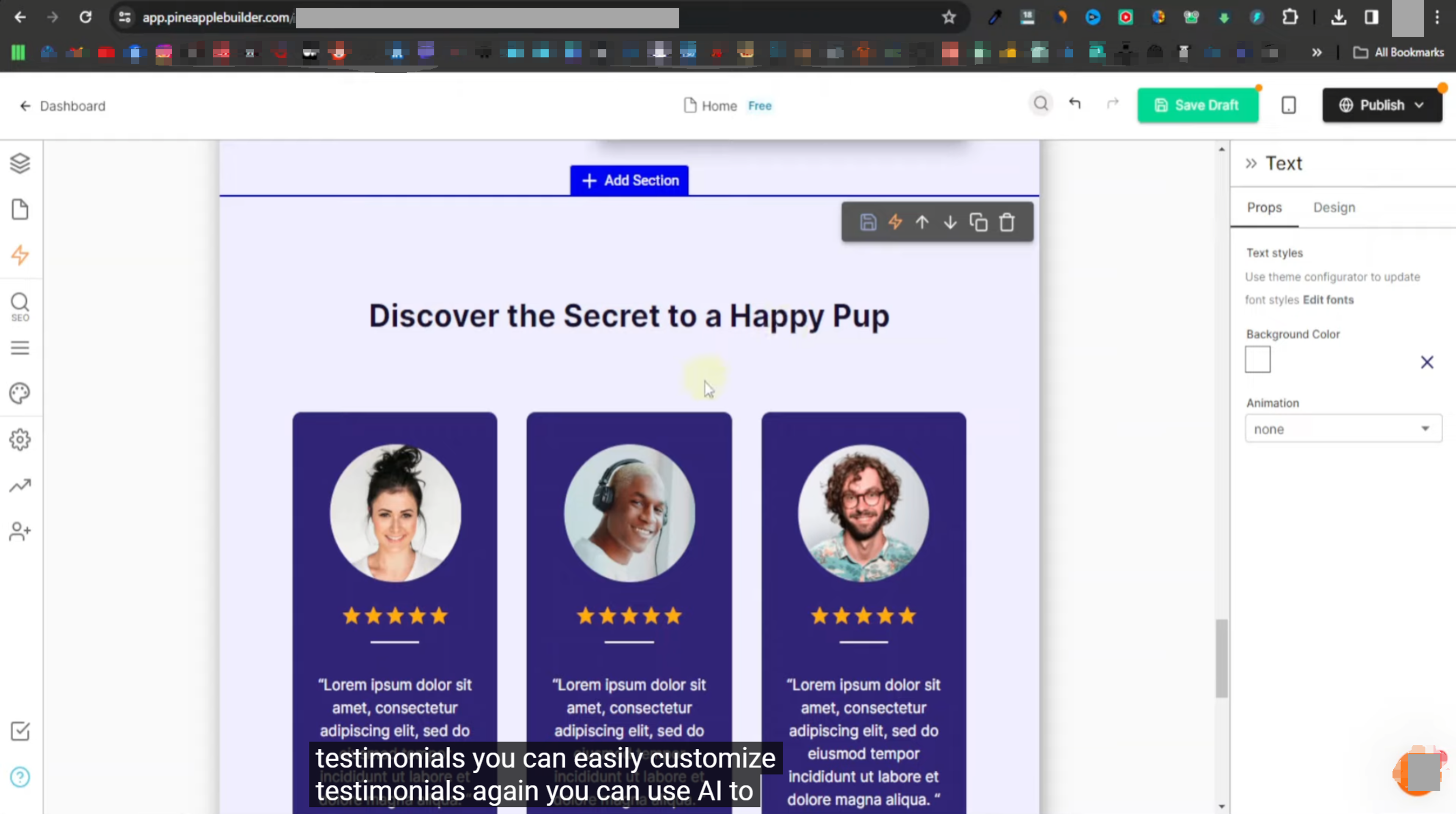}
        \caption{(a) Fake reviews on a product page.} 
        \label{fig.Auth1}
    \end{subfigure}
        &
    \begin{subfigure}[t]{0.4\textwidth}
        \centering
        \includegraphics[width=\linewidth]{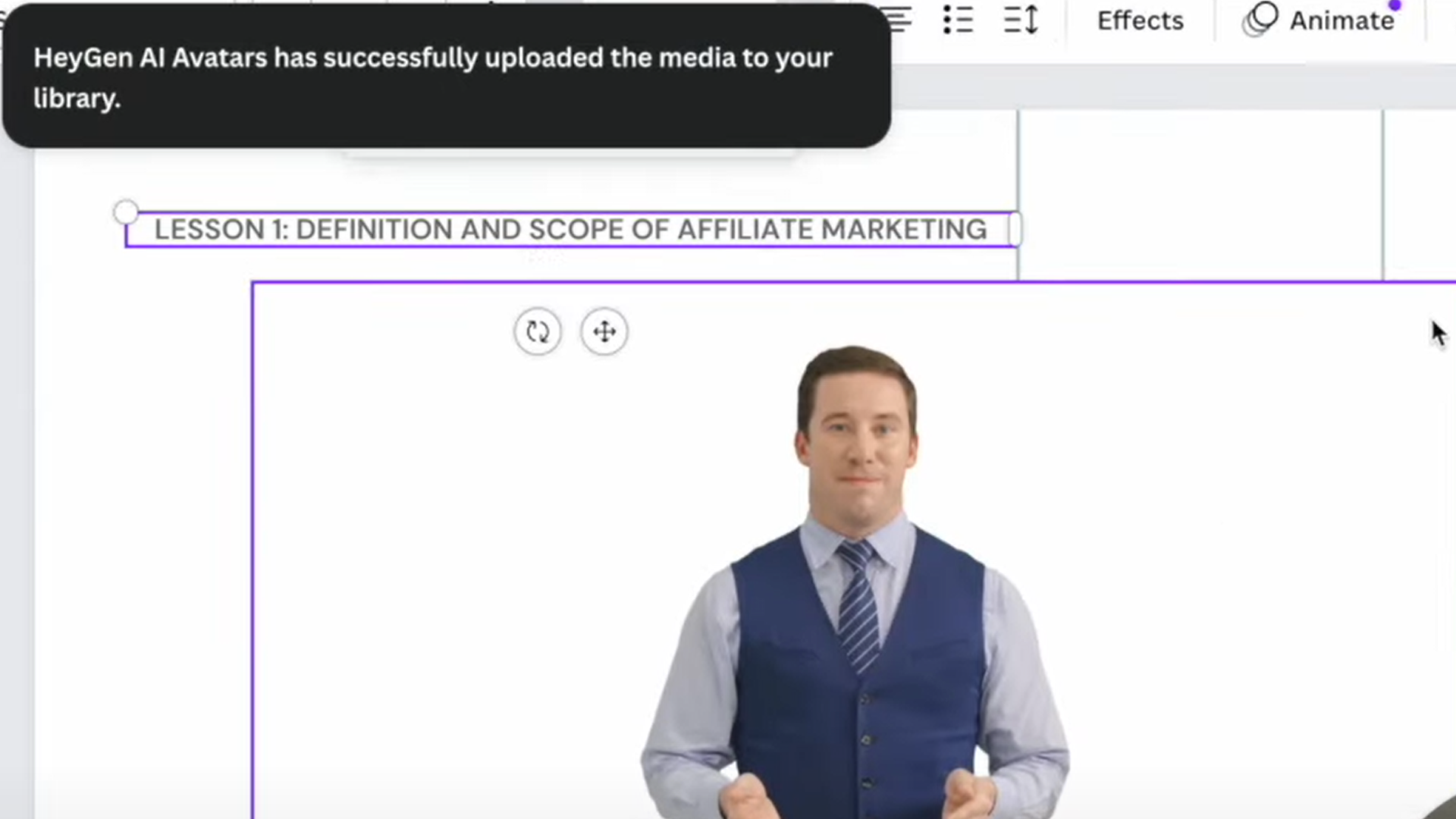}
        \caption{(b) An AI avatar is used to teach a lesson.} 
        \label{fig.Auth2}
    \end{subfigure}
    \end{tabular}
    \caption{Examples of the synthetic human activity challenge and the explicit content challenge.}
    \Description{This image shows Examples of the Synthetic Human Activity Challenge and the Explicit Content Challenge. (1) a: AI-Generated Fake Reviews. This image shows a product page filled with fictional customer reviews. The reviews include invented profile pictures and names that were generated by AI to simulate authentic user engagement. (2) b: AI Avatar Teaching an Online Course. This image shows an educational video with an AI-generated instructor speaking to the camera. The avatar presents course content in a manner designed to resemble a human teacher, intended for upload to online course platforms. }
    \label{fig.Example-Authenticity-Issues}
\end{figure}

We identified 19 videos (5.0\%) demonstrating GenAI being used to fabricate human activities. This knowledge refers to prompting GenAI to generate images or videos that mimic real people, real-world scenes, or human comments. For example, one YouTuber demonstrates how to promote a pet product using an AI-generated website. We annotate this video as containing this challenge because the YouTuber adds fake customer reviews generated by ChatGPT, including AI-generated profile pictures and names (\autoref{fig.Auth1}). In another example, a YouTuber uses a virtual AI avatar to create an online course and claims that the avatar behaves like a human instructor, with the intent of uploading it to platforms such as Udemy to generate income (\autoref{fig.Auth2}).
\par

In this challenge, GenAI content fosters a sense of trustworthiness by generating vivid, real-world personas, scenes, and speeches. Such uses may mislead users into perceiving the content as authentic. AI-generated reviews illustrate how GenAI is employed as a credibility-enhancing tool. However, this type of content can constitute misinformation that deceives other users \cite{Nah2023GenAI}.

\subsubsection{Explicit Content} 
\begin{figure}[!h]
    \centering
    \begin{tabular}{lll}
    \begin{subfigure}[t]{0.44\textwidth}
        \centering
        \includegraphics[width=\linewidth]{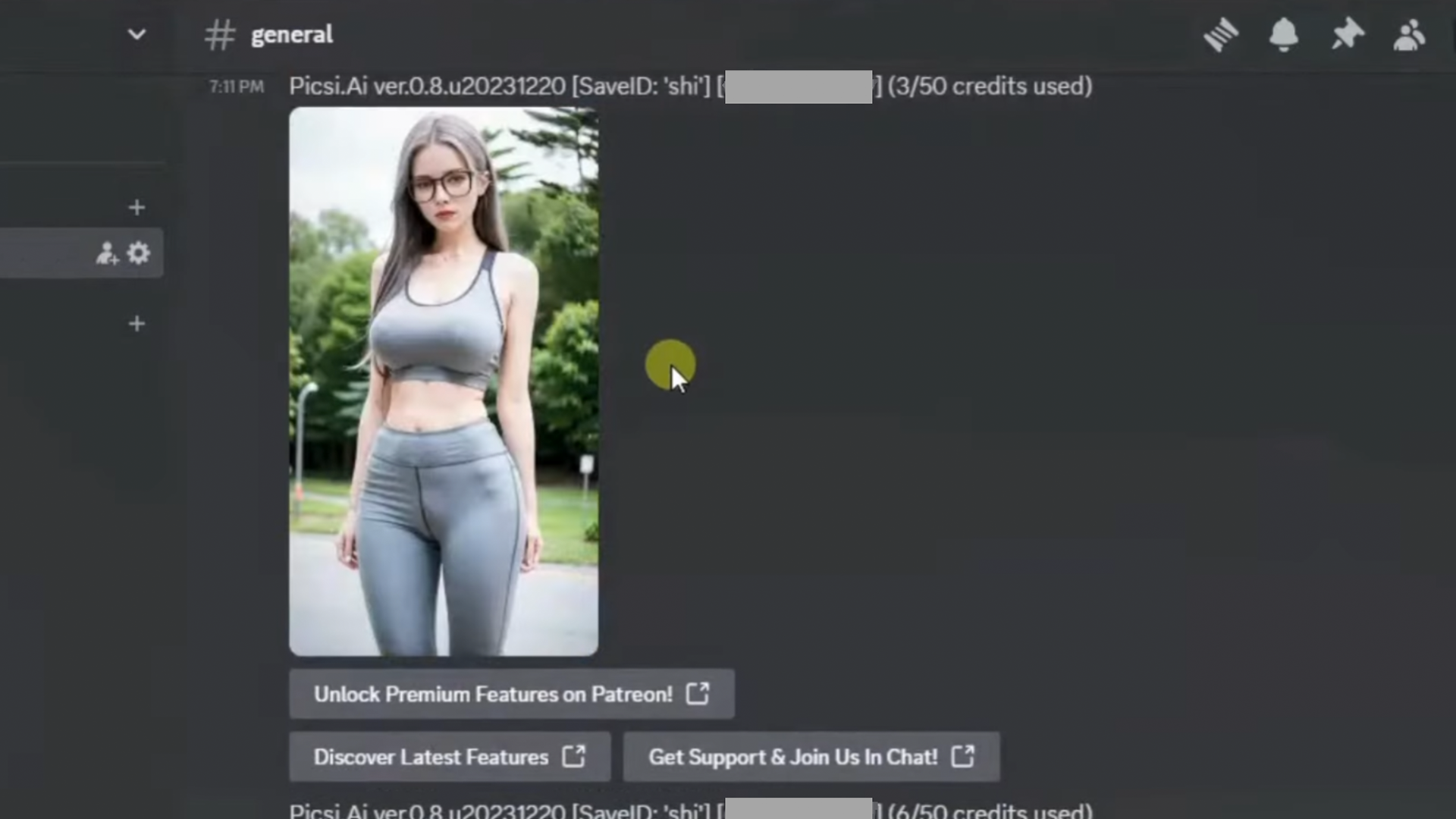}
    \end{subfigure}
    \end{tabular}
    \caption{Create AI figure and swap faces.}
    \Description{This image shows Examples of Create AI figure and swap faces. This image shows an AI-generated female figure whose face has been swapped using explicit imagery. The output is intended for deceptive or inappropriate use, such as impersonation on Tinder or other platforms.}
    \label{fig.HIC2}
\end{figure}

In our data, only two videos include explicit content (0.5\%), both of which involve generating sexualized AI-generated images of women. In one example, a YouTuber demonstrates how to create an AI avatar and perform a face swap. This video is annotated as explicit content because the face-swapping source comes from a Discord channel containing numerous explicit, semi-naked female images, and the YouTuber further suggests that such an AI-generated figure can be used for the Tinder dating app (\autoref{fig.HIC2}).

\section{Discussion}
\begin{figure}[!h]
    \centering
    \includegraphics[width=1\linewidth]{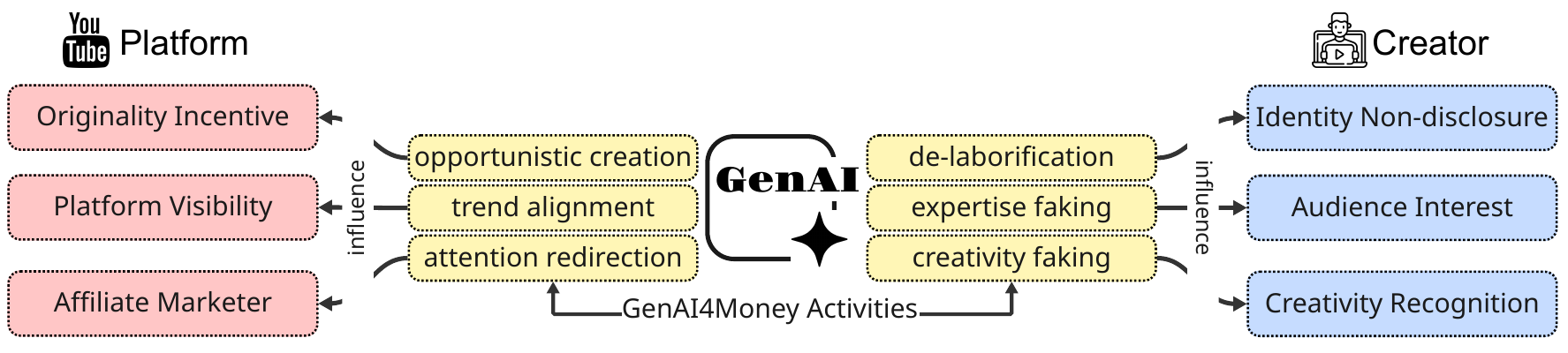}
    \caption{A framework summarizing the collective knowledge of GenAI4Money and its effects on platforms and communities.}

    \label{fig:framework}
    \Description{The diagram illustrates how YouTube's platform factors, creators' motivations, and GenAI-for-money activities influence one another. On the left, three pink boxes represent platform incentives—originality incentive, platform visibility, and affiliate marketing. These influence GenAI-related creator behaviors in the center, shown as yellow boxes: opportunistic creation, trend alignment, and attention redirection. On the right, three blue boxes represent creator-side factors—identity non-disclosure, audience interest, and creativity recognition—which are also influenced by GenAI activities such as de-laborification, expertise faking, and creativity faking. Arrows indicate bidirectional influence between the platform, GenAI activities, and creators.}
\end{figure}

Our analysis reveals the knowledge space formed around GenAI usage for monetization, highlighting discourse surrounding creative labor and platform governance. Ultimately, we present a conceptual framework (\autoref{fig:framework}) that delineates six key GenAI4Money knowledge areas developed and shared within the YouTuber community.

\subsection{GenAI4Money Influences Algorithmic, Monetization, and Moderation Infrastructures}
\subsubsection{GenAI-Driven Opportunistic Creation for Originality Incentives}
Our results indicate that YouTubers develop GenAI knowledge around \textit{opportunistic creation} practices to profit from social media platforms. In contrast to findings in prior HCI research, which emphasize AI's benefits in streamlining creative expression and production pipelines~\cite{Simpson2023CreativeWork, Choi2023Creator, Kim2024ASVG}, some YouTubers instead build knowledge around using GenAI as a shortcut to bypass the creative labor necessary for authentic expression~\cite{BartaAuthenticityTikTok}. These practices are particularly evident in demonstrations of mass-produced videos with GenAI, the reformatting of others' content for misappropriation, and the generation of entire video scripts without verification.
\par

The knowledge of GenAI4Money activities around opportunistic creation calls for rethinking how originality-promotion programs and infrastructures should be reshaped in the age of GenAI. HCI research on ``infrastructures for inspiration'' has noted that monetization programs function as sociotechnical and community structures that support creative identities~\cite{Simpson2025Infrastructures}. Such programs have been found to motivate creators to enhance production skills~\cite{DingBilibili} and foster professionalism~\cite{Ploderer2010}. However, our findings suggest that opportunistic creation may instead be amplified by the ease of producing plausible GenAI content, thereby exploiting incentive programs such as the YouTube Partner Program and TikTok Creator Rewards. In our analysis, GenAI graphics and e-books are marketed as original works on platforms such as Etsy and Kindle Direct Publishing. \textit{Trending video} and \textit{Reformatting} practices are strongly associated with the subscription model, whereas \textit{Graphic Design} and \textit{E-Book} creation align more closely with the transaction model. While prior social computing research has noted that monetization programs encourage high-status creators to contribute and improve overall content quality~\cite{KopfMonetization, Ye2024Monetization}, our findings suggest that, conversely, the involvement of GenAI may have the opposite effect, undermining authenticity and diminishing content quality on social media platforms. These dynamics should prompt platforms to redesign their monetization policies and infrastructures~\cite{Partin2020BitTwitch, Yang2024TheFuture}.

\subsubsection{GenAI for Trend Alignment and Platform Visibility}
While the visibility of content has been identified as a critical concern for creators~\cite{Bishop2019Visibility, KarizatAlgorithmicFolk}, our findings suggest that, as a new form of collective effort in response to platforms' algorithmic management~\cite{Bishop2019Visibility, MacDonald2021AlgorithmicLore, Bishop2020Algorithmic}, creators are building new practical knowledge and heuristics~\cite{Cotter2024Practical} for using GenAI in visibility work. Such knowledge of \textit{trend-alignment} may help other creators boost their professionalism and visibility on the platform. Creators demonstrate GenAI-supported media practices through which they identify content ideas that align with trending topics and algorithmic preferences across platforms. Use cases within subscription models -- such as identifying video topics with GenAI and reformatting content for cross-platform sharing -- illustrate how such GenAI knowledge is used to generate content with high popularity potential. Some creators also demonstrate how to rely on GenAI to produce SEO-optimized titles and tags or to automate newsletters for strengthening audience relationships.

\par

Social computing research has noted that creators appropriate and cope with infrastructures such as algorithmic recommendation systems and monetization policies~\cite{Lyu2025Systematic, Yang2024TheFuture, Eschebach2025Playing}. Creators' knowledge of trend-alignment with GenAI represents a new pathway for making sense of how algorithms work and for attempting to work with algorithmic governance~\cite{Verviebe2026theAlgorithm, Reynolds2024User}. This new practice introduces both positive and negative dynamics into professional infrastructures. On the positive side, creators frame GenAI as a novel affordance for adapting to platform preferences and advertiser-friendly formats~\cite{MaAlgorithmicContent, Omen2022YouTubeMonetization}. GenAI is taught as a means to research audience interests and respond to such demand~\cite{Kojah2025CreativeLabor}. On the other hand, theories of algorithmic creative labor have shown that YouTube's recommendation and monetization algorithms pressure creators to shift away from their original interests in favor of algorithmic preferences~\cite{KopfMonetization, Caplan2020Tiered, MaAlgorithmicContent}. Extending these findings, our analysis shows that GenAI-driven trend analysis may further compel some creators to sacrifice personal distinctiveness, intensifying the risk of content homogenization~\cite{Doshi2024GenAICreativity}.

\subsubsection{GenAI for Attention Redirection and Deception}
We find that some creators build knowledge around using GenAI as a powerful tool to support \textit{attention redirection}, thereby directing users toward embedded affiliate marketing links. Prior studies have noted that undisclosed affiliate marketing activities within user-generated content have created tensions between creators and audiences~\cite{Rieder2023Creator, Mathur2018Endorsements}. Extending these concerns, our study reveals the circulation of GenAI knowledge about how to review products in videos, write ad-embedded blogs, answer questions on Q\&A platforms, and build AI-generated websites -- often with undisclosed intentions of redirecting users to affiliate links. The advertisement model is particularly associated with use cases such as \textit{Blog}, \textit{Product Video}, and \textit{Web Design}. These practices introduce a new practice of endorsement, in which GenAI is leveraged to rapidly generate potentially trending content and monetize through affiliate marketing. 
\par
This GenAI4Money modality within creators' communities contrasts significantly with previously documented endorsement behaviors~\cite{Mathur2018Endorsements}, in which creators primarily added affiliate links within their creative work. GenAI may therefore shift the relationship and norms between creators and affiliate marketers, as creators' GenAI-driven practices often occur outside the direct oversight of marketers. The proliferation of low-quality GenAI content may lead to audience aversion and perceptions of inauthenticity~\cite{Sew2025AIInfluencer, Maeda2024AIParasocial}. GenAI advertisements can further undermine user trust and diminish engagement~\cite{Du2023AIAdvertising}. When affiliate links are embedded in low-quality GenAI content to redirect viewers' attention, they may inadvertently harm marketers' businesses. Furthermore, such practices can disadvantage creators who invest in producing authentic product reviews or sponsored content~\cite{Fitriani2020ReviewVideo}.

\subsection{GenAI4Money Influences Content Creation and Creative Labor}
\subsubsection{Creativity De-laborification through Identity Non-disclosure}
Our results suggest that many creators frame GenAI as a device for reducing the labor (\textit{de-laborification}) associated with publicly disclosing their identities on social media. Addressing creators' intensive social and emotional burdens has been a central focus of social computing research on content-sharing platforms~\cite{Simpson2023CreativeWork, ThomasVSPHarassment, Munoz2022Freelance, Foong2020Freelance}. Our findings highlight GenAI's potential to ease these demands through what creators in our data frequently describe as the creation of ``faceless'' content or ``faceless'' channels. Such GenAI media practices enable monetization without the need to manage a public identity. This knowledge may therefore trend toward new content production practices centered on the benefits of identity concealment, offering an alternative form of publicity that does not require creators to manage their public identities~\cite{KhamisMicroCelebrity, DingBilibili}.

\par
While GenAI is regarded as a new community practice for reducing creative labor, it may simultaneously diminish professionalism and social connection~\cite{Simpson2023CreativeWork}. In creative labor, sharing knowledge has traditionally formed informal learning spaces and helped improve authority, expertise, and community cohesion~\cite{Dezuanni2024BookTok, NiuTeamTrees, Walker2019AScaled, Chen2023MyCulture}. While YouTubers often intend to balance the goals of money-making and supporting informal learning~\cite{Dubovi2019Examining, Eschebach2025Playing}, the GenAI knowledge shared in our data suggests a shift toward reduced creative efforts and stronger monetization motives. For example, creating blogs and trending videos among interest-driven groups may increasingly rely on AI-produced information without verification. The emerging knowledge of reformatting others' content or generating AI-based influencer personas are closely associated with the misappropriation of existing creative works. Such emerging GenAI creation norms may hamper creative expression and educational intentions and limit creators' ability to foster meaningful social connections~\cite{WohnParasocialInteraction, Fitriani2020ReviewVideo}.

\subsubsection{GenAI-Faked Expertise for Interest Pandering}
Our data suggest that creators' collective knowledge depicts AI-generated content as an authoritative source, using it to \textit{fake expertise} in order to satisfy audiences' niche interests. Examples such as trending videos, blogs, merchandise, and e-books produced with GenAI illustrate how creators align their output with popular interests, even when the content lacks personal relevance or authenticity~\cite{Simpson2023CreativeWork, DingBilibili, Xia2022Millions}. Some HCI research has focused on understanding the impact of GenAI's human-likeness~\cite{Sew2025AIInfluencer, Maeda2024AIParasocial}, while creators' collective knowledge frames GenAI as an infrastructure for generating ideas and content to satisfy audience demand~\cite{Xia2022Millions}. While platform monetary incentives shape creators' topic choices and content formats~\cite{Eschebach2025Playing}, our findings indicate that creators' GenAI knowledge may further influence their topic choices and perceived expertise on the topic.
\par
However, expertise-faking with GenAI introduces a new indirect way in which AI-produced content may affect users. In our analysis, YouTubers' suggestions to use GenAI to generate \textit{Trending Video} and \textit{Blog} content are significantly associated with the challenge of non-verification, suggesting that monetization incentives may encourage irresponsible uses of GenAI. Given that GenAI contains biases and hallucinations~\cite{Bandi2023PowerGenAI, Satra2023GenAI, Nah2023GenAI}, such issues may be bypassed by creators and passed on to audiences through user-generated content.

\subsubsection{GenAI Erosion of Creativity Recognition}
Besides faking expertise, the integration of GenAI is also framed as a device for entering handcrafted and stock-image marketplaces through practices of \textit{faking creativity}, which may undermine recognition of artisans' creative work. Our results suggest that some creators recommend using GenAI to sell printed merchandise featuring GenAI graphics, digital graphic designs, and AI photographs. Platforms like Etsy were designed to support artisans in selling handmade crafts~\cite{Luckman2013Etsy, Razaq2022Etsy}, emphasizing originality, authenticity, and distinction from mass-produced products. However, such GenAI-driven practices may reshape such transactional models by altering how creators define, and how audiences perceive, products traditionally regarded as creative, such as handmade goods, digital art, and creative services.
\par
On one hand, many creators integrate GenAI into their workflows as a legitimate source of inspiration and illustration~\cite{Sun2024CreativeWorker, Yildirim2022AIDesign, Hwang2022GenAI}. On the other hand, community knowledge frames GenAI designs as cost-effective substitutes for human artistry, offering comparable visual appeal at a lower cost. Therefore, future research should examine not only how AI is embedded in creative processes~\cite{Kim2024ASVG} but also how creators and buyers assess the value of human creativity in AI-generated artistic work~\cite{Goetze2024ATheft}. A deeper understanding is needed of how GenAI can preserve the uniqueness and rarity valued within artisan communities.

\subsection{Implications}
The YouTube creator community is building a collective knowledge of GenAI4Money that introduces new dynamics into established HCI theories on socio-technical infrastructure and creative labor. Our framework serves as a conceptual guide summarizing how the community tends to disseminate GenAI knowledge for monetization. By making explicit the interplay between monetization and associated GenAI uses, the framework operationalizes key considerations for evaluating future GenAI technologies for creativity support and identifying associated risks for researchers, designers, and policymakers. Based on this conceptual framework (\autoref{fig:framework}), we discuss design implications along three key aspects impacted by GenAI4Money—creative labor, expertise, and creativity.

\begin{itemize}
    \item \textbf{Tradeoff between alleviating creative labor and opportunistic use of GenAI.}
    The emergence of GenAI knowledge in content creation challenges prior creative labor theory, which views human performers' routines as embedded within content-sharing infrastructures~\cite{Simpson2023CreativeWork}. In contrast, we find that GenAI4Money knowledge tends to frame GenAI as a de-laborification technique that conceals creator identity and enables opportunistic monetization. For technologies that integrate GenAI for creative workflow optimization~\cite{Sun2024CreativeWorker, Kim2024ASVG} or replace human labor with AI-generated performance~\cite{Sew2025AIInfluencer, Gamage2022Deepfake}, our conceptualization suggests that these designs should consider whether GenAI tools help alleviate extraneous workloads (such as managing creators' public presentation or selecting trending titles), while also supporting value expression (e.g., topic research and aligning AI-generated content with creators' authentic interests). A future direction is to design evaluation metrics that assess how the tool amplifies creative agency while reducing creators' tendency to use it for opportunistic creation. Our framework also highlights the need to re-examine how revenue-sharing and moderation infrastructures (e.g., subscription and advertisement models) intersect with GenAI use~\cite{Ma2023Moderation, Dunna2022Demonetization}. Platform policymakers need to develop explicit community guidelines that discourage the use of GenAI to misappropriate original content (e.g., through modification or reposting)~\cite{Fiesler2016Copyright}. To reduce opportunistic creation with GenAI, monetization programs should de-emphasize the quantity of production and instead guide creators to assess how GenAI-generated content influences user engagement and what forms of GenAI content are acceptable within the community.

    \item \textbf{Tradeoff between expertise enhancement and expertise faking.} 
    Prior work has shown that creative agency and responsibility are reshaped by GenAI~\cite{Hwang2022GenAI, Yildirim2022AIDesign}. The spread of knowledge about expertise enhancement and expertise faking in GenAI4Money introduces new tensions between novice and expert creators. GenAI lowers skill barriers and enables novices to generate plausible content~\cite{Sun2024CreativeWorker}, but it may also introduce AI hallucinations or inaccuracies~\cite{Kim2024ASVG} into niche communities. Based on the GenAI4Money framework, we propose three design recommendations. First, GenAI tools need to be built with consideration of creators' levels of expertise. For professional creators, GenAI tools should enhance expertise by supporting idea discovery, trend alignment, and topic selection. For novice creators, GenAI tools should be designed to encourage the foregrounding of their expertise, provide guidance on questioning the authority and quality of AI-generated content, and promote responsibility for AI-created content. Second, because AI outputs still exhibit quality problems~\cite{Kim2024ASVG} and we observed cases of non-verification in our study, GenAI tools should prompt creators to verify accuracy and quality -- for example, by highlighting uncertainty in AI-generated scripts or blogs and encouraging human validation. Third, as GenAI tools are rapidly adopted for analyzing trending content (e.g., for SEO) and shaping monetization strategies, platforms should build GenAI tools to support standard methods for working with algorithm with GenAI to reduce visibility inequality~\cite{Weber2021KindArt}.

    \item \textbf{Tradeoff between AI creativity and human creativity.} Our framework should foster design considerations that balance GenAI creativity with human creativity. While prior work argues that AI reshapes rather than replaces creative agency~\cite{Hwang2022GenAI}, the rise of GenAI knowledge within transaction and advertisement models requires stakeholders to reassess the boundaries of GenAI creativity. First, because monetization is a common creator practice~\cite{Rieder2023Creator, Mathur2018Endorsements}, we argue that GenAI production should shift from mimicking human-created content toward forms that are recognizably GenAI yet still creative and trending (e.g., trending AI art). Such media knowledge and practices must be examined to help creators develop healthy GenAI knowledge and skills~\cite{DingBilibili}. For example, dedicated channels and platforms for sharing GenAI-enabled creativity could be developed. Second, beyond examining audiences' ability to identify AI content~\cite{Wang2025Boundary}, HCI research should assess how new forms of GenAI creativity align with users' and advertisers' interests, ensuring that GenAI complements rather than replaces human creativity. Business advertisers should have options to determine whether permitting GenAI content may redirect traffic to their brands. Buyers on handcraft marketplace platforms should also be informed about whether GenAI is used. Third, our research highlights the need for new knowledge to guide creators in being aware of transparency in GenAI usage. Creativity-faking behaviors (e.g., selling AI art as an original piece or modifying others' original work) should be mitigated. Platforms should also establish clear guidelines that allow creators to declare GenAI use in their creative work and that inform buyers when an art piece is generated by AI.

\end{itemize}

\section{Conclusion}
In conclusion, this study offers a foundational conceptual understanding of YouTubers' collective knowledge surrounding the use of GenAI for monetization, contributing valuable insights into the rapid growth of GenAI in user-generated content~\cite{Hua2024GenAIUGC}. We identified 10 GenAI4Money use cases and examined their alignment with monetization models and the creative labor challenges they introduce. We conclude with a framework that highlights how GenAI4Money shapes content-sharing infrastructures and creative labor. This framework guides future GenAI tool design, GenAI adoption, and policymaking by clarifying the tradeoffs between creative labor and opportunistic use, expertise enhancement and expertise faking, and AI creativity and human creativity. Although this work captures the breadth of GenAI in monetization, we acknowledge several limitations and outline directions for future research.
\par

First, our dataset captures on-screen demonstrations and claims about GenAI monetization strategies, not direct logs of creators' off-screen practices or verified income. The videos reveal patterns of YouTubers' collective knowledge and perceived effective ways of using GenAI for monetization, but they do not confirm the actual workflows or any other potentially unethical uses of GenAI. Future interviews and surveys with YouTube creators could further uncover the rationales for GenAI adoption beyond monetization.

\par
Second, our data collection focused primarily on YouTube, though GenAI is also extensively used on other platforms such as TikTok, Instagram, and Twitter. From the audience's perspective, researchers should examine how GenAI-driven, de-identified content reshapes parasocial relationships and influences long-term trust in AI-mediated creators across platforms~\cite{WohnParasocialInteraction}. 

\bibliographystyle{ACM-Reference-Format}
\bibliography{references}

\end{document}